\documentclass[12pt]{article}
\usepackage{axodraw}
\usepackage{epsfig}   
\usepackage{amssymb}
\usepackage{latexsym}
\usepackage{amsmath}

\usepackage{graphics,subfigure} 

\usepackage{color}   

\newcommand{\lsp} {\tilde{\chi}_1^0}
\newcommand{\mlsp} {m_{\tilde{\chi}_1^0}}

\def\chiapm{\tilde{\chi}_1^{\pm}}

\def\br{\mathrm{BR}}

\newcommand{\mtone} {m_{{\tilde t}_1}}

\newcommand{\mbone} {m_{{\tilde b}_1}}

\newcommand{\msql} {m_{{\tilde q}_L}}
\newcommand{\mlepl} {m_{{\tilde \ell}_L}}

\newcommand{\mgl} {m_{\tilde g}}

\newcommand{\mnulepl} {{\tilde m}_{\nu_L}}

\newcommand{\mtauone} {m_{{\tilde \tau}_1}}

\newcommand{\eps} {\epsilon}

\newcommand{\tanb} {\tan\beta}

\newcommand{\hpm} {h^{\pm}}

\def\met        {E\!\!\!\!/_T}

\newcommand{\lsim}{\raisebox{-0.13cm}{~\shortstack{$<$ \\[-0.07cm] $\sim$}}~}
\newcommand{\gsim}{\raisebox{-0.13cm}{~\shortstack{$>$ \\[-0.07cm] $\sim$}}~}
\newcommand{\bea}{\begin{eqnarray}}
\newcommand{\eea}{\end{eqnarray}}
\newcommand{\beq} {\begin{equation}}
\newcommand{\eeq} {\end{equation}}

\newcommand{\nn}{\nonumber}

\begin{document}
\pagestyle{empty}
\begin{flushright}
\today
\end{flushright}
\begin{center}
{\large\sc {\bf Light neutral CP-even Higgs boson within Next-to-Minimal Supersymmetric 
Standard model (NMSSM) at the Large Hadron electron Collider (LHeC)}}
\end{center}
\vspace{1.0truecm}
\begin{center}
{\large Siba Prasad Das \footnote{Email: sp.das@uniandes.edu.co} and Marek Nowakowski$^{1,}$\footnote{Email: mnowakos@uniandes.edu.co}}\\
\vspace*{5mm}
{}$^1${\it  
Department of Physics, Faculty of Science, \\[0.15cm]
Universidad de los Andes, Apartado Aereo 4976-12340, Carrera 1 18A-10, \\[0.16cm]
Bogota -- Colombia.} \\[0.07cm]
{}$^2${\it 
M. Smoluchowski Institute of Physics,
Jagiellonian University,
ul. St. Lojasiewicza 11,
30-348 Krak\'{ o}w -- 
Poland.
} \\[1.0cm]
\end{center}

\begin{abstract}
  We analyze the prospects of observing the light CP-even neutral 
  Higgs bosons ($h_1$) in their decays into $b \bar b$ quarks, in the 
  neutral and charged current production processes $e h_1 q$ and $\nu h_1 q$ 
  at the upcoming LHeC, with $\sqrt s \approx 1.296$ TeV. Assuming that the intermediate 
  Higgs boson ($h_2$) is Standard Model (SM)-like, we study the Higgs production within 
  the framework of NMSSM. We consider the constraints from Dark-matter, Sparticle masses, 
  and the Higgs boson data. The signal in our analysis can be classified as three jets, 
  with electron (missing energy)coming from the neutral (charged) current interaction. We demand 
  that the number of b-tagged jets in the central rapidity region be greater or equal to two. 
  The remaining jet is tagged in the forward regions. With this forward jet and 
  two $b$-tagged jets in the central region, we reconstructed three jets invariant masses. 
  Applying some lower limits on these invariant masses turns out to be an essential criterion 
  to enhance the signal--to--background rates, with slightly different sets of kinematical 
  selections in the two different channels. We consider almost all reducible and irreducible 
  SM background processes. We find that the 
  non-SM like Higgs boson, $h_1$, would be accessible in some of the NMSSM benchmark points, at
  approximately 0.4$\sigma$ (2.5$\sigma$) level in the $e$+3j channel up to Higgs boson masses of 75 GeV  
  and in the $E\!\!\!\!/_T$+3j channel could be discovered with 1.7$\sigma$ (2.4$\sigma$) level 
  up to Higgs boson masses of 88 GeV with 100 fb$^{-1}$ of data in a simple cut-based 
  (with optimization) selection. With ten times more data accumulation at the end of the LHeC run
  and using optimization, one can have 5$\sigma$ discovery in the electron (missing energy) channel 
  up to 85 (more than 90) GeV.
\end{abstract}

\newpage
\setcounter{page}{1}
\pagestyle{plain}

\section{Introduction}
\label{sec:intro}

It is expected since long that the mechanism that triggers the electroweak symmetry breaking
(EWSB) and  generates the fundamental particle masses will involve
at least two experimental parts. The first one is the search and the observation of a spin-zero Higgs particle
that will confirm the scenario of the minimal SM (which has one Higgs isospin doublet)
of Glashow-Weinberg-Salam and most of its extensions. This confirms
a spontaneous symmetry breaking by a scalar field that develops a non-zero vacuum expectation
value (vev). This part has recently been closed by the ATLAS and CMS experiments \cite{ATLASdiscovery,CMSdiscovery} at
Large Hadron Collider (LHC) with the spectacular observation of a new boson with 
present central mass value around $125.09 \pm  0.21 \pm  0.11$ GeV. The width,
and the couplings to all SM particles and the CP-quantum numbers are also known. All this seems consistent   
with the symmetry breaking mechanism in the SM and opens up the second part: are there any other scalars from 
beyond the SM model, which would participate in the symmetry breaking ? This second part is mandatory in order to 
establish the exact nature of the electroweak symmetry breaking (EWSB) mechanism and, eventually, 
identify the effects of new physics beyond the SM.

The original idea of having scalars in the model is, of course, the spontaneous 
breaking of the electroweak gauge group. A detailed overview has been given in \cite{Wells:2009kq}  
and in particular the non-standard way of the EWSB in \cite{Grojean:2006wr}. However, scenarios where 
the scalar do not participate in the EWSB do exist (see for instance, neutrino models where the vevs  
of singlet scalars breaks spontaneously the lepton number \cite{Dev:2015vra}).

Also worth mentioning are the higher dimensional theories, based on the Standard Model gauge group 
\cite{Seidl:2004xw} where the electroweak constraints can be consistent with experimental 
results, even without Higgs boson. In this kind of models, the electroweak symmetry 
is broken by the boundary conditions and the choices of compactification scales  
lead to the masses of the gauge bosons.

The available theoretical models at our disposal are many: generic two Higgs doublet Model (2HDM) 
\cite{Branco:2011iw,Gunion:2002zf} and various flavor violating Yukawa-textured models \cite{Cordero-Cid:2013sxa}, 
Minimal Supersymmetric Standard Model (MSSM)\cite{Kane:1993td}, non-minimal realization of supersymmetric models,
and models with additional singlets, the Next-to-Minimal Supersymmetric SM (NMSSM), 
doublets and/or triplets \cite{nmssm,Drees:1988fc, Franke:1995tc,Maniatis:2009re,review} and
non-minimal NMSSM type of models, e.g., in \cite{Diessner:2015iln}, \cite{Hagimoto:2015tua}. 
Hence, one of the most important tasks for experimentalists and theorists is to find ways to 
either exclude or confirm aspects of these models which may have varieties of 
signatures for the different Higgs production and decay channels at the present and the upcoming 
collider experiments. 

From the perspective of model building, the NMSSM \cite{nmssm,Drees:1988fc,Franke:1995tc,Maniatis:2009re,review} 
is ideally suited to search for new physics as its gauge group is the same as in the SM  and thus it can easily 
accommodate the SM-like discovery without any unnatural fine-tuning of its parameters. Although the MSSM 
contains less free parameters than the NMSSM, the SM Higgs boson type signal can also be easily 
accommodated in the latter model whereas some amount of fine tuning is necessary for the MSSM. 
Some variants of the NMSSM models also have nice features of the Higgs sector. Worth mentioning 
is the model with a slightly broken PQ-symmetry \cite{Miller:2005qua} and the 
so-called $\lambda$-NMSSM \cite{Barbieri:2013hxa}.

It is worth pointing out that in the NMSSM the upper limit of the lightest SM-like Higgs boson mass
is lifted up to 155 GeV (as compared to 139 GeV in the MSSM). Secondly, the problem with the absence of Sparticle
signals may easily be explained by different Supersymmetry (SUSY) cascade decays occurring in the NMSSM,
owing to an additional singlino entering as the last step and thereby inducing topologies to
which present SUSY searches are less sensible. Furthermore, as is well known,
the MSSM suffers from the so-called $\mu$-problem, i.e., the Higgs(ino) mass term entering the soft SUSY
Lagrangian ought to be manually set at the EW scale in order to achieve EWSB, while SUSY itself would
require it to be at the Planck scale (or else be zero, in virtue of some postulable additional symmetry)
\cite{Kim:1983dt,Ellis:1988er}. This is elegantly remedied in the NMSSM, since herein the aforementioned 
soft term is replaced by the vev of an additional Higgs singlet state, which appears naturally at the  
EW scale. In turn, this implies that the Higgs sector of the NMSSM is very rich. In fact, while only one
Higgs boson exists in the SM and five Higgs bosons in the MSSM, there are seven such states
in the NMSSM: three CP-even Higgses $h_{1, 2, 3}$ ($m_{h_1} < m_{h_2} < m_{h_3}$), two CP-odd
Higgses $a_{1, 2}$ ($m_{a_1} < m_{a_2} $) and two charged Higgses $h^\pm$.

As the SM is embedded within any two--Higgs doublet model, the recently discovered SM-Higgs boson 
can be part of the spectrum. Generically, this SM-type Higgs is either the lightest CP-even neutral
one or the second-lightest one. Light as well as heavy Higgs boson phenomenology within the MSSM
has been studied extensively in \cite{Bechtle:2016kui}. Having many free parameters in NMSSM, the masses
of the Higgs bosons vary in a wide range so that their decay branching ratios in various modes can also
vary widely. The Higgs boson masses together with their couplings to gauge boson and/or fermions
are most important to identify the viable modes to look for the Higgs boson in any collider experiment.
From the theoretical perspective, the two-loop corrected Higgs boson masses   
and couplings to quarks and gauge bosons within NMSSM have been carried out in \cite{Drechsel:2016jdg,Goodsell:2016udb}.

The NMSSM Higgs boson phenomenology at high energy colliders has been studied over a decade 
\cite{Ellwanger:2004gz,Dreiner:2012ec,Ellwanger:2013rsa,Domingo:2015eea,Barducci:2015zna,cpodd,
Potter:2015wsa,Guchait:2015owa} and direct experimental searches are reported in \cite{Khachatryan:2015nba}.

From the upcoming experiment perspective, the LHeC facility \cite{cern:lhec} is expected to be operational 
at CERN around 2020. It will be a Deep Inelastic Scattering (DIS) experiment at the TeV scale, with the 
center-of-mass energy of around 1.3 TeV. In comparison, another recently closed (in 2007) DIS experiment 
(the Hadron-Electron Ring Accelerator (HERA) \cite{heraphys} at DESY had a center-of-mass energy of around 320 GeV with
an integrated luminosity of around 0.5 fb$^{-1}$). The LHeC might deliver data samples of approximately 100 fb$^{-1}$
and at the end of full data accumulation with 1000 fb$^{-1}$ (with a higher detector coverage). Taking into considerations
all kinematical and detector aspect details, the overall kinematic range (in $x$ and $Q^2$) at LHeC is 
20 times larger than the HERA experiment. Other than the in-depth studies of QCD, the LHeC also has an
enormous scope to probe electroweak and Higgs boson physics \cite{Han:2009pe, Sarmiento-Alvarado:2014eha, Das:2015kea}.

One of the nicest features of almost all SUSY models, is that the neutral lightest Sparticle state is 
naturally the dark matter candidate \cite{Drees:2012ji,Sanabria:2014yva}. Within 
the standard cosmological scenario, we assumed that the dark matter candidate is the lightest  
neutralino, $\lsp$ with the correct abundance of relic density consistent with 
recent Planck measurement \cite{Planck:2015xua}. We refer to \cite{Belanger:2005kh,Badziak:2015exr} 
where the dark matter phenomenology within NMSSM has been studied. 
Some variant models are discussed in \cite{Horiuchi:2016tqw} and a discussion on low mass 
weakly interacting massive particle (WIMP) searches consistent with the Higgs boson data 
at LHC has been studied recently in \cite{Alvares:2012qv}. Sparticle co-annihilation consistent  
with the dark matter relic-density and related SUSY phenomenology in the electroweak gaugino 
sector are discussed in \cite{Das:2014kwa,Florez:2016lwi}.

To the best of our knowledge no study has been done to find non-SM type Higgs boson signals
within the NMSSM at the LHeC. In our analysis, we assumed the second intermediate Higgs boson is the SM-type 
($h_2$-SM scenario). This, of course, refers to the mass, coupling ratios and signal-strength from the recent LHC results.
The lightest non-SM Higgs boson, $h_1$, offers itself to be looked for at the upcoming LHeC.  
We identify two main production processes, namely the neutral current one $ep \to e h_1 q$ and  
the charged current one $ep \to \nu h_1 q$. We are particularly motivated by the possible branching 
ratio enhancement in the $b$-quark decay mode, i.e., $h_1 \to b \bar b$. Finally the reconstructed 
invariant mass of the two $b$-tagged jets ensures the evidence of the non-SM like Higgs boson.

The plan of this paper is as follows. In the next section we will describe briefly the NMSSM model.  
In Sec.2, we will randomly vary the NMSSM model parameters and identify the allowed parameter space  
consistent with most up-to-date theoretical, phenomenological and experimental constraints. 
For the allowed model space, we then estimate the number of non-SM type Higgs boson signal 
events, $ep \to e h_1 q$ and $ep \to \nu h_1 q$ with the decay channel $h_1 \to b \bar b$  and identify 
few high event--rated benchmark points to carry out the phenomenological analysis in Sec.3. 
In doing so, we estimate all the reducible and irreducible SM backgrounds for both of the 
signal channels under considerations. In Sec.4, we carry out a simple cut-based optimization  
to isolate the Higgs boson signals in both the channels. We summarize our findings in Sec.5.

\section{The NMSSM models}
\label{sec:NMSSM}

The NMSSM model has been described in many reviews \cite{nmssm,Drees:1988fc,Franke:1995tc,Maniatis:2009re,review}. 
However, for completeness let us mention the part relevant for our analysis (we will follow \cite{review}).

The general NMSSM contains the MSSM superfields with most general renormalizable couplings as in 
the MSSM superpotential. There is, however, one additional gauge singlet chiral superfield $\widehat{S}$.

The Higgs superpotential $W_\mathrm{Higgs}$ reads   
\beq\label{higgssuper}
W_\mathrm{Higgs} = (\mu + \lambda \widehat{S})\,\widehat{H}_u \cdot
\widehat{H}_d + \xi_F \widehat{S} + \frac{1}{2} \mu' \widehat{S}^2 +
\frac{\kappa}{3} \widehat{S}^3
\eeq
where $\lambda$, $\kappa$ are 
dimensionless Yukawa couplings. The bi-linear $\mu$, $\mu'$ terms are 
the supersymmetric mass terms, and $\xi_F$ with mass-dimension two  
is the supersymmetric tadpole term.

Assuming R-parity and CP-conservation (scenarios without these requirements have 
been studied in \cite{JeanLouis:2009du} and \cite{Goodsell:2016udb}) the corresponding soft 
supersymmetry breaking terms, ${\cal L}_\mathrm{soft}$ are the following:

\bea\label{2.5e}
-{\cal L}_\mathrm{soft} &=&
m_{H_u}^2 | H_u |^2 + m_{H_d}^2 | H_d |^2
+ m_{S}^2 | S |^2+m_Q^2|Q^2| + m_U^2|U_R^2| \nn \\
&&+m_D^2|D_R^2| +m_L^2|L^2| +m_E^2|E_R^2|
\nn \\
&&+ (h_u A_u\; Q \cdot H_u\; U_R^c - h_d A_d\; Q \cdot H_d\; D_R^c
- h_{e} A_{e}\; L \cdot H_d\; E_R^c\nn \\ &&
+\lambda A_\lambda\, H_u \cdot H_d\; S + \frac{1}{3} \kappa A_\kappa\,
S^3 + m_3^2\, H_u \cdot H_d + \frac{1}{2}m_{S}'^2\, S^2 + \xi_S\, S
+ \mathrm{h.c.}) \; ,
\eea

where all the parameters have the standard meanings.

The dimensionful supersymmetric parameters $\mu$, $\mu'$ and 
$\xi_F$ ((with mass dimension two) in the superpotential and the 
associated soft SUSY breaking parameters $m_3^2$, $m_{S}'^2$ and 
$\xi_S$ (with mass dimension three) have to be of the order 
of the weak or SUSY breaking scale. 

In general, these terms are non-vanishing, however a simplified 
version requiring scale invariance leads to $\mu$ = $\mu'$ = $\xi_F$ = 0 and
the superpotential takes the simple form  
\beq\label{scaleinv}
W_\mathrm{scale-invariant} = \lambda \widehat{S}\,\widehat{H}_u \cdot
\widehat{H}_d + \frac{\kappa}{3} \widehat{S}^3
\eeq
together with the parameters $m_3^2$, $m_{S}'^2$ and $\xi_S$ in
(\ref{2.5e}) also set to zero. An effective $\mu$-term of the 
order of weak scale is generated from the vev $s$ of $\widehat{S}$:
\beq\label{2.7e}
\mu_\mathrm{eff} = \lambda s\; ,
\eeq

From the supersymmetric gauge interactions, the $F$-term and the soft supersymmetry 
breaking terms one can obtain the Higgs potential:
\bea
V_\mathrm{Higgs} & = & \left|\lambda \left(H_u^+ H_d^- - H_u^0
H_d^0\right) + \kappa S^2 + \mu' S +\xi_F\right|^2 \nn \\
&&+\left(m_{H_u}^2 + \left|\mu + \lambda S\right|^2\right)
\left(\left|H_u^0\right|^2 + \left|H_u^+\right|^2\right)
+\left(m_{H_d}^2 + \left|\mu + \lambda S\right|^2\right)
\left(\left|H_d^0\right|^2 + \left|H_d^-\right|^2\right) \nn \\
&&+\frac{g_1^2+g_2^2}{8}\left(\left|H_u^0\right|^2 +
\left|H_u^+\right|^2 - \left|H_d^0\right|^2 -
\left|H_d^-\right|^2\right)^2
+\frac{g_2^2}{2}\left|H_u^+ H_d^{0*} + H_u^0 H_d^{-*}\right|^2\nn \\
&&+m_{S}^2 |S|^2
+\big( \lambda A_\lambda \left(H_u^+ H_d^- - H_u^0 H_d^0\right) S +
\frac{1}{3} \kappa A_\kappa\, S^3 + m_3^2 \left(H_u^+ H_d^- - H_u^0
H_d^0\right) \nn \\
&& +\frac{1}{2} m_{S}'^2\, S^2 + \xi_S\, S + \mathrm{h.c.}\big)
\label{2.9e}
\eea
where $g_1$ and $g_2$ are $U(1)_Y$ and $SU(2)$ gauge couplings, respectively.

The physical neutral Higgs fields (with index~R for the CP-even, index~I
for the CP-odd states) are obtained by expanding the full 
scalar potential (\ref{2.9e}) around the real neutral vevs 
$v_u$, $v_d$ and $s$: 
\beq\label{2.10e}
H_u^0 = v_u + \frac{H_{uR} + iH_{uI}}{\sqrt{2}} , \quad
H_d^0 = v_d + \frac{H_{dR} + iH_{dI}}{\sqrt{2}} , \quad
S = s + \frac{S_R + iS_I}{\sqrt{2}}\; ;
\eeq
where the vevs have to be obtained from the minima of
\bea
V_\mathrm{Higgs} & = & \left(-\lambda v_u v_d + \kappa s^2 + \mu' s
+\xi_F\right)^2 +\frac{g_1^2+g_2^2}{8}\left(v_u^2 - v_d^2\right)^2
\nn \\
&&+\left(m_{H_u}^2 + \left(\mu + \lambda s\right)^2\right) v_u^2
+\left(m_{H_d}^2 + \left(\mu + \lambda s\right)^2\right) v_d^2
\nn\\
&&+m_{S}^2\, s^2 -2 \lambda A_\lambda\, v_u v_d s +  \frac{2}{3}
\kappa A_\kappa\, s^3 - 2m_3^2\, v_u v_d + m_{S}'^2\, s^2 +
2\xi_S\, s \;,
\label{2.11e}
\eea

The minimization of (\ref{2.11e}) with respect the the three vevs and the  
proper electroweak symmetry breaking (generating the correct $Z$-boson 
mass) leads to the following input parameters: 

\beq\label{2.17e}
\lambda,\ \kappa,\ A_\lambda,\ A_\kappa,\ \tan\beta,\
\mu_\mathrm{eff},
\eeq
to which one has to add the (in the convention $\mu
= 0$) five parameters of the NMSSM
\beq\label{2.18e}
m_3^2,\ \mu',\ m_{S}'^2,\ \xi_F\ \mathrm{and}\ \xi_S\; .
\eeq

The tree level Higgs mass matrices are obtained by expanding the full
scalar potential~(\ref{2.9e}) around the real neutral vevs $v_u$, $v_d$
and $s$ as in (\ref{2.10e}). The elements of the $3 \times 3$ CP-even mass matrix ${\cal
M}_S^2$ are conveniently written in the basis $(H_{dR}, H_{uR}, S_R)$ after the elimination
of $m_{H_d}^2$, $m_{H_u}^2$ and $m_{S}^2$. 

The basis for the elements of the $3 \times 3$ CP-odd mass matrix ${\cal M'}_P^2$ are 
$(H_{dI}, H_{uI}, S_I)$. Dropping the Goldstone mode, in the remaining $2 \times 2$ CP-odd 
mass matrices one can use the doublet ($M_A$) and singlet component ($M_P$) mass parameters 
as inputs together with the $\mu_{eff}$.

Our model under consideration is not the $Z_3$ invariant NMSSM, but rather a 
general phenomenological NMSSM. However, by setting: $m_3^2 = m_{S}'^2 = \xi_S = \mu = \mu' =\xi_F = 0$   
in the general phenomenological NMSSM, one recovers the $Z_3$ invariant NMSSM. 

\section{The NMSSM parameter spaces}
\label{sec:NMSSMparam}

We used the package \texttt{NMSSMTools~5.0.1}~\cite{Ellwanger:2004xm} to obtain the Sparticle spectrum, 
decay branching ratios and various low energy observables. 

We randomly scanned approximately $10^6$ points. The varied parameters and their ranges are 
{\footnote{All the masses and mass parameters in our analysis are in GeV.}} 
tabulated in Table  \ref{tab:param}.

\begin{table}[t!]
\centering
{\scriptsize
\begin{tabular}{||c||c|c||}
\hline
Parameters&Min&Max\\
\hline
\hline
$\lambda$&0.001& 0.7\\
$\kappa$& 0.001& 0.7\\
$A_{\lambda}$&100.0& 2500.0\\
$A_{\kappa}$& -2500.0& 100.0\\
$\tanb$&1.5 & 60.0 \\
$\mu_{eff}$& 100.0& 500.0\\
$M_1$& 50.0& 400.0\\
$M_2$& 50.0& 500.0\\
$\msql$& 300.0& 1500.0\\
$A_t$=$A_b$& -4000.0& 1000.0\\
$M_A$& 100.0& 500.0\\
$M_P$& 100.0& 3000.0\\
\hline
\hline
\end{tabular}
}
\caption{The minimum and maximum values of varied NMSSM parameters. 
The following parameters remain fixed: $M_3$ = 1800.0 GeV (this allows the gluino mass $\mgl$ 
to be above the mass limits from recent LHC-run2); $m_{\tilde \ell}$ = 300.0 GeV 
(for all three generation as well as left and right state) and $A_{\tau}$=$A_e$=$A_{\mu}$ = 1500.0 GeV.
Here $M_A$ ($M_P$) is the Doublet(Singlet) component of the CP-odd Higgs mass matrices.  
}
\label{tab:param}
\end{table}

\begin{figure}[ht!]
\begin{center}
\raisebox{0.0cm}{\hbox{\includegraphics[angle=-90,scale=0.33]{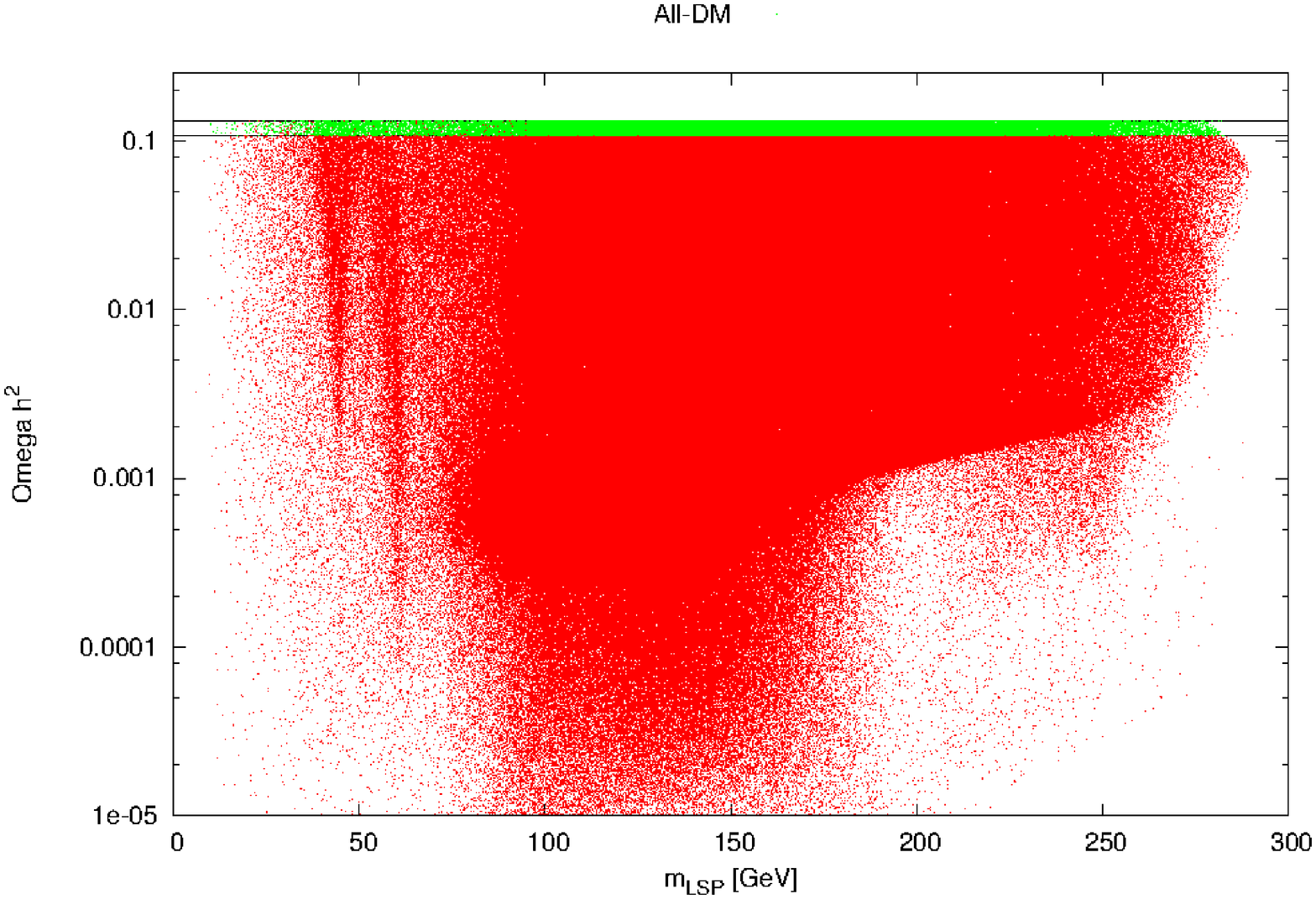}}}
\raisebox{0.0cm}{\hbox{\includegraphics[angle=-90,scale=0.33]{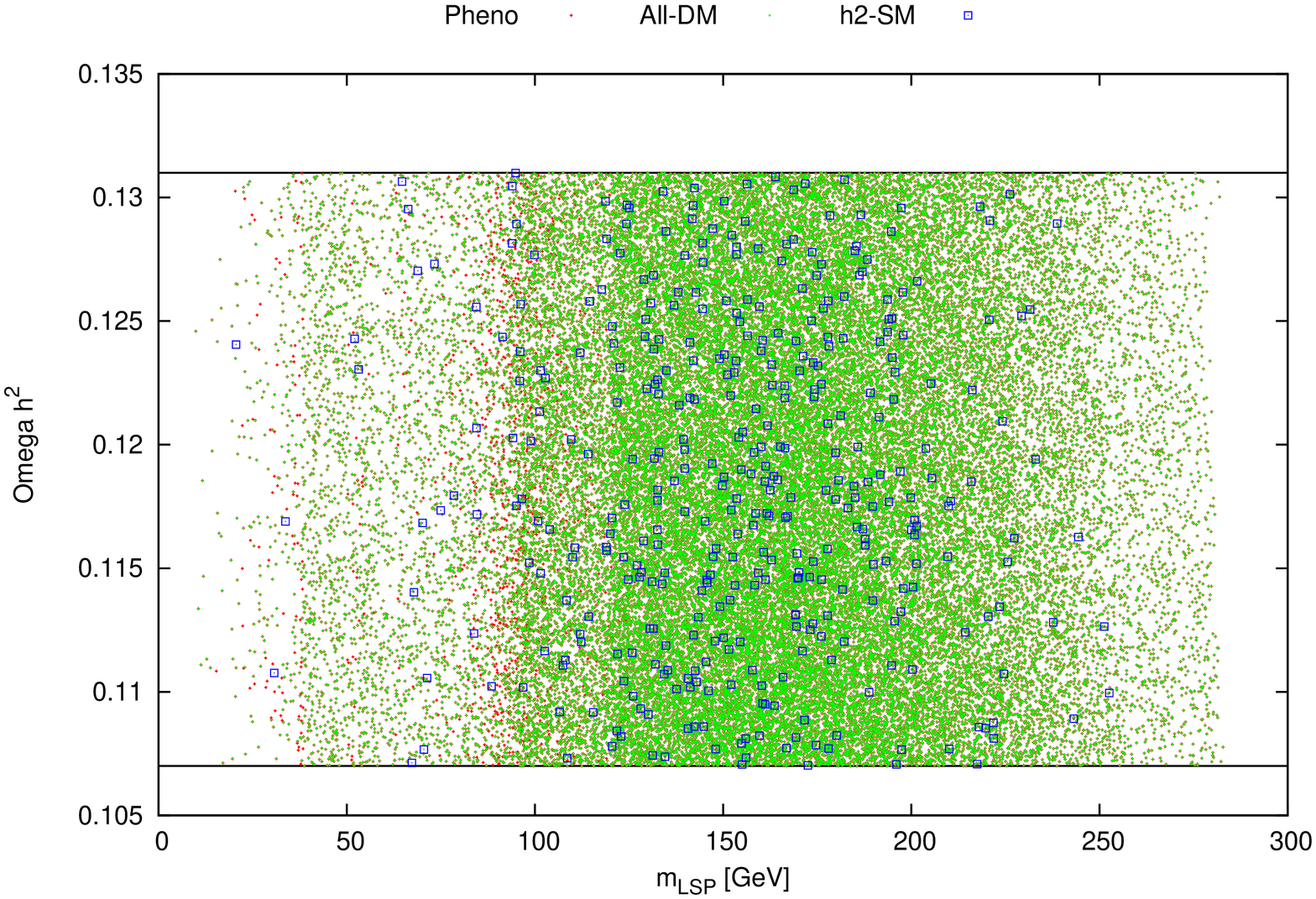}}}
\caption{Left-panel: The Dark matter relic density ($\Omega_{\lsp} h^2$) as a function 
of lightest neutralino masses($\mlsp$) within the standard cosmology. The two lines 
represent the upper ($\Omega_{\lsp} h^2$=0.131) and lower ($\Omega_{\lsp} h^2$=0.107) 
bounds from the Planck measurements \cite{Planck:2015xua}. The first deep around 45 GeV 
is due to the Z-boson exchange (annihilation diagram) and the second around 63 GeV, 
from the Higgs-boson ($h_2$-SM) exchange annihilation within the NMSSM parameter spaces. 
The green point within the strips satisfy the direct and indirect Dark Matter searches 
results and termed as ``All-DM''. Right-panel: we only show the relic-density allowed parameter 
space together with various other constraints (in the legends), see the text for details.
}
\label{dm12}
\end{center}
\end{figure}

\begin{figure}[ht!]
\begin{center}
\raisebox{0.0cm}{\hbox{\includegraphics[angle=-90,scale=0.33]{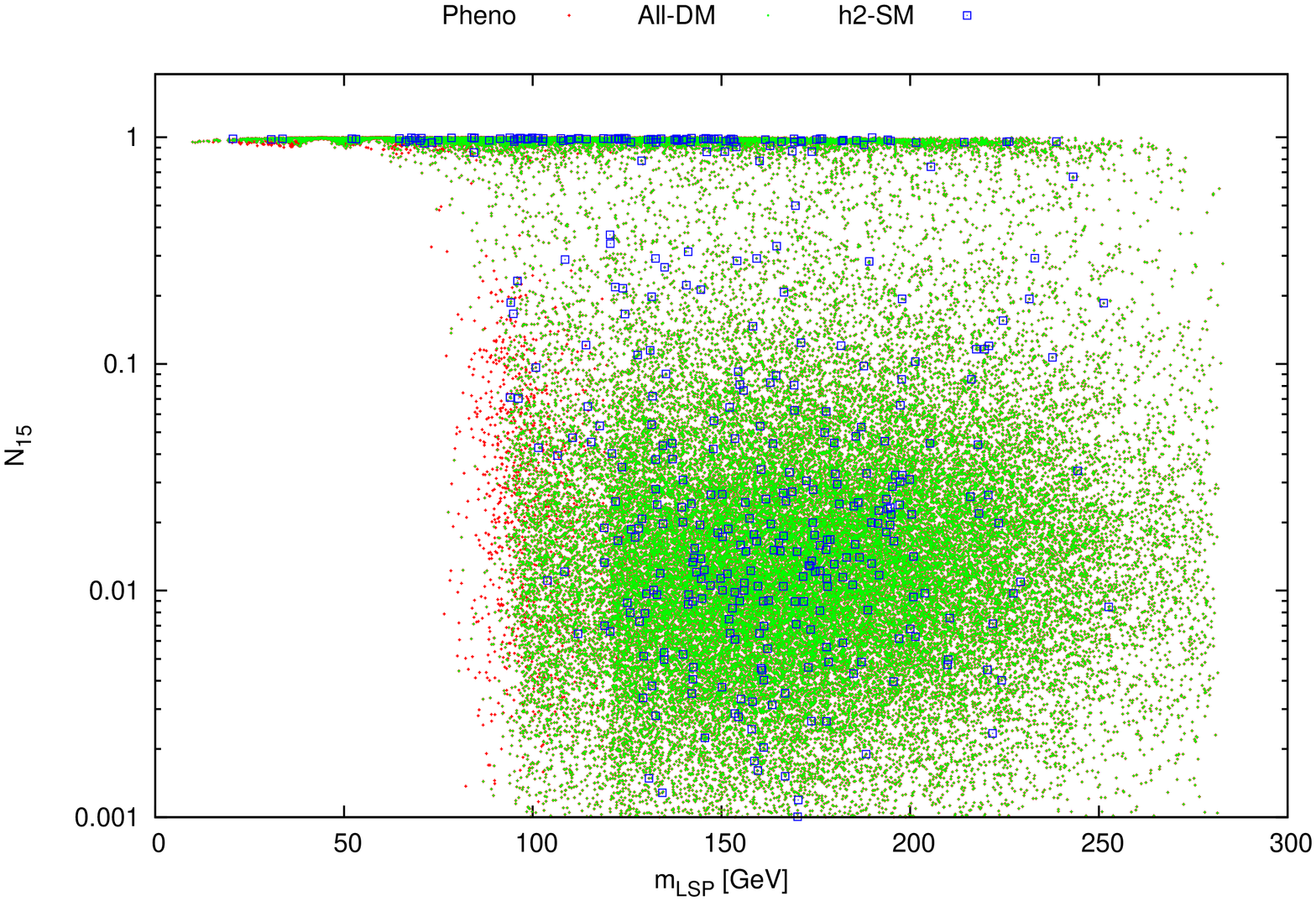}}}
\raisebox{0.0cm}{\hbox{\includegraphics[angle=-90,scale=0.33]{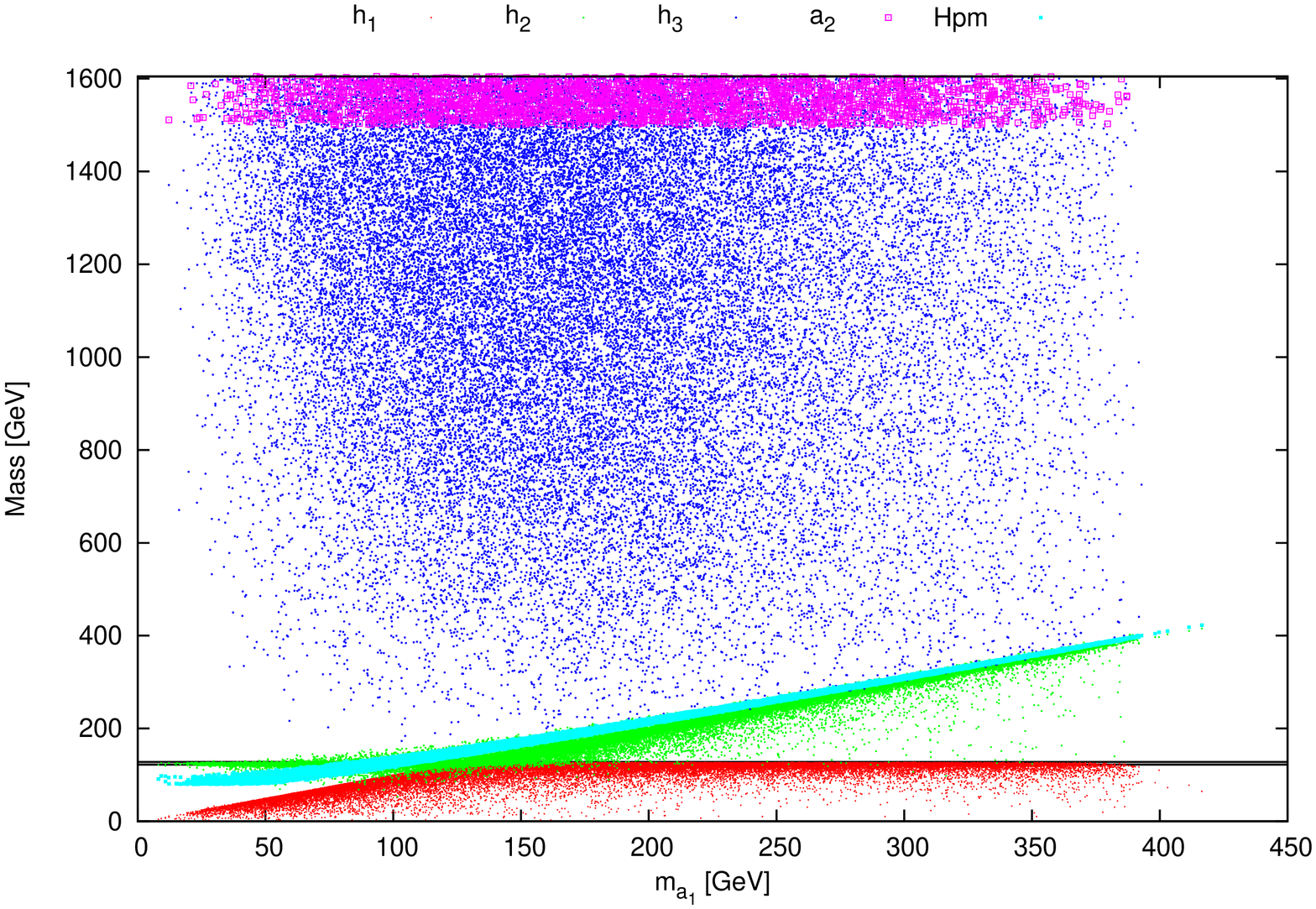}}}
\caption{Left-panel: The Singlet component ($N_{15}$) of the lightest neutralino ($\lsp$) consistent 
with the dark matter relic density ($\Omega_{\lsp} h^2$) and various other constraints (see text for details). 
Right-panel: The masses of the Higgs boson masses, $h_1$, $h_2$, $h_3$, $a_2$ and $\hpm$ (``Hpm'') as a function 
of the lightest CP-odd Higgs bosons masses ($m_{a_1}$). The mass of the $a_2$ is shown up to 1600.0 GeV,  
extends, however, even beyond this point in this allowed NMSSM model parameter spaces.}
\label{lspcomp}
\end{center}
\end{figure}

\begin{figure}[ht!]
\begin{center}
\raisebox{0.0cm}{\hbox{\includegraphics[angle=-90,scale=0.33]{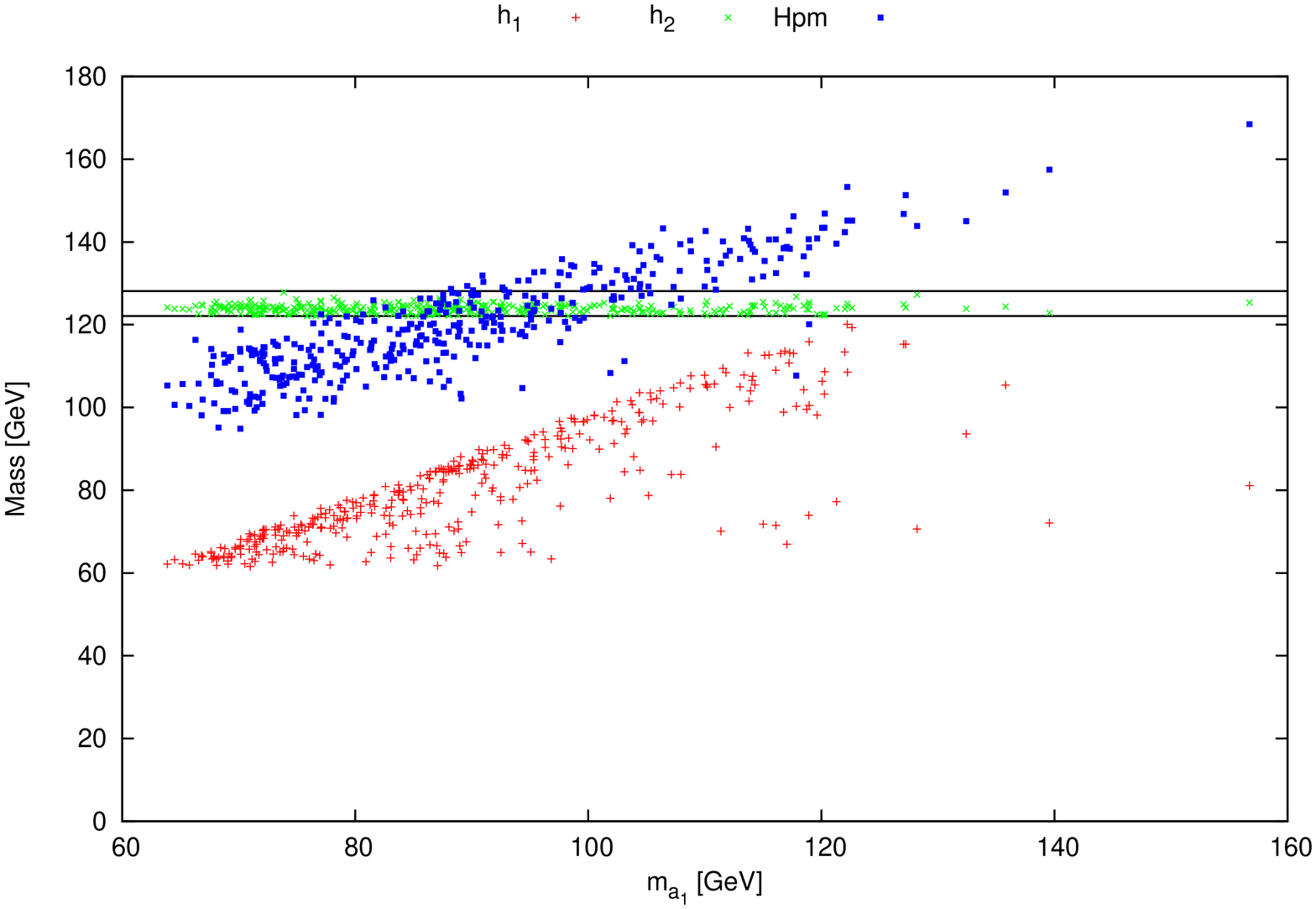}}}
\raisebox{0.0cm}{\hbox{\includegraphics[angle=-90,scale=0.33]{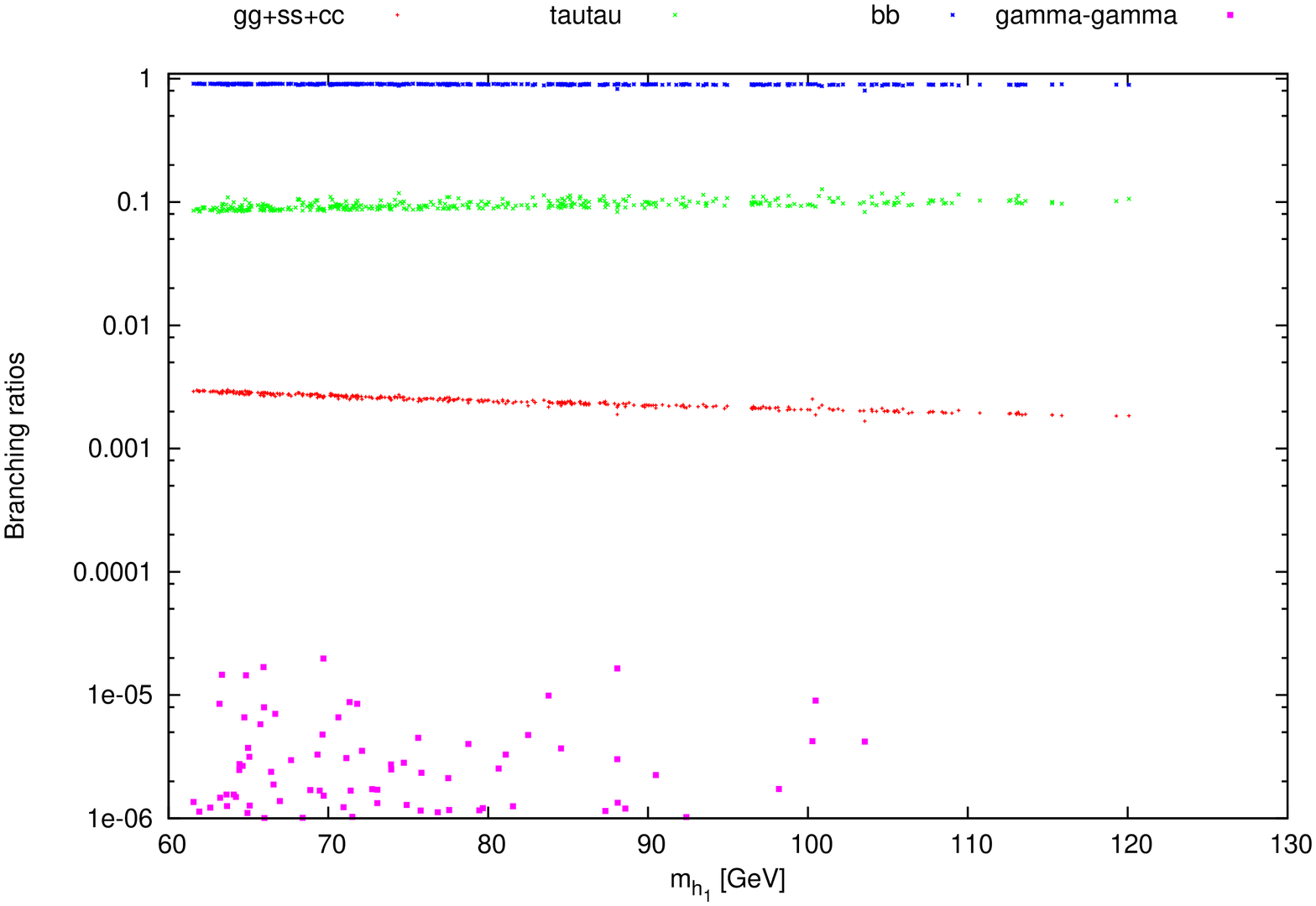}}}
\caption{Left-panel: The Higgs boson masses, $h_1$, $h_2$, and $\hpm$ as a function of the lightest CP-odd Higgs 
bosons mass ($m_{a_1}$) for $h_2$-SM scenario and consistent with all other constraints (see details in the text). 
In the right panel, we show the branching ratio of the $h_1$ in different channels and it is clear that in  large 
region of allowed parameter space the branching ratio of $h_1 \to b \bar b$ is above 90\%.}
\label{mhcons}
\end{center}
\end{figure}

For each randomly generated parameter spaces, we invoke the following constraints: 

\begin{description}

\item[Perturbative bounds:] We first imposed, 
$\lambda^2 + \kappa^2$ $\lsim$ $(0.7)^2$ \cite{Zheng:2014loa} and if not satisfied we discard the 
parameter space and generate the next random model space. 

\item[Dark Matter relic density:] 
We required that the lightest neutralino relic density will be: 0.107 $<$ $\Omega_{\lsp} h^2$ $<$ 0.131, consistent 
with the Planck measurement~\cite{Planck:2015xua} within standard cosmology.  
The estimated relic density ($\Omega_{\lsp} h^2$) as a function of the $m_{\lsp}$ 
has been shown in the left panel of Fig.\ref{dm12} with constraining only the upper 
limits, i.e., $\Omega_{\lsp} h^2$ $<$ 0.131. The green-marked points within the upper and lower strips 
are consistent with the direct--detection and indirect--detection bounds (in the legends termed as: ``All-DM''). 

The \texttt{NMSSMTools~5.0.1} is interfaced with \texttt{micrOMEGAs v4.3} ~\cite{micr43,Belanger:2013oya} 
to estimate the observed dark matter relic density, their direct detection, and indirect detection limits. 
It is to be noted that the standard and non-standard cosmological implication in the dark matter 
relic density has been analyzed within the NMSSM in \cite{Barducci:2015zna,Belanger:2005kh,Badziak:2015exr}.  

\item[Higgs bounds:]
We demand that the intermediate Higgs boson($h_2$) should be SM-like and 
its mass should be within the range of $ 125.09\ \mathrm{GeV} < m_{h_2} < 128.09\ \mathrm{GeV} $ 
(taken into consideration the 3 GeV error in theoretical estimates). 
Its coupling ratios to other SM particles should also be consistent with 
the LHC-run1 ATLAS and CMS combined study \cite{Sanabria:2015mxh,Khachatryan:2016vau}. 
The allowed coupling ratios and the signal strengths considered in our analysis 
has been tabulated in Table ~\ref{tab:coupsigma}.
We have also taken the constraint on the invisible branching ratio: $BR(h_{SM} \rightarrow invisible)$ $\lsim$ 0.25  
\cite{Khachatryan:2016whc,Aad:2015pla} {\footnote{The SM-like Higgs boson invisible decay within the NMSSM has been 
studied recently in \cite{Butter:2015fqa}.}}. Furthermore, we required $m_{\hpm} > 80.0\ \mathrm{GeV} $.

\begin{table}[t!]
\centering
{\scriptsize
\begin{tabular}{||c||c|c||c||c|c||}
\hline
Parameters&Min&Max & Parameters&Min&Max \\
\hline
\hline
$\kappa_W$&0.81&0.99 & $\mu_{VBF}^{\tau \tau}$&0.50&2.10 \\
$\kappa_{t}$&0.99&1.89 & $\mu_{ggF}^{\tau \tau}$&-0.20&2.20\\
$|\kappa_{\gamma}|$&0.72&1.10 & $\mu_{VH}^{bb}$&0.00&2.00\\
$|\kappa_{g}|$&0.61&1.07 & $\mu_{ttH}^{bb}$&-0.90&3.10\\
$|\kappa_{\tau}|$&0.65&1.11 & $\mu_{VBF}^{WW}$&0.40&2.00\\
$|\kappa_{b}|$&0.25&0.89 &$\mu_{ggF}^{ZZ}$&0.51&1.81\\
$Br(h_{SM} \to inv.)$& & 0.25 &$\mu_{VBF}^{\gamma\gamma}$&0.30&2.30\\
&& &$\mu_{ggF}^{\gamma\gamma}$&0.66&1.56\\
\hline
\hline
\end{tabular}
}
\caption{The couplings ($\kappa$) and signal strength ($\mu$) have been allowed within 
2$\sigma$ ranges (except $\kappa_W$) from the combined ATLAS and CMS measurements 
\cite{Khachatryan:2016vau}, following Table 17 (upper panel) and Table 8 respectively. 
 }
\label{tab:coupsigma}
\end{table}

\item[LEP bounds:]
Direct SUSY searches at the LEP have set bounds on superpartners, 
in particular the lighter chargino should satisfy $m_{\chiapm} >$ 103.5 GeV.
The other one refers to the $Z$ invisible width and should satisfy 
$\Gamma_Z^\mathrm{inv}$ $< 2~$ MeV at 95\% C.L.~\cite{ALEPH:2005ab}.  
When the decay channel into $\lsp\lsp$ opens, this width may exceed the experimental value 
{\footnote{A light Higgs would significantly affect the muon anomalous magnetic moment $a_\mu=(g_\mu-2)/2$, 
whose most accurate measurement comes from the E821 experiment~\cite{Bennett:2004pv}.  
Having the large theoretical uncertainties with the measurements, we have not consider 
this constraints in our numerical scan and subsequent analysis.}}. 

\item[B physics bounds:] The rare decays of $B$ meson, such as $B_s\to \mu^+\mu^-$, $B^+ \to \tau^{+} \nu$, and $B_s\to X_s\gamma$ 
lead to the flavor constraints. In our analysis, we set the recent experimental results 
at $95\%$ C.L.: $1.7 \times 10^{-9} < \br(B_s \to \mu^{+} \mu^{-}) < 4.5 \times 10^{-9}$~\cite{Amhis:2014hma}, 
$0.85 \times 10^{-4} < \br(B^+ \to \tau^{+} \nu) < 2.89 \times 10^{-4}$~\cite{Lees:2012ju}, and 
$2.99 \times 10^{-4} < \br(B_s\to X_s\gamma) < 3.87 \times 10^{-4}$~\cite{Amhis:2014hma}.

\item[Sparticle masses:] We have set the following lower bounds 
from the superparticle masses following \cite{Aaboud:2016tnv}: 
$\mgl$ $\gsim$ 1600.0 GeV, 
$\mtone$ $\gsim$ 95.0 GeV, $\mbone$ $\gsim$ 325.0 GeV, 
$\msql$ $\gsim$ 600.0 GeV, $\mlepl$ $\gsim$ 100.0 GeV,  
$\mnulepl$ $\gsim$ 90.0 GeV and $\mtauone$ $\gsim$ 87.0 GeV.  

\end{description}
  
If the randomly generated NMSSM model space satisfy all the above constraints, we consider them for further 
phenomenological studies. In the left-panel of Fig.\ref{dm12}, we calculated $\Omega_{\lsp} h^2$  
using the {\tt micrOMEGAs} as a function the mass of the cold-dark matter candidate ($m_{\lsp}$). 
All the points satisfy the upper bounds coming from the recent Planck measurements \cite{Planck:2015xua} 
and the upper and lower bounds are 0.131 and 0.107, respectively referring to the standard cosmological scenario.
Within this strips, the green point satisfy the direct and indirect Dark Matter searches which we term as ``All-DM'' . 
The lightest neutralino annihilation would occur via the Z-boson exchange diagram -- this shows a dip 
around the $M_Z$/2, i.e, 45 GeV. This annihilation rate would get enhanced via the SM-Higgs boson ($h_2$) 
exchange diagram and this leads to another dip around $m_{h_2}$/2, i.e., 63 GeV. 

In the right panel of Fig.\ref{dm12}, we invoke other constraints discussed above. If the constraints on the 
Sparticles masses and B-physics and other phenomenological limits are satisfied, we indicated  
it by ``Pheno''; and if $h_2$-SM type scenario is satisfied, we indicate it by ``$h2$-SM''. All the constraints 
are imposed cumulatively. Finally, the $h2$-SM  model is interesting per se to look for as 
it is an unusual scenario.

In the left-panel of Fig.\ref{lspcomp}, we displayed the singlino component of the 
lightest neutralino, i.e., $N_{15}$ as a function of $m_{\lsp}$ with all the constraints 
mentioned in the legend. This shows that lightest neutralino with $m_{\lsp}$ $\gsim$ 80 GeV are more 
favorable with the $h_2$-SM type scenario. It can be seen that the singlino domination would  
occur mostly between 25 GeV and 250 GeV, while non-singlino type $\lsp$ would also 
be possible (would go up to 1\%) but the mass ranges are rather squeezed 
ranging between 100 -- 250 GeV.

In the right-panel of Fig.\ref{lspcomp}, we showed the masses of all the Higgs bosons satisfying 
all constraints. It turns out that the heavy CP-odd Higgs boson mass (we showed it up to 1600.0 GeV) 
is quite heavy, as it mainly depends upon the values of $M_P$.

In the left-panel of Fig.\ref{mhcons} we showed all the Higgs boson masses as a function of $m_{a_1}$.
However, in comparison to the right-panel of Fig.\ref{lspcomp} here we additionally imposed 
the ``All-DM'', ``Pheno'' and ``$h2$-SM'' criterion. This particular parameter spaces are of 
our phenomenological interest. In the right-panel of Fig.\ref{mhcons}, we show the decay branching ratios of the 
non-SM type $h_1$ in the light flavor quark mode: $h_1 \to gg + c \bar c + s\bar s$, tau-lepton: $h_1 \to \tau \bar \tau $, 
$b$-quark: $h_1 \to b \bar b$ and the much suppressed two-photon decay: $h_1 \to \gamma\gamma$.

For all these points we estimated the event rates for the two signal processes  
under consideration. We will describe the details in the following section.

\section{Numerical Analysis}
\label{sec:numana}

In our analysis we consider the $h_2$-SM scenario following Fig. \ref{mhcons}, such 
that the non-SM like $h_1$ is light enough to have some appreciable production rates at LHeC. 

In this section, we first describe the main two different production mechanism of Higgs bosons 
at the LHeC collider. These are the neutral and charged current production, 
which lead to the $e$ +3j and $\met$ +3j signals, respectively. As a result we will have two 
different sets of SM backgrounds which we will address in the subsequent section.

\begin{figure}[ht!]
\begin{center}
\raisebox{0.0cm}{\hbox{\includegraphics[angle=-90,scale=0.33]{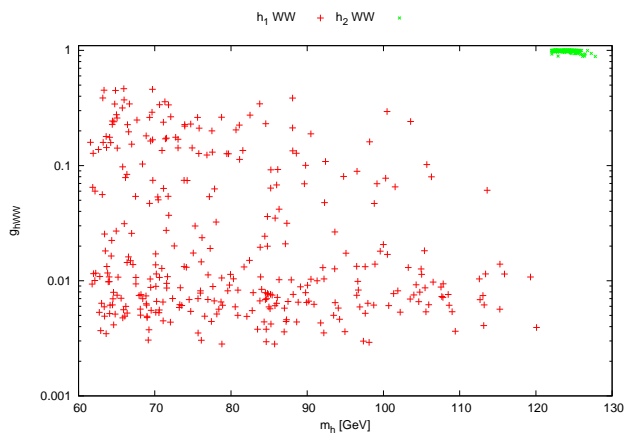}}}
\caption{The coupling $g_{h WW}$ (for $h$ = $h_1$ and $h_2$) is shown as a function of the Higgs boson mass. 
The top right corner (with populated green points) shows the masses of the $h_2$ and the 
corresponding coupling constraints (that we imposed following Table \ref{tab:coupsigma}) 
from the recent LHC measurements.}
\label{gh1VV}
\end{center}
\end{figure}

\begin{figure}[ht!]
\begin{center}
\raisebox{0.0cm}{\hbox{\includegraphics[angle=-90,scale=0.33]{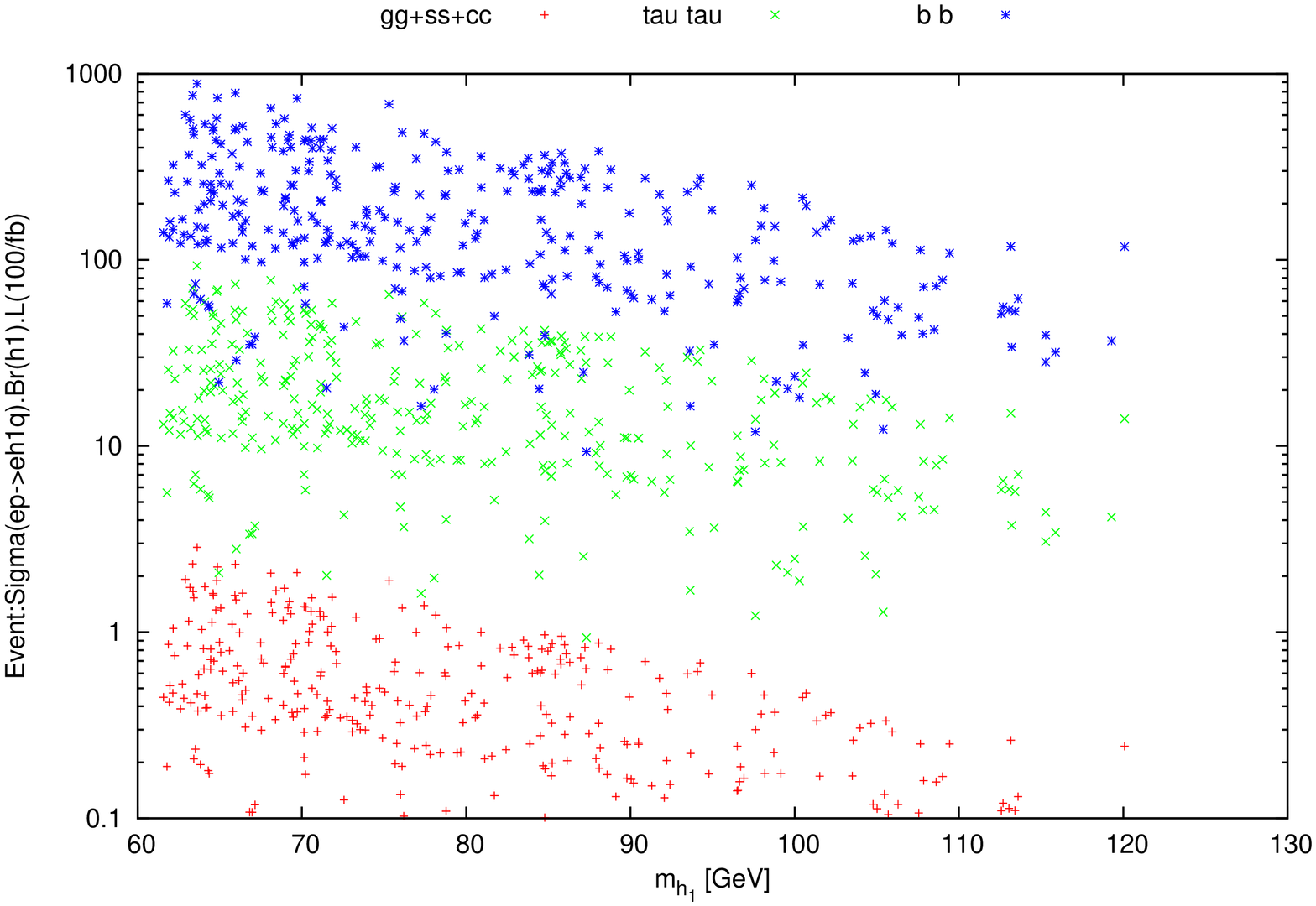}}}
\raisebox{0.0cm}{\hbox{\includegraphics[angle=-90,scale=0.33]{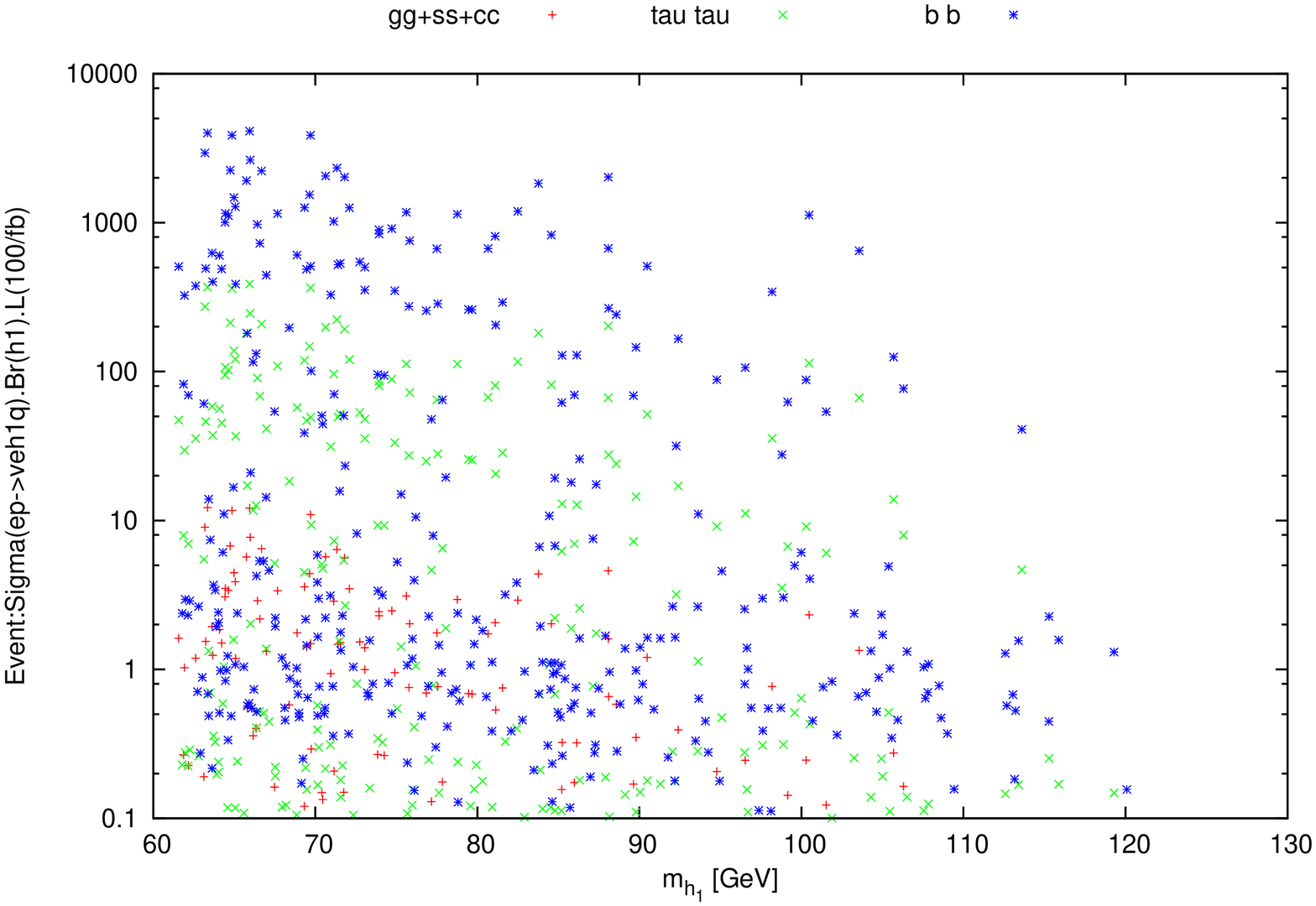}}}
\caption{Number of event in $e$ $+$ 3-jets (left-panel) and $\met$ $+$ 3-jets (right-panel) 
channels at LHeC with integrated luminosity of 100$fb^{-1}$.}
\label{evtrate}
\end{center}
\end{figure}

\begin{figure}[ht!]
\begin{center}
\raisebox{0.0cm}{\hbox{\includegraphics[angle=-90,scale=0.33]{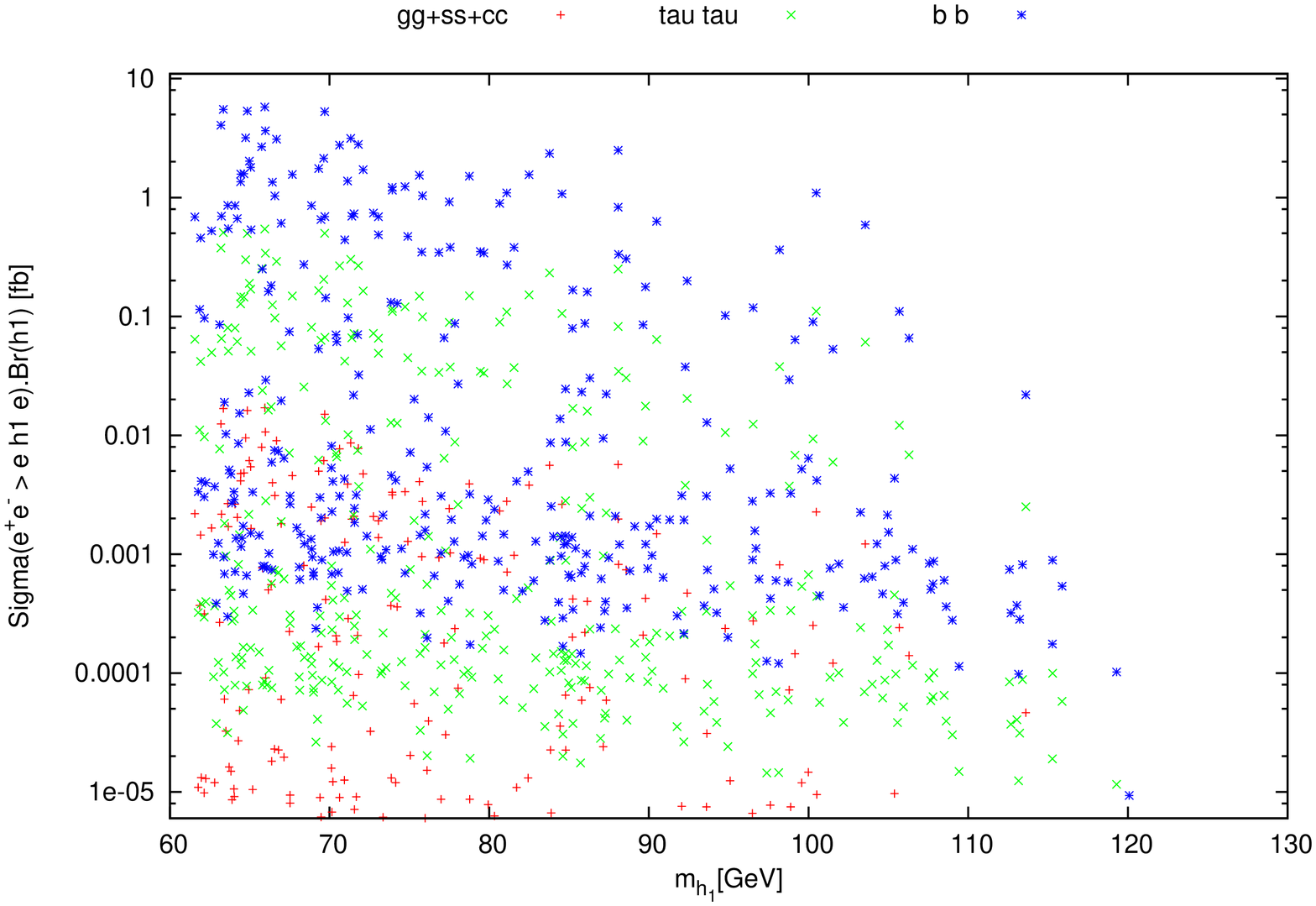}}}
\raisebox{0.0cm}{\hbox{\includegraphics[angle=-90,scale=0.33]{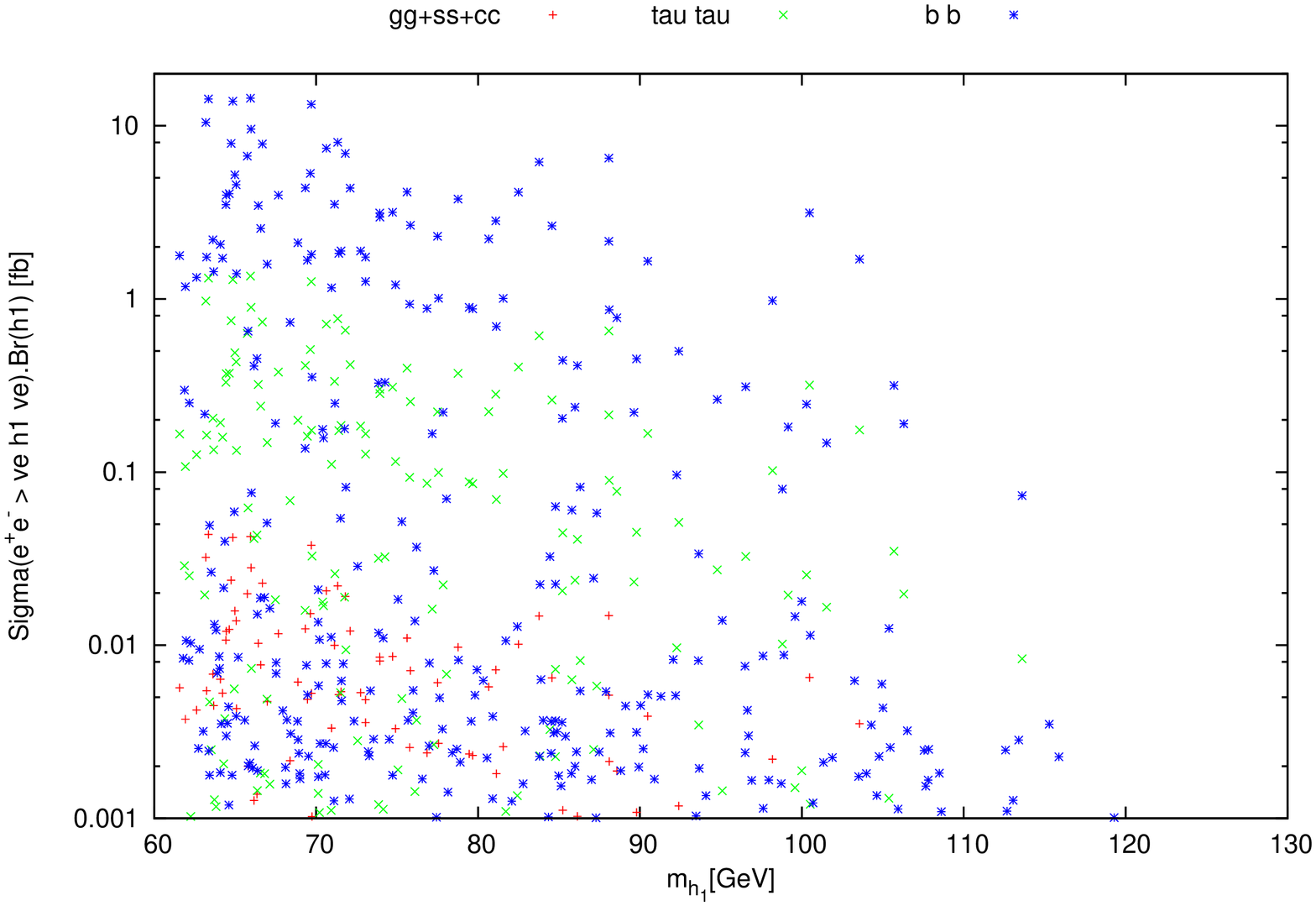}}}
\caption{The cross-section multiplied with the Higgs branching ratios in units of $[fb]$ in the $e^{+}e^{-}$ 
LEP collider with center of mass energy 209 GeV, for the $\sigma$($e^+ e^- \to  e^+ h_1 e^-$) and 
 $\sigma$($e^+ e^- \to  \nu_e h_1 \bar \nu_e$) in the left and right panel respectively.}
\label{evtLEP}
\end{center}
\end{figure}

\begin{figure}[ht!]
\begin{center}
\raisebox{0.0cm}{\hbox{\includegraphics[angle=-90,scale=0.33]{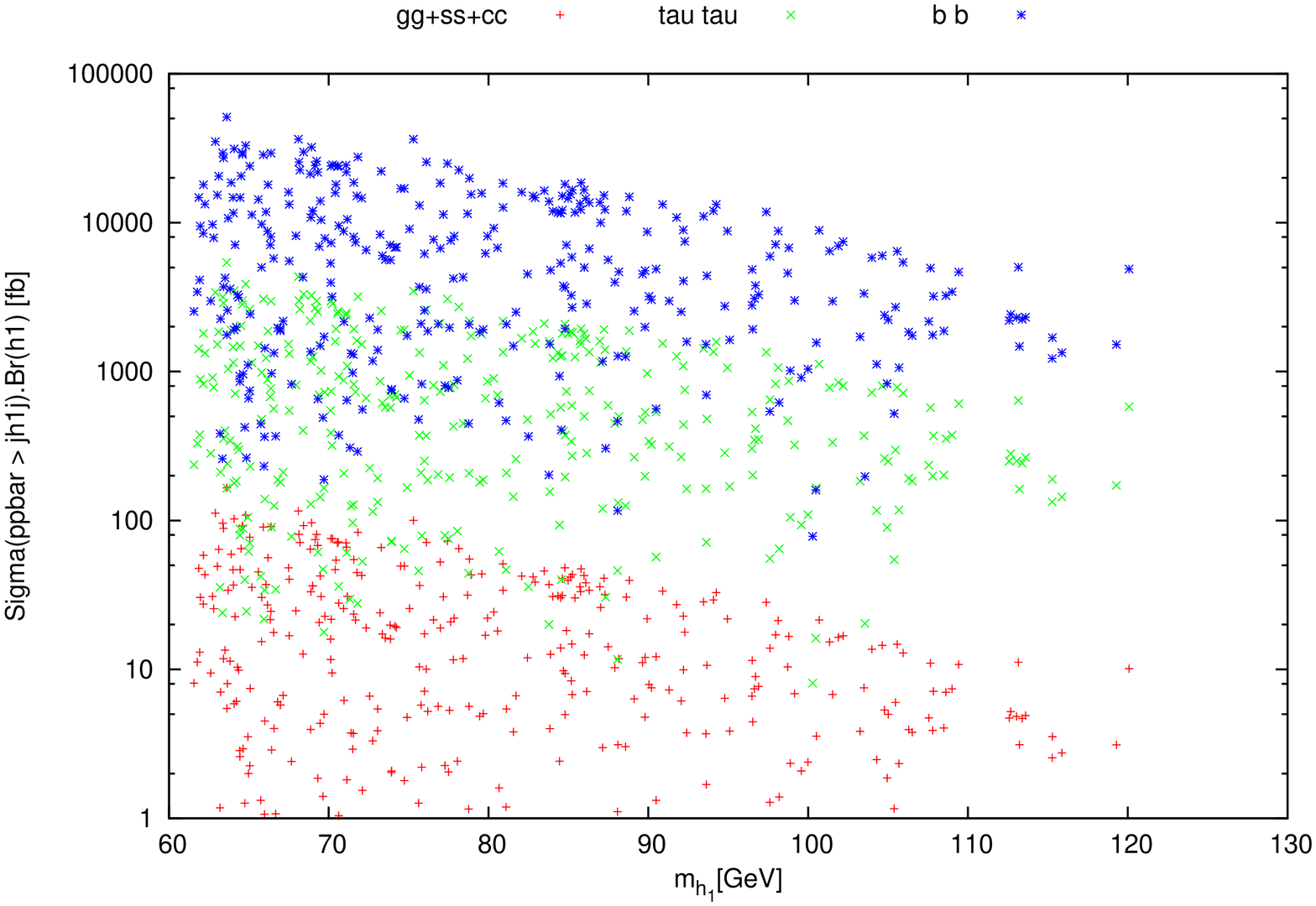}}}
\raisebox{0.0cm}{\hbox{\includegraphics[angle=-90,scale=0.33]{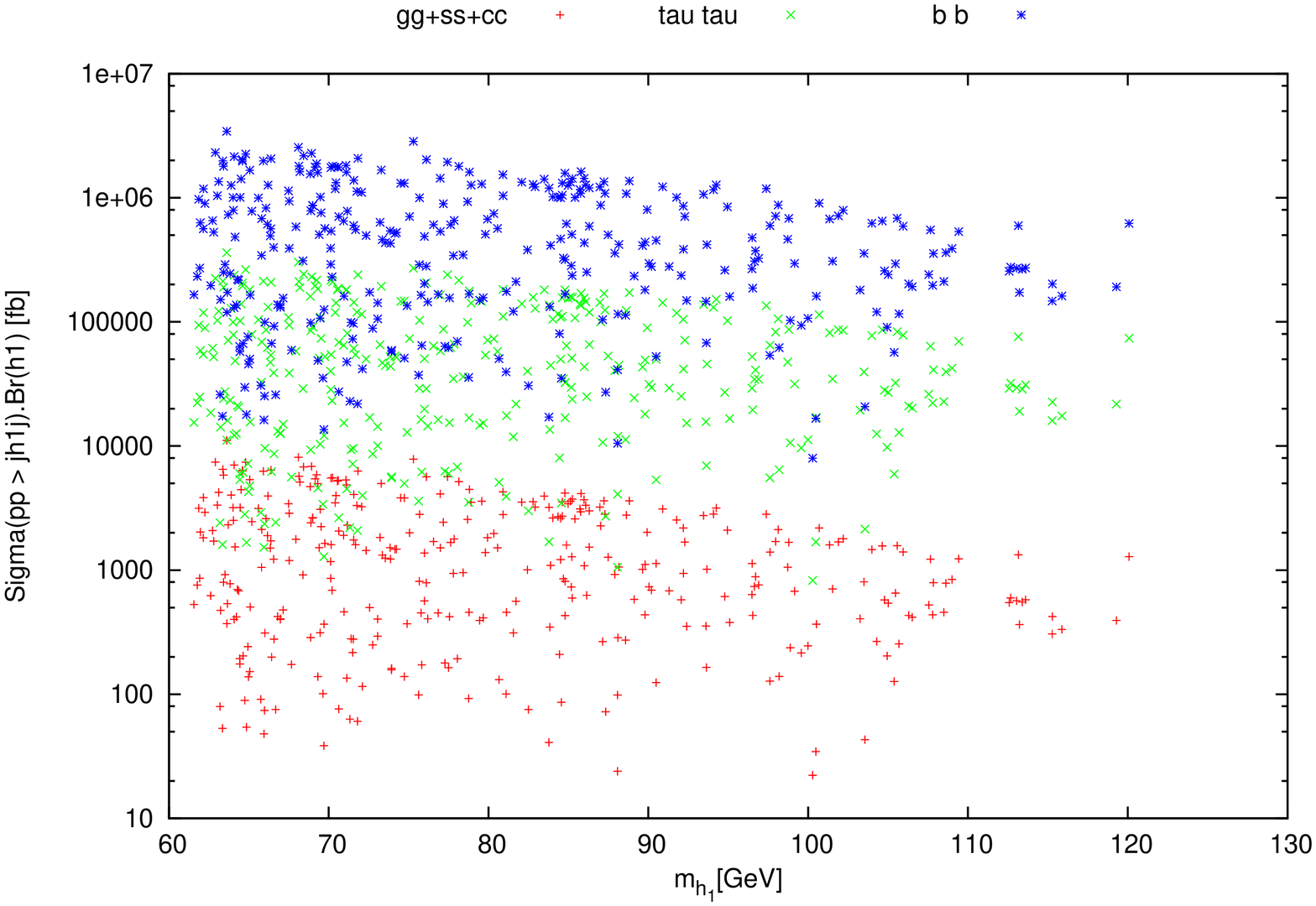}}}
\caption{Left-panel: The cross-section with Higgs branching ratios in units of [fb] at the Tevatron 
($\sigma$$(p \bar p \to j h_1 j$)) with center of mass energy 1.96 TeV. 
Right-panel: $\sigma$$(p p \to j h_1 j$) at LHC collider with center of mass energy 14TeV, 
where $j$=u,d,c,s,b,g and their charge-conjugation.}
\label{evtHad}
\end{center}
\end{figure}

\subsection{Higgs bosons signals}

In our analysis, we consider the leading production processes via the neutral 
current as well as charged current. As we consider an $ep$-collison, the neutral (charged) 
current gives charged (neutral) lepton, electron (neutrino) in the final 
states. Generally, the charged current production cross-sections are larger 
than the neutral currents as shown in Fig.\ref{evtrate} which is mainly due to the 
isospin couplings. The production processes of the Higgs bosons are :  
$e h_1 q$ and $\nu_e h_1 q$,  \footnote{Please recall that we are working 
in the $h_2$-SM scenario. This is to say that the second CP-even Higgs boson 
is the SM-type consistent with all the couplings and observables and in particular compatible 
with the recent Higgs discovery at the LHC experiments.} where $q$ (or its charge conjugated version)  is 
the light parton, i.e., $u,d,s,c$ or $g$. 

The Higgs boson production in our analysis is mainly dominated by the t-channel 
vector-boson fusion (VBF) processes. 

The couplings (the neutral and charged gauge-boson fusion vertex) relevant at LHeC are the following:

\beq\label{vbfcoup}
g_{h_1 Z Z} = \frac {{g_1}^2 + {g_2}^2}{\sqrt 2} (v_d S_{11} + v_u S_{12})~,~~
g_{h_1 W W} = \frac {{g_2}^2}{\sqrt 2} (v_d S_{11} + v_u S_{12})
\eeq

Here $S_{11}$ and $S_{12}$ are the mixing parameters in the singlet--doublet Higgs mixing matrices. 
The $v_u$ and $v_d$ are the up-type and down-type Higgs doublet vevs and $g_1$ and $g_2$ 
are the ${U(1)}_Y$ and ${SU(2)}_L$ gauge couplings. We have plotted the allowed couplings, $g_{h_1 W W}$, 
(using red color plus points) complying all the constraints discussed earlier in 
Fig.\ref{gh1VV} (see the caption for details). The green points in the top right corner are 
for $g_{h_2 W W}$ and the masses and the corresponding coupling values consistent with the coupling  
ratios for the $h_2$-SM scenario, following Table ~\ref{tab:coupsigma}.

It is to be noted that the same couplings given in Eq.\ref{vbfcoup} are also responsible 
for the Higgs boson production (t-channel VBF) at the LEP and Hadron colliders. We would like to 
estimate the event rates for this kind of signal at the large electron positron (LEP) colliders 
and also at the hadron colliders, like the recently closed Tevatron and presently operating LHC. 

We find that the exact signal hard--processes at LHeC under considerations are not possible 
at LEP, Tevatron and LHC. However, we estimated the closest processes that we could have 
in these three colliders. At these colliders, we have 
estimated the event rates (in $fb$) with inclusion of Higgs branching ratios at the $e^+e^-$ LEP collider 
with center of mass energy 209 GeV, for the $\sigma$($e^+ e^- \to  e^+ h_1 e^-$) and 
$\sigma$($e^+ e^- \to  \nu_e h_1  \bar \nu_e$) processes in Fig.\ref{evtLEP} in the 
left-panel and right-panel respectively.

In hadron colliders the most closet process is $\sigma$$(pp,p \bar p \to j h_1 j$), with $j$=$u,d,c,s,b,g$ 
and their charge-conjugation. The estimated event rates (in $fb$) at Tevatron and LHC with center 
of mass energy 1.96 TeV and 14 TeV, is shown in Fig.\ref{evtHad}, in the left-panel and 
right-panel respectively. The LHC collider is presently operating 
and the number of events rates are substantial. In spite of the huge SM backgrounds contamination, it would be worth 
to look for this non-standard Higgs signal in this ongoing machine.

We see from the right-panel of Fig.\ref{mhcons} that $h_1$ is dominantly decaying into $b \bar b$. 
So our signals in both channels contain three jets (one is forward light flavored and two central $b$-tagged) 
and an electron (missing transverse energy) in case of neutral (charged) current.

We estimated the parton level signal cross sections using the \texttt{MadGraph v 2.4.3} \cite{Alwall:2014hca}. 
The allowed NMSSM model parameter spaces from the \texttt{NMSSMTools~5.0.1}~\cite{Ellwanger:2004xm} are 
written in SLHA format and fed to \texttt{MadGraph v 2.4.3} \cite{Alwall:2014hca}. The Branching ratios (BRs) 
of the Higgs boson in all the decay modes is estimated by using \texttt{NMHDECAY}~\cite{Ellwanger:2004xm}.

\begin{table}[!ht]
\vspace*{-5mm}
\vspace{3mm}
\footnotesize
\begin{center}
\scalebox{0.9}{
\begin{tabular}{|l||r|r|r|r|r||r|r|r|r|r||}
\hline
{\bf Benchmark Points}&$e$1&$e$2 & $e$3 &$e$4&$e$5&$\nu$1 & $\nu$2 & $\nu$3 &$\nu$4=$e$5&$\nu$5
\\\hline
$\lambda$& 0.241 &0.168 &0.171 &0.237& 0.384&0.208 &0.263 &0.296& 0.384&0.498 
\\\hline
$\kappa$&0.371 &0.0567&0.0324&0.0384&0.0152&0.0577&0.143&0.147&0.0152&0.291
\\\hline
$\tan \beta$&56.71&48.62&56.32&56.27&4.41&3.81&4.95&5.15&4.41&6.18
\\\hline
$A_\lambda$(GeV)&974.0&1007.8&1230.8&964.1&1222.4&1130.3 &1105.8&1078.3&1222.4&1107.8
\\\hline
$A_\kappa$(GeV)&-1139.2&-1105.6&-1214.2&-1108.3&-1062.3&-1230.3&-835.9&-803.6& -1062.3&-1732.7
\\\hline
$\mu_{\rm eff}$&466.4&272.9&297.2&454.6&381.1&392.6&374.5&342.7&381.1&193.8
\\\hline 
$M_1$ (GeV)&274.6&347.5&241.7&272.5&335.3&173.3&146.5&172.7&335.3&112.1
\\\hline
$M_2$ (GeV)&293.4&482.3&462.5&277.5&352.3&283.9&257.3&495.3&352.3&430.7
\\\hline
$M_{\tilde q}$&1010.4&941.9&700.6&1036.6&734.0&807.1&812.9&772.5&734.0&762.0
\\\hline
$A_t$ = $A_b$ (GeV)&-2661.8&-1695.4&-1735.6&-2679.4&-1585.1&-1946.3&-1883.7&-1676.3&-1585.1&-1894.7
\\\hline
$M_A$ (GeV)&316.9&282.2&239.8&319.9&125.6&102.9&106.9&131.7&125.6&134.5
\\\hline
$M_P$ (GeV)&2015.5&1650.2&1995.9&2019.2&2396.3&1660.2&2429.6&1474.9&2396.3&1858.8
\\\hline\hline
$\xi_F$ ($10^6$$GeV^2$)&-2.01 &-1.77&-2.23&-1.98&-1.21& -2.32 &-1.85&-1.43&-1.21&-0.47
\\\hline\hline
$\xi_S$ ($10^9$$GeV^3$)&-6.79 &-3.27&-6.06&-6.77&-5.56&-3.42&-6.15&-1.06&-5.56&-0.89
\\\hline\hline
$m_{\lsp}$ (GeV)&142.2&181.9&112.1&146.3&33.7&164.5&139.5&163.6&33.7&94.2
\\\hline\hline
$m_{h_1}$ (GeV)&63.59&70.59&75.29&82.24& 88.07& 65.93&71.32&83.77&88.07& 100.47
\\\hline
$m_{h_2}$ (GeV)&122.9&122.7&122.8&123.6&126.1&127.8&126.5&124.6&126.1&125.4
\\\hline
$m_{h_3}$ (GeV)&1858.7&1394.5&1852.1&1861.1&2365.0&1315.5 &2078.4&950.1&2365.0&1467.3
\\\hline
$m_{a_1}$ (GeV)&67.8&73.4&77.8&85.3&89.0&73.9&78.2&107.1&89.0&118.9
\\\hline
$m_{a_2}$ (GeV)&2014.3&1649.4&1995.2&2018.0&2393.3&1659.6&2427.6&1473.5&2393.3&1848.9
\\\hline
$m_{h^\pm}$ (GeV)&112.4&114.9&121.6&124.1&103.3&102.9&101.4&124.8&103.3&120.1
\\\hline
BR($h_1 \to b \bar b$)& 0.902&0.910&0.909&0.901&0.828&0.910&0.909&0.906&0.828&0.907
\\\hline
$\sigma$ [fb]&9.783&5.627&7.535&4.815&4.628&45.209&25.561&20.205&24.371&12.403
\\\hline
$\sigma$.BR[fb]&8.824&5.121&6.850&4.338&3.832&41.141&23.235&18.306&20.180&11.250
\\\hline
\end{tabular}
}
\end{center}
\caption{The selected NMSSM benchmark points obtained using \texttt{NMSSMTools~5.0.1}~\cite{Ellwanger:2004xm} 
to find the $h_1$ signal at LHeC. The values displayed are at the electroweak scale. 
The following parameters are fixed: $M_3$ =1800 GeV, $A_{\tau}$ = $A_{\ell}$=1500 GeV and 
$M_{\tilde \ell}$ = 300 GeV. Please note that we used $M_A$ and $M_P$ as inputs -- 
thus our scenario is not the $Z_3$-NMSSM, and for that $\xi_F$ and $\xi_S$ are non-zero 
and also given in the table. We mention the cross section $\sigma$  $\times $ BR($h_1 \to b \bar b$) 
(as $\sigma$.$BR$) at LHeC. Please note that the $e$5 benchmark for $e$+3-jets and 
$\nu$4 benchmark for $\met$+3-jets is identical.} 
\label{table:nmssmbp}
\end{table}

To obtain the cross sections at the LHeC \cite{cern:lhec,Bruening:2013bga,AbelleiraFernandez:2012ty,
AbelleiraFernandez:2012cc,lheclumi}, we consider an electron beam, of energy $E_{e^-}$= 60 GeV and 
a proton beam of energy $E_{p}$= 7000 GeV, corresponding to a center-of-mass energy of 
approximately $\sqrt s = 1.296$ TeV. 

To estimate the signal event rates at parton level we applied the following basic pre-selections:
\beq \label{presel}
p^{q,e}_T >  10~{\rm GeV},  \qquad \eta^{q,e} < 5.0 \qquad  \Delta R (qq,qe) > 0.2 \,
\eeq
with $\Delta R^2 = \Delta \eta^2 + \Delta \phi^2$, where $\eta$ and $\phi$ are the
pseudo-rapidity and azimuthal angle, respectively. 
We take $m_t$=173.3 GeV as the top-quark pole mass.

We have set the renormalization and factorization scales at $\sqrt {\hat s}$, the center-of-mass (CM) 
energy at the parton level, and adopted NN23LO Parton Distribution Functions 
(PDFs) \cite{Ball:2013hta,Pumplin:2002vw}. 
including the $b$-flux, with $\alpha_{\rm s}$ (the strong coupling constant with four-flavor schemes) evaluated 
consistently at all stages (i.e., convoluting PDFs, hard scattering and decays). 
Parton shower (both initial and final), hadronization, heavy hadron decays, etc. have been 
dealt with by {\tt PYTHIA} v.6.428 \cite{pythia}.  

We consider all the light-flavor quarks, $b$-quark and gluon in the proton flux. The flavor-mixing, 
wherever appropriate, is also considered for the allowed diagrams. Following this, it was realized that the 
signal processes have unique kinematic profiles and we will discuss it below. 

The cross section of the scattering $fa\rightarrow f^{\prime} X$ via the gauge boson ($V$)  exchange can 
be expressed as
\begin{equation}
\sigma(fa\rightarrow f^{\prime} X) \approx \int dx\ dp_T^2\ P_{V/f}(x,p_T^2)\ \sigma(Va\rightarrow X)
\label{eq:effWapp}
\end{equation}

where the fermion $f$ with a c.m.~energy $E$ is radiating a gauge boson $V$ ($s \gg M_V^2$), $\sigma(Va\rightarrow X)$ is the 
cross-section of the $Va\rightarrow X$ scattering and $P_{V/f}$ can be viewed as the probability 
distribution for a weak boson $V$ of energy $xE$ and transverse momentum $p_T$.
The dominant kinematical feature is a nearly collinear radiation of $V$ off $f$, termed as 
``Effective $W$-Approximation" \cite{effw}. The probability distributions of the weak bosons 
with different polarizations can be approximated by (in the limit of $s \gg M_V^2$)
\begin{eqnarray}
P_{V/f}^T(x,p_T^2) &=&
{g_V^2 + g_A^2 \over 8\pi^2} {1 + (1-x)^2 \over x } { p_T^2 \over \left(p_T^2 + (1-x)M_V^2\right)^2 }
\label{eq:PT} \\
P_{V/f}^L(x,p_T^2) &=&
{g_V^2 + g_A^2 \over 4\pi^2} {1-x \over x } { (1-x) M_V^2 \over \left(p_T^2 + (1-x)M_V^2\right)^2 }, 
\label{eq:PL}
\end{eqnarray}
where $g_V$($g_A$) is the vector (axial) vector couplings of fermion-gauge boson vertices.

From these equations we can understand that the final state quark $f'$ typically has transverse momentum, 
$p_T\sim\sqrt{1-x}M_V\leq M_V$, i.e, less than the mass of the vector boson. Secondly, due to the $1/x$ 
behavior for the gauge boson distribution, the out-going
parton energy $\left(1-x\right)E$ tends to be high and leads to very high energetic forward jet,    
with small angle (i.e., high forward rapidity) with respect to the beam direction. Finally, at high $p_T$, 
the probabilities of the gauge bosons can be approximated as: $P_{V/f}^T\sim 1/p_T^2$ and $P_{V/f}^L\sim 1/p_T^4$. 
At high $p_T$ the longitudinally polarized gauge bosons is relatively suppressed as compared to the 
transversely polarized one. In particular, the first two criterions serve as a guidance in our 
event selection for exploiting the kinematical features. Also, in both signals under 
consideration, the final state forward jet could be also a $b$-jet. However, as it is mostly in the 
forward region, with the tighter constraints of the rapidity of $b$-taggable jet, it hardly 
qualifies  as $b$-tagged.

\subsection{Backgrounds}

There are mainly two groups of SM backgrounds in our Higgs signals.
The charged-current backgrounds consisting of $\nu t \bar b$, $\nu b \bar b j$, $\nu b2j$, $\nu 3j$,
and the neutral-production ones identified as $e^{-} b \bar b j$, $e^{-} t \bar t$,  $e^{-} b jj$ and $e^{-} jjj$.
In all of these backgrounds the charge-conjugated processes are naturally implied.

For estimating the cross sections of these SM backgrounds, we used the same set of 
pre-selections, identical conventions and parameter sets as for the  signal. The 
expected number of the background events for 100 fb$^{-1}$ of integrated luminosity are  
given in the second column of Tables \ref{tab:lepori} and \ref{tab:met}.

\subsection{Signal-to-Background analysis}

We generated the SM backgrounds at the parton level using \texttt{MadGraph v 2.4.3} \cite{Alwall:2014hca} 
and then fed them to {\tt PYTHIA} v.6.428 \cite{pythia} for parton showering 
(both initial and final), hadronization, heavy hadron decays etc. The 
initial state radiation (ISR) will reduce the total center-of-mass energy of
the collision, however at the LHeC with the main dynamics along the t-channel, 
the center-of-mass energy loss due to ISR has less impact.
The top-quark and $W$-boson were allowed to decay freely within {\tt PYTHIA} program.  
The four-momentum of the jets are different as compared to the parton level 
quark due to the final state radiation (FSR) and in our analysis 
we considered the Gaussian type of smearing effects. 

The LHeC detectors and their parameters considerations follow 
one of our recent analysis \cite{Das:2015kea}. However to be complete, 
let us describe it here briefly. 

We have considered the experimental resolutions of the jet angles and energy  using 
the toy calorimeter {\tt PYCELL}, in accordance with the LHeC detector parameters, 
given in {\tt PYTHIA}. As the invariant mass has been used to isolate the Higgs signal -- 
this has some non-trivial effect. We considered LHC type of calorimeter for the LHeC. 
To be explicit, we set a somewhat symmetric detector coverage, however in reality 
the electromagnetic and the hadronic calorimeter at LHeC, unlike ATLAS and CMS, 
are not exactly symmetric. 

Since we are not doing detector simulation and also not considering the 
cracks in the detectors, we applied symmetric large rapidity coverage for 
jets and leptons\footnote{Here the lepton means only electron unless mentioned otherwise.}. 
We expect that these assumptions hardly alter our numerical findings. The detector parameters 
in {\tt PYCELL} are set according to the LHeC detector \cite{AbelleiraFernandez:2012cc}. 
Specifically, we assume large calorimeter coverage $|\eta| < 5.5$, with segmentation (the 
number of division in $\eta$ and $\phi$ are 320 and 200, respectively) 
$\Delta \eta \times \Delta \phi = 0.0359 \times 0.0314$. Furthermore, we have used 
Gaussian energy resolution \cite{Bruening:2013bga} for electron and jets (labeled as $j$), with
\beq \label{lhc_res}
{{\Delta E \over E} = { a \over \sqrt{E} } \oplus b},
\eeq
where $a=0.32$, $b=0.086$ for jets and $a=0.085$, $b= 0.003$ for leptons and
$\oplus$ means addition in quadrature.

\begin{figure}[ht!]
\begin{center}
\raisebox{0.0cm}{\hbox{\includegraphics[angle=0,scale=0.42]{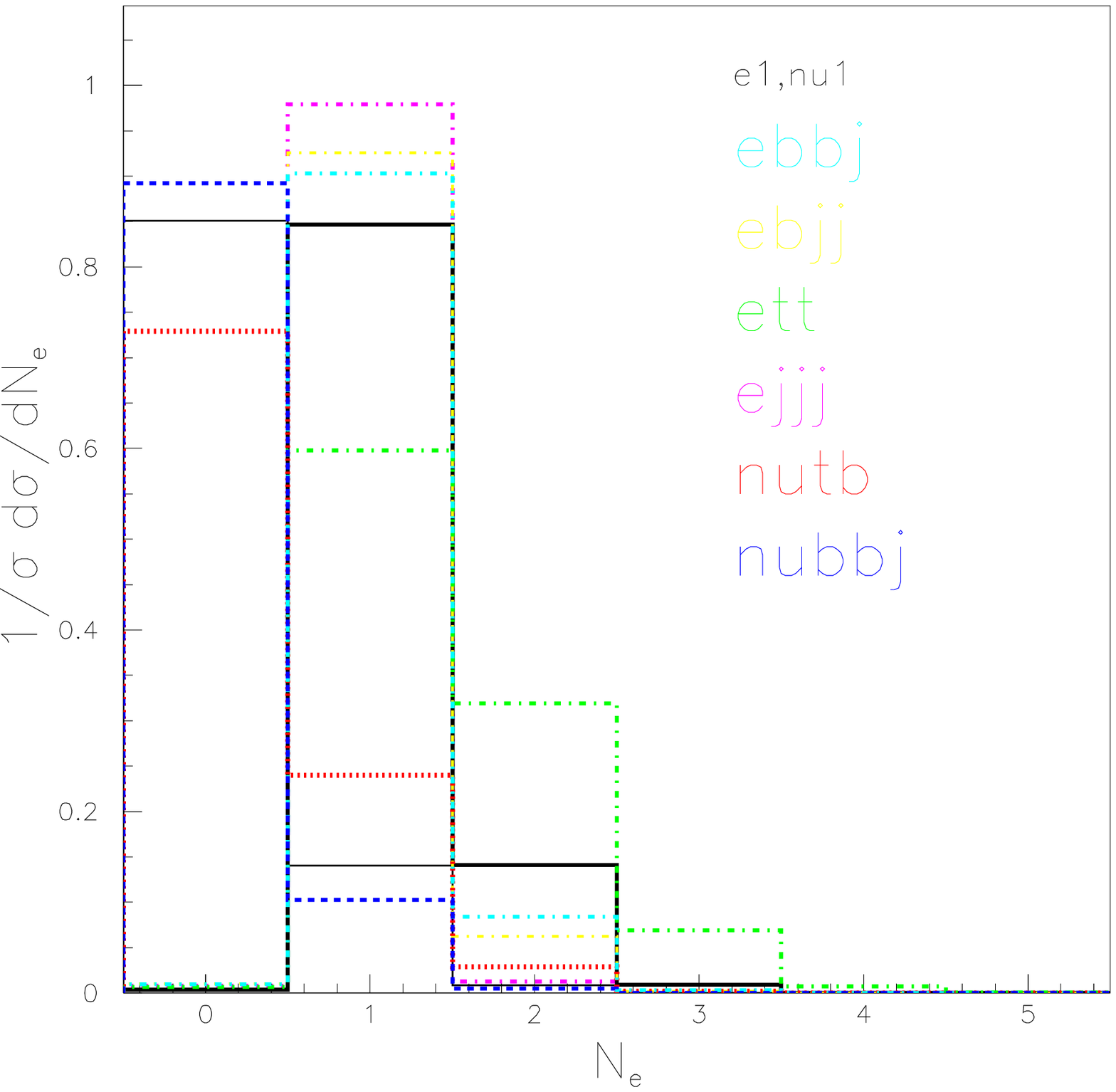}}}
\raisebox{0.0cm}{\hbox{\includegraphics[angle=0,scale=0.42]{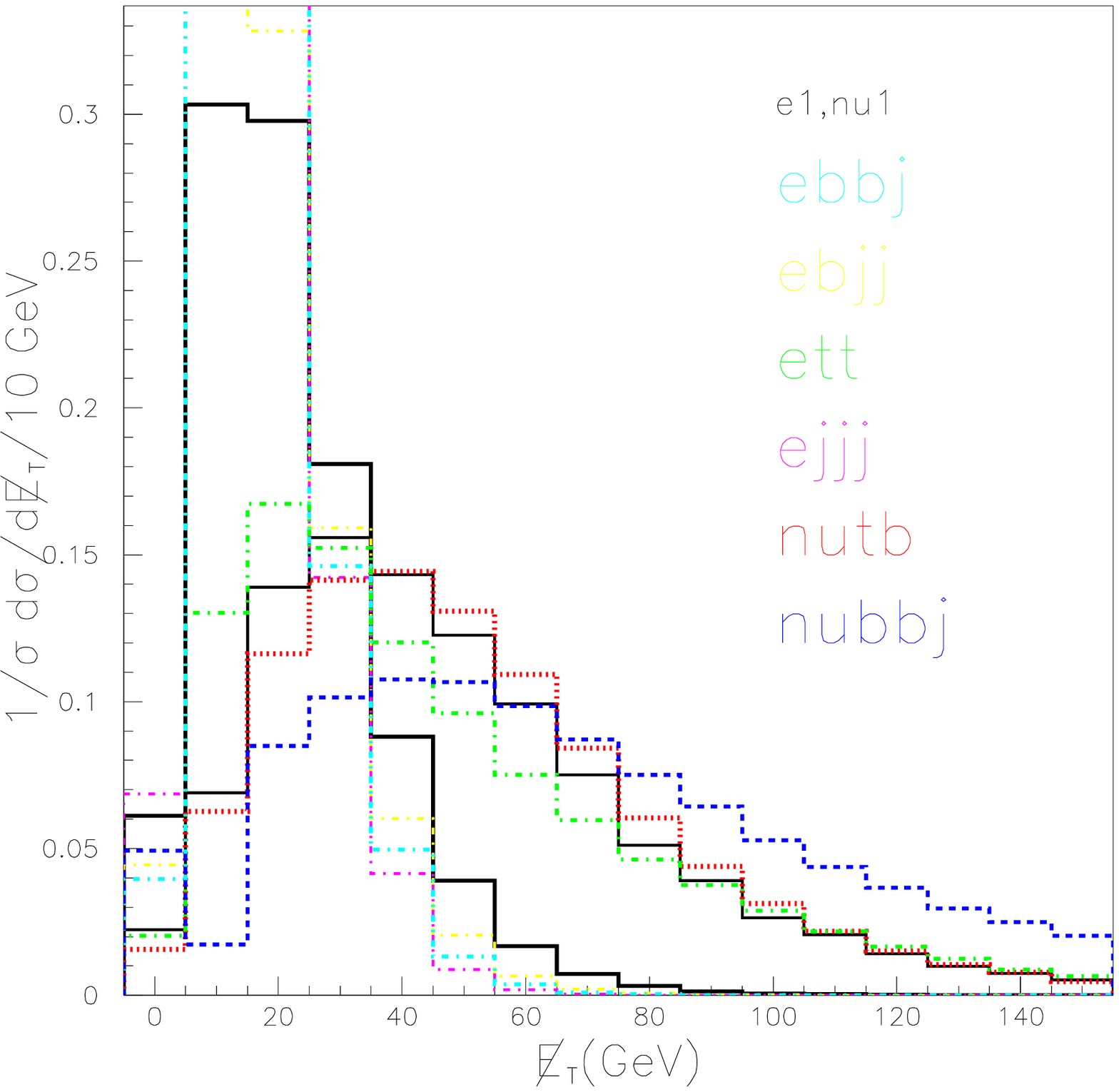}}}
\caption{Number of electron ($N_e$) distribution for $e$1 benchmark ($\nu$1 benchmark) using thick (thin) 
black line in the left panel. The distribution of missing energy ($\met$) for $e$1($\nu$1) benchmark 
using thick (thin) black line in the right panel.
}
\label{nlepmet}
\end{center}
\end{figure}

We have used a cone algorithm for the jet-finding, with jet radius 
$\Delta R(j) = \sqrt{\Delta\eta^{2}+\Delta\phi^{2}} = 0.5$. Calorimeter cells 
with $E_{T,min}^{\rm cell} \ge 5.0$ GeV are considered to be potential candidates 
for jet initiators. All cells with $E_{T,\rm min}^{\rm cell} \ge 1.0$ GeV were 
treated as part of the would--be jet. A jet is required to have minimum summed energy $E_{T,min}^{j} \ge 15$ GeV
and the jets are ordered in $E_{T}$.
The leptons ($\ell = e $ only ) are selected if they satisfy the requirements: 
$E_T^{\ell} \ge 15$ GeV and $\left| \eta^{\ell} \right| \le 3.0$.
In our jet finding algorithm we include leptons as parts of jets.
Finally we separate them, putting some isolation criterion as follows:
if we find a jet near a lepton, with $\Delta R (j-\ell) \le 0.5$ and $ 0.8 \le
E_{T}^{j}/E_{T}^{\ell} \le 1.2$, i.e. if the jet $E_T$ is nearly
identical to that of this lepton, the jet is removed from the list of
jets and treated as a lepton. However, if we find a jet within $\Delta
R (j-\ell) \le 0.5$ of a lepton, whose $E_T$ differs significantly
from that of the lepton, the lepton is removed from the list of leptons.
This isolation criterion mostly removes leptons from $b$ or $c$ decays.
We reconstructed the missing (transverse) energy ($\met$) from all observed particles 
and for the charge current the signal is shown in right panel of Fig. \ref{nlepmet}. 
We have also calculated the same from the energy deposition in the calorimeter 
cells and found consistency between these two methods. 
Only jets with $|\eta^j| < 2.5$ and  $E_{T}^j \geq 15$ GeV
``matched'' with a $b-$flavored hadron ($B-$hadron), i.e.  with
$\Delta R(j,B-{\rm hadron}) < 0.2$ is considered to be ``taggable''.
We assume that these jets are actually tagged with probability
$\eps_b = 0.50$.
We also adopted mis-tagging of non$-b$ jets as $b-$jets and treated $c-$jets
differently from the gluon and light-flavor jets. A jet with
$\left| \eta^j \right| \le 2.5$ and $E_{T}^j \geq 15$ GeV matched
with a $c-$flavored hadron ($C-$hadron, e.g., a $D-$meson or $\Lambda_c-$baryon),
i.e., with $\Delta R(j,C-{\rm hadron}) < 0.2$, is again
considered to be taggable, with (mis-)tagging probability $\eps_c = 0.10$.
Jets that are associated with a $\tau-$lepton, with $\Delta R(j,\tau) \le 0.2$, and all jets
with $\left| \eta^j \right| > 2.5$, are taken to have vanishing tagging
probability. All other jets with $E_{T}^j \geq 15$ GeV and
$\left| \eta^j \right| \leq 2.5$ are assumed to be (mis-)tagged with
probability $\eps_{u,d,s,g} = 0.01$, following \cite{Han:2009pe}. The overall 
analysis strategy has been adopted from \cite{Agrawal:2013qka,Agrawal:2013owa} and \cite{Das:2010ds}.

\begin{figure}[ht!]
\begin{center}
\raisebox{0.0cm}{\hbox{\includegraphics[angle=0,scale=0.42]{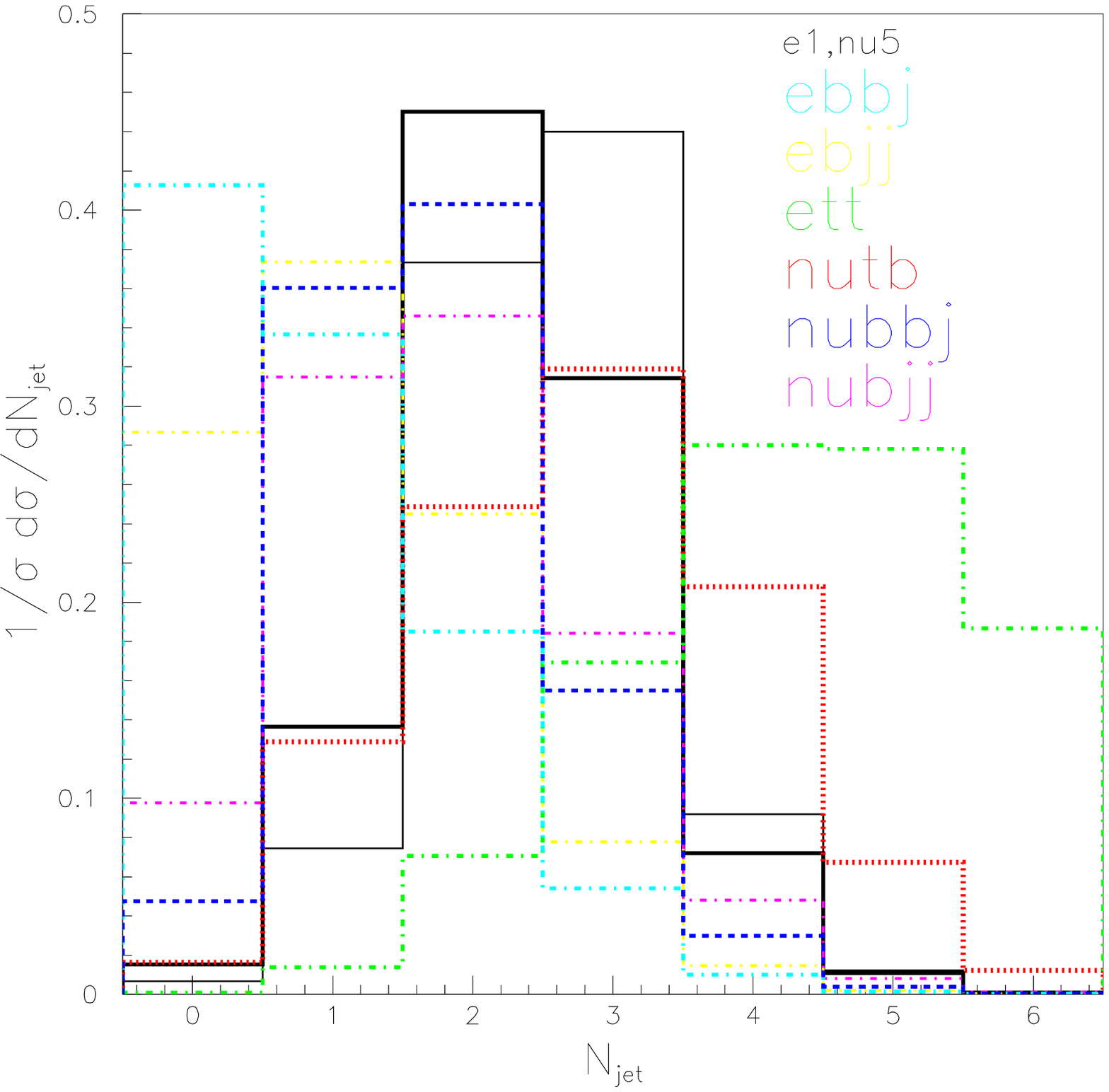}}}
\raisebox{0.0cm}{\hbox{\includegraphics[angle=0,scale=0.42]{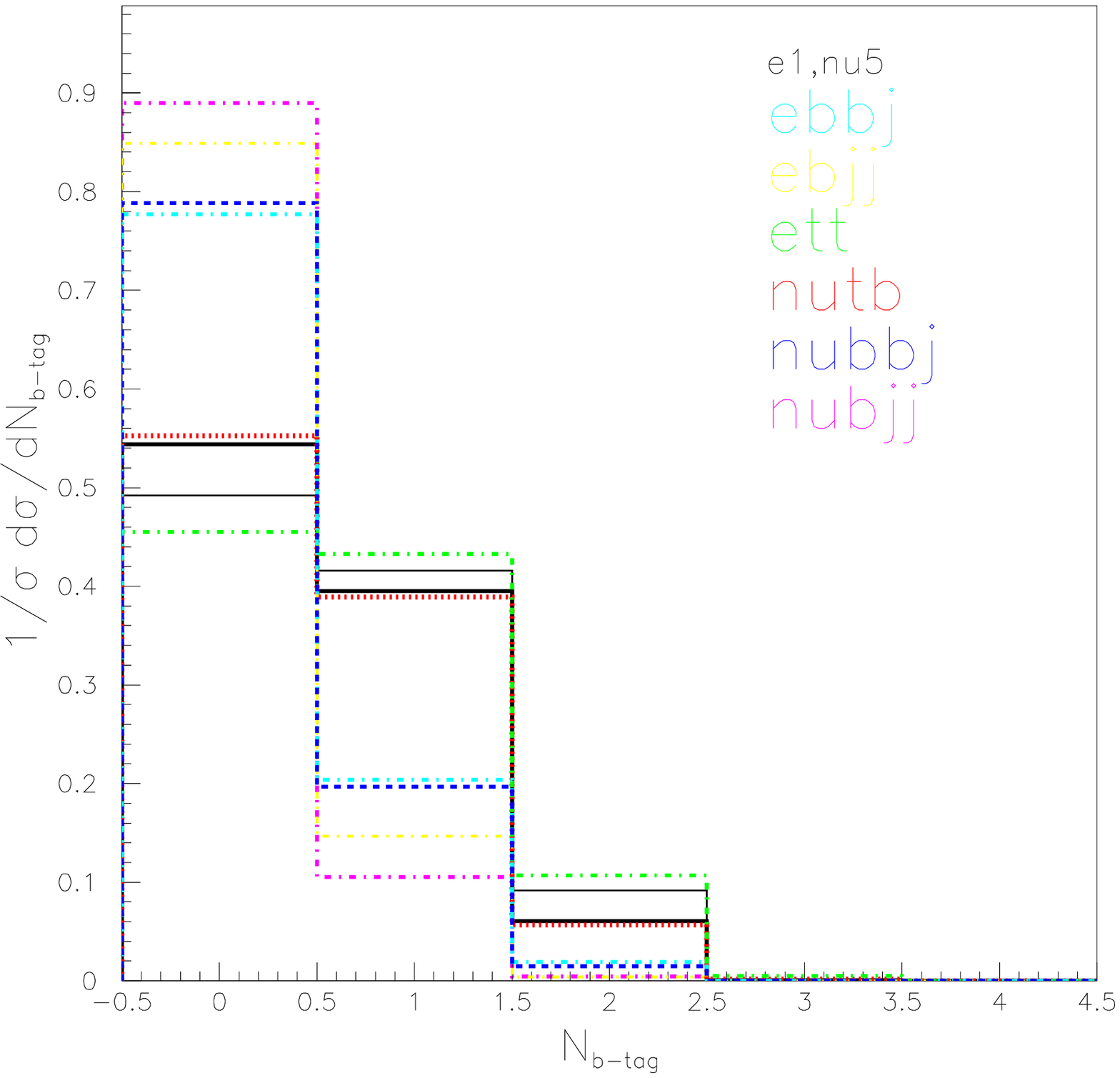}}}
\caption{Left-panel: Number of jet distribution for the $e$1 benchmark ($\nu$5 benchmark) in the 
$e$$+$3-jets ($\met$$+$3-jets) channel using the thick (thin) black lines. 
Right-panel: The number of $b$-tagged jet distribution is shown. 
Both the distributions depends mainly upon the masses of the Higgs boson in two signal 
channels, i.e., the more massive the Higgs, the more the number of jets peaks towards 
higher values. For all other signal benchmarks, the distributions follow a similar pattern 
and can be understood from the numerics in the corresponding columns 
in Tables \ref{tab:lepori} and \ref{tab:met}.} 
\label{njetbtag}
\end{center}
\end{figure}

\begin{table}[t!]
\centering
\scalebox{0.7}{
\begin{tabular}{||c|c||c|c|c|c|c|c|c|c|c|c|c||}
\hline
\hline
&&&&&&&&&&&&\\
Proc,$m_h$&RawEvt& a & b & c & d& e & f &g &h& i&j& ${\cal S}$\\
\hline
\hline
$e1$,63.59& 882.4 & 351.2 &330.0 &45.3 &24.5 &23.9 &12.8& 12.5 &6.3 &6.0 &2.2& 0.40(1.3) \\
\hline
$e bbj$&2688390.0 &176102.0 &    135388.5 &15530.5 &3759.5 &3461.2 &2655.6  &2476.5 & 343.1 &179.0 & 29.8&\\
$e bjj$&330834.0 &31317.5 & 23381.6 &575.6 &146.4 &110.8 &74.4 & 73.6 &3.3 & 2.5 &  0.8&\\
$e t \bar t$&1425.5 &1313.8 &1136.9 &131.8 &    20.2 &19.9 &7.9 &7.8 &0.4 &0.4 & 0.2&\\
$e jjj$&37224100.0 &4943049.0 & 3940097.5 &8862.8 &1611.4&966.9 & 805.7 & 805.7 &161.1 &0.0 & 0.0& B=30.8\\
$\nu bbj$& 21385.4 &4040.5 &112.0 &10.9 &2.2 &1.8 &0.0 &0.0 &0.0 &0.0 & 0.0&\\
$\nu bjj$&4077.8 &   985.1 & 13.3 & 0.4 & 0.1 & 0.1 & 0.0 & 0.0 &  0.0 & 0.0 & 0.0&\\
$\nu t b$& 84395.2 &51227.5 & 3848.3 & 405.4 &73.3 & 70.6 & 0.7 &  0.7 & 0.0 & 0.0 &   0.0&\\
$\nu jjj$&3870920.0 & 718974.7 & 1675.5 &0.0 &0.0 &0.0 & 0.0 &0.0  &0.0 &0.0 & 0.0&\\
\hline
\hline
$e2$,70.59&512.1 & 219.2 &206.2 &28.4 & 13.6 &   13.2 & 7.3 & 7.2 & 3.8 & 3.6 & 1.3&0.33(1.0)\\
\hline
$e bbj$& 2688390.0 &    176102.0 &    135388.5 &15530.5 &3505.9 & 3222.5 &2357.2 &2222.9 &402.8 &223.8 & 14.9&\\
$e bjj$&330834.0 & 31317.5 &23381.6 &575.6 &142.3 & 100.9 & 64.5 & 64.5 &3.3 &0.8 & 0.8&\\
$e t \bar t$&1425.5 & 1313.8 & 1136.9 & 131.8 &20.9 & 20.7 &  8.2 &  8.1 & 0.5 & 0.5 & 0.3&\\
$e jjj$&37224100.0 &4943049.0 &3940097.5 &8862.8 &1611.4 & 1128.0 & 1128.0 &1128.0&161.1 & 0.0 & 0.0&B=16.0\\ 
$\nu bbj$& 21385.4 & 4040.5 &112.0 & 10.9 & 1.7 &1.3 & 0.0 & 0.0 & 0.0 & 0.0 & 0.0&  \\
$\nu bjj$& 4077.8 &  985.1 & 13.3 &0.4 &0.1 &0.1 &0.0 &0.0 & 0.0 &   0.0 & 0.0&\\
$\nu t b$&84395.2 &51227.5 & 3848.3 &  405.4 &  74.7 & 71.9 & 2.1 &  2.1 & 0.0 & 0.0 & 0.0&\\
$\nu jjj$&3870920.0 &718974.7 &1675.5 & 0.0 & 0.0 & 0.0 & 0.0 & 0.0  & 0.0 & 0.0 & 0.0&\\
\hline
\hline
$e3$,75.29&685.0 &310.1 & 291.6 & 42.0 &18.2 & 17.9 &10.0 & 9.8&4.9 &4.7&1.5&0.38(1.2)\\
\hline
$e bbj$&2688390.0 & 176102.0 &135388.5 &15530.5 &3147.9 &2894.3 &2103.6 &1999.1 &373.0 & 238.7 &14.9&\\
$e bjj$&330834.0 & 31317.5 &23381.6 & 575.6 &134.0 & 97.6 & 63.7 & 63.7 & 2.5 & 0.8 & 0.8&\\
$e t \bar t$& 1425.5 & 1313.8 & 1136.9 &131.8 & 21.4 & 21.1 & 8.4 &8.3 &0.6 &0.5 &0.3&\\
$e jjj$&37224100.0 &4943049.0 &3940097.5 & 8862.8 &1611.4 &      1128.0 &966.9 &966.9 & 161.1 & 0.0 &0.0&B=16.0\\
$\nu bbj$&21385.4 &      4040.5 & 112.0 & 10.9 &1.4 & 1.0 & 0.0 & 0.0 & 0.0 & 0.0 & 0.0&\\
$\nu bjj$&4077.8 & 985.1 & 13.3 &0.4 & 0.1 &0.1 & 0.0 &0.0 & 0.0& 0.0 & 0.0&\\
$\nu t b$& 84395.2 & 51227.5 & 3848.3 & 405.4 &78.2 &75.4 &2.1 & 2.1 &0.0 &  0.0 & 0.0&\\
$\nu jjj$&3870920.0 &  718974.7 & 1675.5 &0.0 & 0.0 & 0.0 & 0.0 & 0.0 &0.0 &0.0 & 0.0&\\
\hline
\hline
$e4$,82.24& 433.8 &210.9 &198.6 & 28.0 & 10.8 &10.6 &6.2 &6.0 &3.3 &3.1 &1.1&0.28(0.9)\\
\hline
$e bbj$&2688390.0 &176102.0 & 135388.5 & 15530.5 &2745.1 & 2566.0 &1879.8 &1849.9 &432.6 &283.5 &14.9&\\
$e bjj$&330834.0 & 31317.5 & 23381.6 &575.6 &130.7 &93.5 & 60.4 &60.4 &     1.7 &  0.0 & 0.0&\\
$e t \bar t$&1425.5 & 1313.8 &  1136.9 &131.8 & 21.3 & 21.1 &  8.3 &  8.2 & 0.6 & 0.6 & 0.4&\\
$e jjj$& 37224100.0 &   4943049.0 &3940097.5 & 8862.8 &1450.3 & 1128.0 & 966.9 &966.9 &322.3 &161.1 & 0.0 &B=15.3\\
$\nu bbj$&21385.4 &4040.5 &  112.0 & 10.9 &  1.2 & 0.7 & 0.0 & 0.0 &  0.0 & 0.0 & 0.0&\\
$\nu bjj$& 4077.8 & 985.1 &  13.3 & 0.4 & 0.1 & 0.1 & 0.0 & 0.0 & 0.0 & 0.0 & 0.0&\\
$\nu t b$&84395.2 & 51227.5 &3848.3 & 405.4 & 78.9 &78.2 & 2.1 & 2.1 & 0.0 & 0.0 & 0.0&\\
$\nu jjj$&3870920.0 & 718974.7 & 1675.5 &  0.0 & 0.0 & 0.0 & 0.0 & 0.0  & 0.0 & 0.0 & 0.0&\\
\hline
\hline
$e5$,88.07&383.2 &198.0 & 186.2 &27.2 &9.5 &9.3 &5.2 & 6.1 & 2.6 & 2.6 &0.83&0.12(0.4)\\
\hline
$e bbj$& 2688390.0 &176102.0 &135388.5 &15530.5 &2700.3 & 2506.4 &1894.7 &1820.1 &343.1 &253.6 & 44.8&\\
$e bjj$&330834.0 &     31317.5 &23381.6 &575.6 &  112.5 &  80.2 & 49.6 &  48.8 &  3.3 &  2.5 &  0.0&\\
$e t \bar t$&1425.5 &      1313.8 & 1136.9 &131.8 & 21.4 & 21.2 &  8.3 &  8.2 & 0.7 & 0.7 & 0.4&\\
$e jjj$&37224100.0 &4943049.0 &3940097.5 & 8862.8 & 966.9 &966.9 & 805.7 &805.7 &  322.3 &161.1 &  0.0&B=45.2\\
$\nu bbj$&  21385.4 & 4040.5 & 112.0 &10.9 &1.1 & 0.8 &0.0 & 0.0 & 0.0 & 0.0 & 0.0&\\
$\nu bjj$& 4077.8 & 985.1 &  13.3 &  0.4 & 0.1 & 0.1 & 0.0 & 0.0 & 0.0 & 0.0 & 0.0&\\
$\nu t b$&84395.2 &51227.5 & 3848.3 &405.4 & 86.5 &  85.8 &  1.4 &  1.4 &0.0 & 0.0 &  0.0&\\
$\nu jjj$&3870920.0 &718974.7 &1675.5 & 0.0 & 0.0 & 0.0 & 0.0 & 0.0 &0.0&0.0 & 0.0&\\
\hline
\hline
\end{tabular}
}
\caption{The number of signal and backgrounds events $e$ $+$ 3-jet channel after cumulative set of 
selections at the LHeC collider with 100 fb$^{-1}$ integrated luminosity  
for selective Benchmark points consistent with all the Phenomenological, Sparticle masses, 
Relic density of cold-dark matter, direct and indirect dark matter searches and 
most up-to-dated Higgs boson data. We simulated each benchmark signal with 200K and each 
background with 400K Monte Carlo simulated events. In the second column ``RawEvt'' stands 
for the number of events with only the generator--level cuts (Eq.\ref{presel}) imposed  
for the signal as well as for backgrounds. For signal benchmarks, the proper branching factor, 
$h_1 \to b \bar b$ has been multiplied with the total cross section and  
for backgrounds with $W$-boson and $t$-quark we assume a free decay. 
In the final column we list the significances(${\cal S}$) defined as 
${\cal S} = S / \sqrt{B}$, where $S$($B$) stands for signal (background) events for 100 fb $^{-1}$ 
of data after all cuts mentioned in the ``j'' column are implemented. 
The Significances for 1000 fb $^{-1}$ are shown in the parenthesis.}
\label{tab:lepori}
\end{table}

\begin{table}[t!]
\centering
\scalebox{0.7}{
\begin{tabular}{||c||c|c|c|c||c|c|c|c||}
\hline
\hline
&&&&&&&&\\
BP,$m_h$& $\eta_{l}$,$\Delta {\eta_{jl}}$,$m_{\phi j}$,$H_T$,$\vec H_T$&S&B&${\cal S}$&$\eta_{l}$,$\Delta {\eta_{jl}}$,$m_{\phi j}$,$H_T$, $\vec H_T$ &S&B&${\cal S}$\\
\hline
$e$1,63.59&$(1.0,-1.0),(0.0,-4.3),180,130,60$&4.9&162.3&0.38(1.2)&$(1.6,-2.5),(0.3,-6.0),100,140,30$&12.8&412.7&0.63(1.99)\\
$e$2,70.59&$(1.0,-1.0),(0.0,-3.0),180,140,60$&2.7&1.3&2.36(7.5)&$(1.1,-2.5),(0.2,-5.7),90,90,30$&10.1&1295.3&0.28(0.89)\\
$e$3,75.29&$(1.0,-2.5),(0.4,-3.4),180,140,60$&3.1&1.5&2.53(8.0)&$(1.0,-2.1),(0.4,-6.0),120,110,30$&11.6&565.2&0.49(1.54)\\
$e$4,82.24&$(1.0,-1.4),(0.0,-3.4),180,140,60$&1.6&0.6&2.09(6.6)&$(1.0,-2.1),(0.1,-3.4),110,140,30$&4.1&154.1&0.32(1.0)\\
$e$5,88.07&$(1.0,-1.8),(0.0,-3.0),180,140,60$&1.3&2.4&0.85(2.7)&$(1.3,-2.1),(0.1,-5.9),150,140,30$&4.8&340.0&0.26(0.82)\\
\hline
\hline
\end{tabular}
}
\caption{
The optimization of the signal channel with different sets of kinematical selection cuts, e.g., 
$\eta_l$, $\Delta {\eta_{jl}}$, $m_{\phi j}$, $H_T$, magnitude of vector $\vec {H_T}$ (see text for details) 
with the best significance obtained for 100fb$^{-1}$. In the right--sided columns we 
required strictly that the number of signal events must be greater than 10 or at least 
approximately 5 for the low luminosity options. The significances in the parenthesis are for 1000fb$^{-1}$.
}
\label{tab:optlep}
\end{table}

\begin{table}[t!]
\centering
\scalebox{0.7}{
\begin{tabular}{||c|c||c|c|c|c|c|c|c|c|c||c||}
\hline
&&&&&&&&&&&\\
Proc,$m_h$&RawEvt&A&B&C&D&E&F&G&H&I&${\cal S}$\\
\hline
\hline
$\nu$-1,65.93&4114.1 &1540.7 & 1475.6 & 200.6 &   112.5 & 111.2 & 102.0 & 49.8 & 47.8 &26.0&3.34(10.6)\\
\hline
$\nu bbj$&21385.4 & 4040.5 & 3928.5 &       244.7 &        43.0 &    33.1 &    32.1 &     5.3 &     5.2 &  4.1 &\\
$\nu bjj$&4077.8 &985.1 &  971.8 &        16.4 &         4.6 &     3.0 &     2.8 &     0.1 &     0.1 &  0.0 &\\
$\nu t b$&84395.2 & 51227.5 & 47379.2 &      3671.2 &       709.1 &   661.3 &   598.4 &    27.7 &    26.3 & 12.5&\\
$\nu jjj$&3870920.0 & 718974.7 & 717299.2 &       722.2 &       173.3 &   144.4 &   144.4 &    28.9 &    28.9 & 28.9& B=60.5\\
$e bbj$& 2688390.0 & 176148.3 &40728.9 &      5302.1 &      1389.0 &  1254.6 &   507.8 &   149.4 &   104.5 & 14.9&\\
$e bjj$& 330834.0 &31317.5 & 7935.9 &       214.2 &        58.7 &    34.7 &    11.6 &     0.8 &     0.8 &  0.0&\\
$e t \bar t$&1425.5 &1313.8 & 176.9 &        21.7 &         3.0 &     3.0 &     2.5 &     0.1 &     0.1 &  0.1&\\ 
$e jjj$&37224100.0 &   4943049.0 &    1002951.5 &     3706.3 &  644.6 &   644.6 &   0.0 &     0.0 &  0.0 &  0.0&\\
\hline
\hline
$\nu$-2,71.32&2323.5 &917.2 &879.1 &123.5 & 61.4 &60.9 &55.8 &27.3 & 26.1 & 14.2&1.84(5.8)\\
\hline
$\nu bbj$&21385.4 & 4040.5 &      3928.5 &       244.7 &        35.9 &    28.2 &    27.2 &     4.5 &     4.5 &  3.5 & \\
$\nu bjj$&4077.8 & 985.1 &       971.8 &        16.4 &         4.3 &     2.9 &     2.6 &     0.1 &     0.1 &  0.0 & \\
$\nu t b$&84395.2 &  51227.5 &     47379.2 &      3671.2 &       727.0 &   681.4 &   626.0 &    31.8 &    30.4 & 17.3 & \\
$\nu jjj$&3870920.0 & 718974.7 &    717299.2 &       722.2 &       202.2 &   173.3 &   173.3 &    28.9 &    28.9 & 28.9 &B=59.7\\
$e bbj$&  2688390.0 & 176148.3 &     40728.9 &      5302.1 &      1344.2 &  1209.8 &   433.1 &    89.6 &    59.7 &  0.0 &\\
$e bjj$& 330834.0 & 31317.5 &      7935.9 &       214.2 &        59.5 &    36.4 &    15.7 &     0.8 &     0.8 &  0.0 &\\
$e t \bar t$& 1425.5 & 1313.8 &       176.9 &        21.7 &         3.3 &     3.3 &     2.8 &     0.1 &     0.1 &  0.0 & \\
$e jjj$&37224100.0 & 4943049.0 & 1002951.5 &  3706.3 &  483.4 &   322.3 &     0.0 &     0.0 &     0.0 &  0.0 &\\
\hline
\hline
$\nu$-3,83.77&1830.6 & 845.3 &809.8 &116.0 &        44.0 &    43.4 &    40.0 &    21.2 & 20.7 & 11.1 &1.54(4.9)\\
\hline
$\nu bbj$& 21385.4 &4040.5 &      3928.5 &       244.7 &        24.2 &    20.0 &    19.5 &     3.8 &     3.8 &  3.0&\\
$\nu bjj$& 4077.8 &  985.1 &       971.8 &        16.4 &         3.2 &     2.2 &     2.0 &     0.1 &     0.1 &  0.1&\\
$\nu t b$&  84395.2 & 51227.5 &     47379.2 &      3671.2 &       736.0 &   695.9 &   643.3 &    36.0 &    35.3 & 20.1&\\
$\nu jjj$&  3870920.0 &718974.7 &    717299.2 &       722.2 &       202.2 &   173.3 &   173.3 &    28.9 &28.9 & 28.9& B=52.1\\
$e bbj$& 2688390.0 &176148.3 &     40728.9 &      5302.1 &      1015.6 &   896.1 &   388.3 &   104.5 &    74.7 &  0.0& \\
$e bjj$&   330834.0 &31317.5 &      7935.9 &       214.2 &        38.9 &    25.6 &    13.2 &     0.0 &     0.0 &  0.0&\\
$e t \bar t$&1425.5 & 1313.8 &       176.9 &        21.7 &         3.6 &     3.6 &     2.9 &     0.2 &     0.2 &  0.0&\\ 
$e jjj$& 37224100.0 &   4943049.0 &   1002951.5 &  3706.3 &   322.3 &     0.0 &     0.0 &  0.0 & 0.0 &  0.0&\\
\hline
\hline
$\nu$-4,88.07& 2018.0 & 975.7 & 932.1 &133.8 & 47.0 &46.5 & 42.8 & 23.9 &    23.5 & 12.6&1.75(5.5)\\
\hline
$\nu bbj$& 21385.4 & 4040.5 & 3928.5 & 244.7 &        22.0 &    18.3 &    18.0 &     3.6 &     3.6 &  2.8&\\
$\nu bjj$& 4077.8 &  985.1 &  971.8 &  16.4 &         3.0 &     2.0 &     1.9 &     0.1 &     0.1 &  0.1& \\
$\nu t b$&  84395.2 &51227.5 &47379.2 & 3671.2 &       720.1 &   680.0 &   624.7 &    34.6 &    33.2 & 20.1&\\
$\nu jjj$& 3870920.0 &718974.7 &717299.2 & 722.2 &       144.4 &   115.5 &   115.5 &    28.9 &    28.9 & 28.9&B=52.0\\
$e bbj$& 2688390.0 &176148.3 &40728.9 & 5302.1 &       970.8 &   866.3 &   403.3 &   104.5 &    74.7 &  0.0&\\
$e bjj$& 330834.0 & 31317.5 &      7935.9 & 214.2 &        41.4 &    28.1 &    16.5 &     0.8 &     0.8 &  0.0&\\
$e t \bar t$& 1425.5 &1313.8 & 176.9 & 21.7 &         3.8 &     3.8 &     3.0 &     0.2 &     0.2 &  0.1&\\ 
$e jjj$&37224100.0 & 4943049.0 & 1002951.5 &3706.3 & 161.1 & 0.0 & 0.0 &  0.0 &  0.0& 0.0&\\
\hline
\hline
$\nu$-5,100.47&1125.0 &613.9 &585.0 &  88.2 &  23.5 &    23.3 &    21.5 &12.0 &    11.9 &6.1&1.41(4.5)\\
\hline
$\nu bbj$& 21385.4 & 4040.5 &      3928.5 &       244.7 &        15.0 &    11.8 &    11.6 &     2.4 & 2.4 &  1.9&\\
$\nu bjj$&  4077.8 & 985.1 &       971.8 &        16.4 &         1.9 &     1.3 &     1.3 &     0.1 & 0.1 &  0.1&\\
$\nu t b$& 84395.2 &51227.5 &     47379.2 &      3671.2 &       644.0 &   615.0 &   562.4 &    39.4 &38.7 & 16.6&\\
$\nu jjj$& 3870920.0 &718974.7 &    717299.2 &       722.2 &       115.5 & 86.7 &    86.7 &    28.9 &28.9 &  0.0&B=18.7\\
$e bbj$& 2688390.0 &176148.3 &     40728.9 &      5302.1 &       806.5 &   672.1 &   313.6 &    59.7 &29.9 &  0.0 &\\
$e bjj$& 330834.0 &31317.5 &      7935.9 &       214.2 &        25.6 &    17.4 &    11.6 &     2.5 &2.5 &  0.0&\\
$e t \bar t$& 1425.5 &1313.8 &       176.9 &        21.7 &         4.0 &     3.9 &     3.0 &     0.2 &0.2 &  0.1&\\ 
$e jjj$&37224100.0 & 4943049.0 &   1002951.5 &  3706.3 &   483.4 & 322.3 & 0.0 & 0.0 &  0.0 &  0.0&\\
\hline
\hline
\end{tabular}
} 
\caption{Same as of previous table but for $\met + $ 3-jet channel. In the final column 
we mention the significances(${\cal S}$) defined as ${\cal S} = S / \sqrt{B}$, where $S$($B$) for 
signal (background) events for 100 fb $^{-1}$ of data after all cuts mentioned in the ``I'' column are  
implemented. The significances for 1000 fb $^{-1}$ are shown in the parenthesis.}
\label{tab:met}
\end{table}

\subsubsection{$e$ $+$ 3-jet: electron channel}

In this subsection we will analyze the neutral current signal electron channel, i.e., $e$ $+$ 3-jet 
and apply different kinematical cuts to isolate the signal from the backgrounds.

\begin{itemize} 

\item {\bf a:} We first selected events containing at least three jets, i.e., $N_{jet} \gsim 3$, 
with $Pt_j > 15.0$ and $\eta_j < 5.5$. The distribution of the number of jet 
($N_{jet}$) is shown in the left panel of Fig. \ref{njetbtag}. The efficiency of having 
$N_{jet} \gsim 3$, are approximately, 40.0\%, 42.8\%, 45.3\%, 45.6\% and 51.7\% 
\footnote{The efficiencies quoted here are with respect to the previous set of selections.} for the 
signal benchmarks of $e$1 to $e$5, respectively. Since the two central jets 
originate directly from the Higgs boson itself, heavier masses lead to 
higher efficiencies. The jet selection criterions are the same for all the signals and corresponding 
backgrounds. Thus the jet efficiencies are identical in all backgrounds.  
Among all the backgrounds, $e t \bar t$ leads to a total of six-jets when both the top-quarks  
decay hadronically -- here the jet efficiency is maximal, 
about 92.2\%.  The immediate next high efficiency is from $\nu t b$ -- where the maximal 
number of jets is four. This leads to an efficiency around 60.7\%. The jet--efficiency for the 
$ebbj$, $ebjj$ and $ejjj$ are in between 7.0 - 10.0\% whereas for the 
$\nu bbj$, $\nu bjj$ and $\nu jjj$ channels they range from 18.0 to  24.0\%. In the neutrino cases, first 
of all there is no strict selection of neutrinos momentum unlike  
minimum transverse momentum requirements of the lepton, i.e., $e$ in this channel. Moreover the lepton-jet 
isolation criterion reduces the number of jets in the backgrounds with explicit lepton. 
This leads to the lower efficiency. 

\item {\bf b:}  We required one lepton ($e$) with $p_T$ $>$ 15 GeV and $\eta$ $<$ 3.0 as  
our signal is generated from the neutral current interaction. Since the lepton is 
originating from the $e$-beam, we required somewhat larger rapidity. The distribution 
of number of leptons is shown in the left panel of Fig. \ref{nlepmet}. 
The lepton efficiency for all the signal benchmarks is approximately 94.0\% and for the benchmarks 
where the Higgs mass is smaller, this lepton efficiency is larger as more collison energy 
is transferred to the lepton -- however small it is, it  does get reflected in the corresponding 
efficiencies. Among the backgrounds, the larger efficiency is for $e t \bar t$ with approximately 86.5\%    
-- as one top quark is allowed to decay leptonically whereas the other top quark must decay hadronically 
satisfying the three jet criterion. The efficiencies for $ebbj$, $ebjj$ and $ejjj$ are  
between 76.9\% -- 79.7\%. The efficiency of $\nu t b$ is approximately 8.0\%. This is close to 
the electron channel branching fraction, i.e., 10\%. There are also secondary sources of electrons, 
like semi-leptonic $b$-decays or meson decays. Taking this into account, the transverse momentum and rapidity criterion reduces 
the efficiency to approximately 2\%.  
The efficiency for $\nu bbj$ is approximately 2.8\% where the source of the lepton 
is only from the semi-leptonic $b$-decay satisfying the isolation criterion or from secondary sources 
like meson or photon. The efficiency for $\nu bjj$ being 1.3\% is just the half of $\nu bbj$ as it is clear from the 
presence of b-quark in these two cases. In case of $\nu jjj$, the lepton would only be coming 
from the secondary sources (meson or photo) during fragmentation and hadronization and the efficiency 
turns out to be approximately 0.002\% \footnote{Please note that we have not considered 
here the lepton mis-tagging efficiency from the jets. This is approximately 0.001\% and 
having the three (or more) jets explicitly after considering the ISR and FSR, this efficiencies 
are somewhat consistent with the mis-tagging numbers with proper combinatorics.}.

\item {\bf c:} We demanded at least two $b$-tagged jets with the inclusion of proper 
mis-tagging. The distributions of the number of $b$-tagged jets ($N_{b-tag}$) are  
shown in the right panel of Fig. \ref{njetbtag}. The efficiencies are approximately, 
13.7\%, 13.8\%, 14.4\%, 14.1\%, and 14.6\% for the $e$1 to $e$5 benchmark points, 
respectively. In fact, all our signal benchmarks contain at least two $b$-quark. Since we considered 
$\eps_b$=0.50, an approximately 10\% lowering for the double $b$-tag is quite realistic  
due to the fact that not all $b$-quarks in the signal are eligible for the $b$-taggable 
criterion adopted in our analysis (another possibility is having a $c$-quark with $\eps_c$=0.10 
and a light flavored quark with $\eps_q$=0.01).

Among the backgrounds, the irreducible one $e b b j$ has an efficiency of approximately 
11.5\%, roughly 2\% less than the signal and this is mainly due to the rapidity 
acceptance of the taggable $b$-jet. Unlike the signal, the rapidity distribution of 
the two $b$-taggable jet for this background are not very central (i.e., we imposed the $\eta < 2.5$ a
for taggable $b$-jet). For $e t b$ and $\nu t b$ the efficiencies are approximately 11.6\% and 10.0\%, 
respectively whereas $\nu b b j$ and $\nu b j j$  have efficiencies of 8.2\% and 2.4\%.  
This $b$-tagged ratios with neutrino follow very closed the corresponding number 
efficiencies of $e b b j$ and $e b j j$. The efficiency of $e j j j$ is approximately 
0.002 where low-flavor mis-tagging efficiencies with $\eps_c$=0.10 and 
$\eps_q$= 0.01 ($q$=$u,d,s,g$) have been taken care of. We would expect similar $b$-tagged efficiencies 
(i.e, 0.002) for $\nu j j j$. However, as the lepton selection criterion (b) above severely reduced 
the number of events -- the survived events hardly pass this $b$-tag criterion and 
$\nu jjj$ goes to zero-level.

\begin{figure}[ht!]
\begin{center}
\raisebox{0.0cm}{\hbox{\includegraphics[angle=0,scale=0.42]{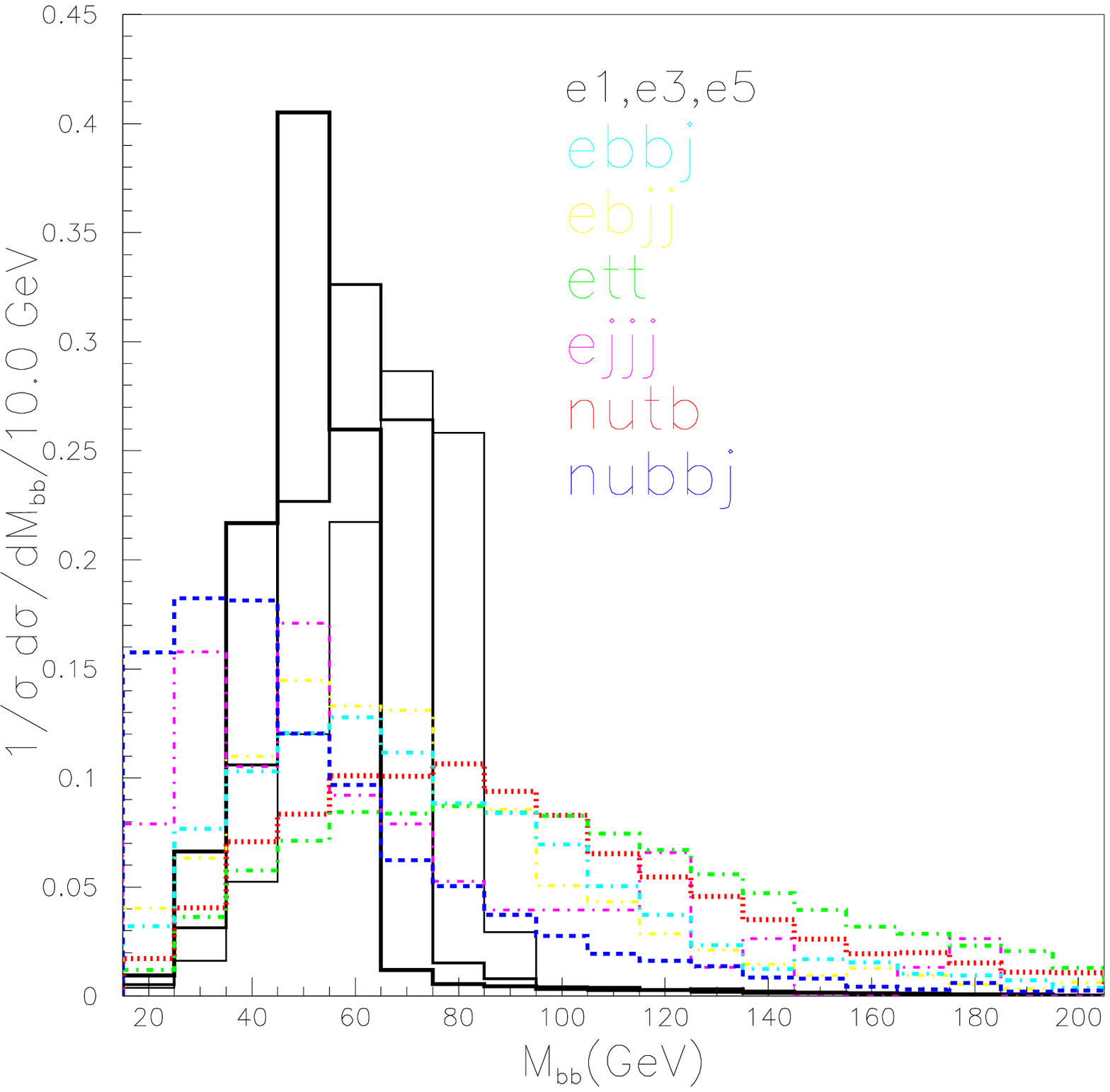}}}
\raisebox{0.0cm}{\hbox{\includegraphics[angle=0,scale=0.42]{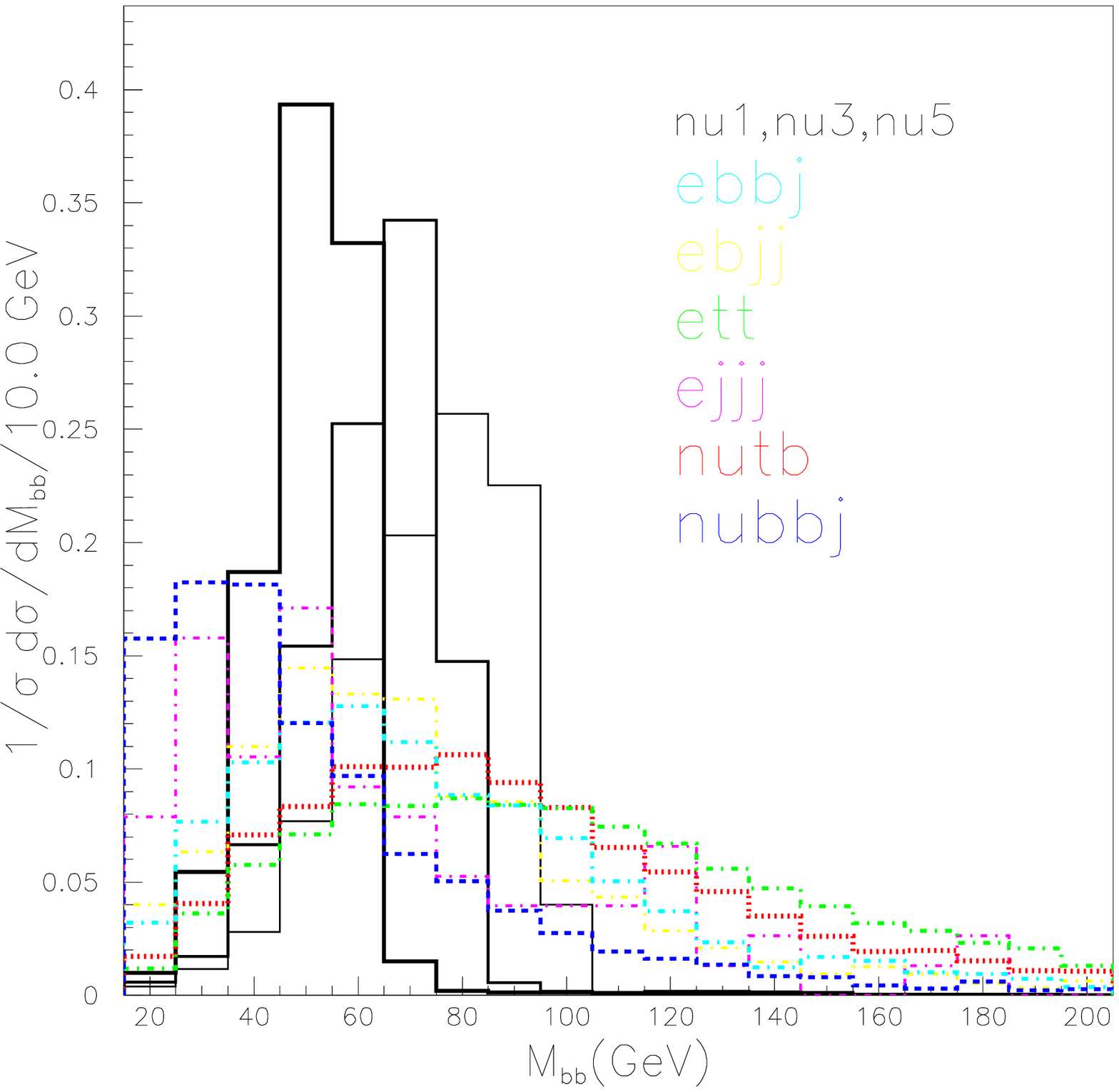}}}
\caption{Left-panel: The di-$b$-tagged invariant masses, $M_{bb}$, distribution for 
the $e$$+$ 3-jets channel, for $e$1, $e$3 and $e$5 benchmark points from left to right  
(thick to thinner) respectively. Right-panel: Similar for the $\met$ $+$ 3-jets channel, 
with the mass peaks for $\nu$1, $\nu$3, and $\nu$5 from left to right 
(thick to thinner) respectively. See the Table \ref{table:nmssmbp} for 
the corresponding Higgs boson masses.}
\label{mbs}
\end{center}
\end{figure}

\item {\bf d:} In the central region, defined above in the selection (c), having the number 
of b-tagged jets greater than or 
equal to two (i.e., $N_b$ $\gsim$2 with mis-tagging), we reconstruct all possible
combination of di-jet invariant masses, i.e., $M_{bb}$. Out of all possible combinations, we have 
chosen the best combination for which the absolute difference, $|M_{bb} - M_{h_1}|$ is minimum. 
We identified this as the correct di-$b-$jet candidates for the Higgs boson. The distribution of 
$M_{bb}$ is shown in the left panel of Fig.\ref{mbs} for $e$1, $e$3 and $e$5 benchmark points. 
We have not shown explicitly the mass peaks of $e$2 and $e$4 benchmark points, since it lies 
between the displayed peaks as the Higgs boson mass lies between the respective benchmarks 
(see Table \ref{table:nmssmbp}). The peaks of all the signal benchmarks always show up to the 
left side of the actual masses due to jet energy smearing. Moreover, the shift depends on the 
jet-cone size under considerations --  the larger the cone size the more the peak moves to 
the right. The price is having a relatively less $N_{jet}$ efficiency in (a).
The $M_{bb}$ distributions in Fig.\ref{mbs} show a rapid fall on the higher side.  
Hence we have selected events with some asymmetric mass windows: $ M_{h_1} -15.0$ GeV $< M_{bb} < M_{h_1}+5$ GeV 
\footnote{ Please note that the $M_{bb}$ distributions are shown without imposing the mass window selection. 
The numbers start to differ from this column onwards as the Higgs mass window selection 
depends upon the Higgs boson masses of the corresponding benchmarks.}. 

For the signal the efficiencies are approximately, 54.0\%, 47.7\%, 43.3\%, 38.7\%, and 35.0\% for 
$e$1 to $e$5 benchmark points, respectively. The reconstructed Higgs boson mass tends to lie in the lower 
regions which depends mainly on the jet-cone size and the showering effects. Thus, the  lower the 
benchmark Higgs boson mass, the larger is the efficiency as can be seen from the second  
benchmark point.

The SM $ebbj$ processes have a $Z$-boson exchange resonant diagram with $Z\to b\bar b$. 
This leads to the appearances of the mass peaks around 60 GeV (approximately 30 GeV  
less than the $M_Z$ due to jet energy smearing). The $\nu bjj$ mass peaks are
somewhat similar to the $ebjj$, as this has a $W$ and/or $Z$-boson 
exchange resonant diagrams. As the Higgs boson mass is very close to 
the $W$-boson mass at least in the $e$1 benchmark, the efficiencies 
are 32.5\% and 25.4\% for $\nu bjj$ and $ebjj$, respectively.

In the cases of $ett$ and $\nu t b$, the pure di-$b$-tagged jet 
is uncorrelated and as a result the mass distributions are flat. 
However, it is not always the case 
that the primary hard $b$-jet or the secondary 
$b$-jet from the top quark decay are tagged. 
In the cases, of the hadronic top decay, i.e., W-hadronic decay,  both can be mis-tagged  
and those combinations generally show the mass peaks around 80 -- 90 GeV.
The efficiencies of  this selections are approximately around 15.3\% and 18.1\% 
for $ett$  and $\nu t b$, respectively. 

The efficiency for $ejjj$ is approximately 18.2\% and the mass distribution has a peak 
in the lower side. Due to this peak shift in the lower side, among all the benchmarks 
the $ejjj$ events survive better than of the benchmarks with minimum Higgs boson mass.

\begin{figure}[ht!]
\begin{center}
\raisebox{0.0cm}{\hbox{\includegraphics[angle=0,scale=0.42]{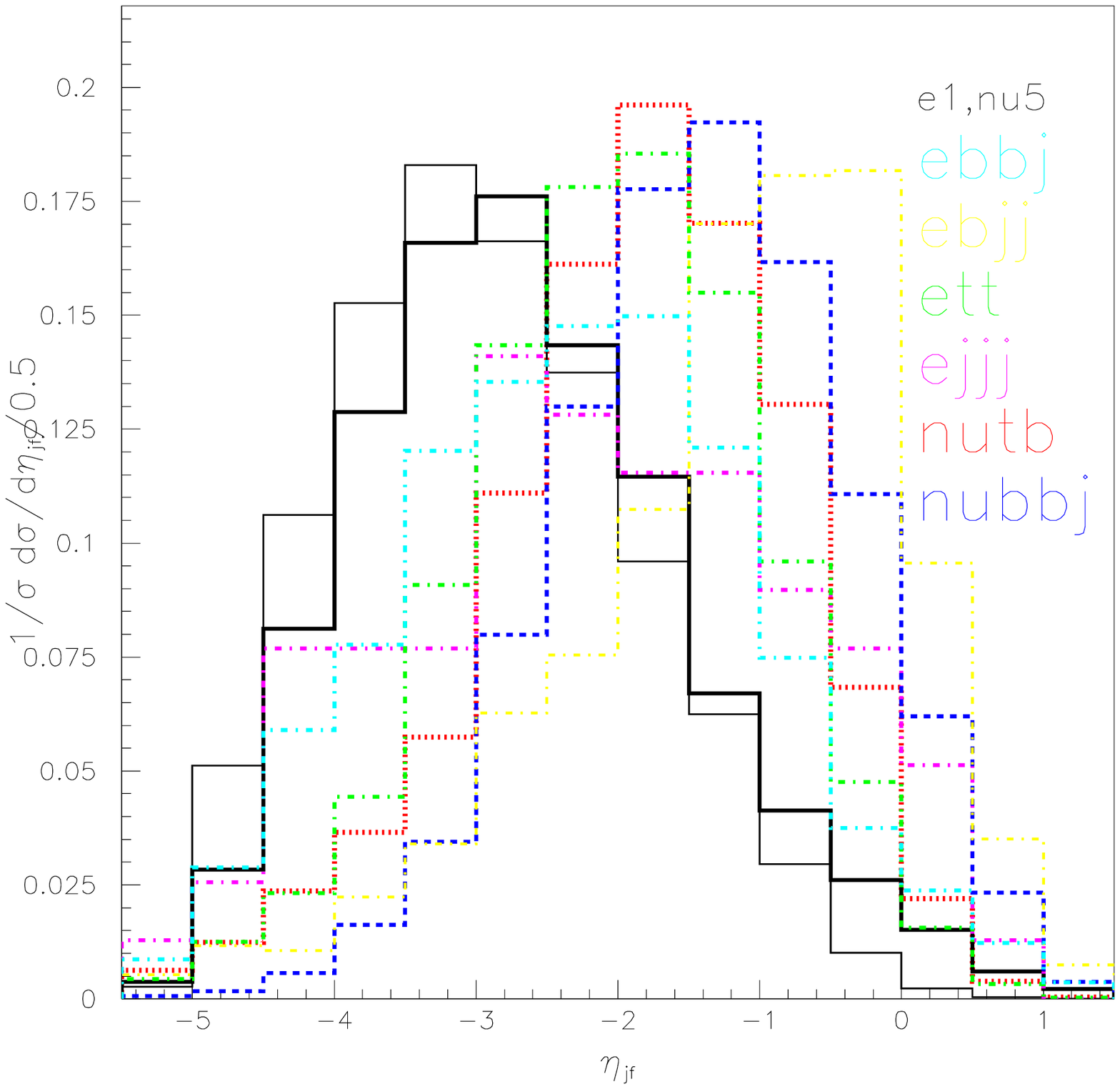}}}
\raisebox{0.0cm}{\hbox{\includegraphics[angle=0,scale=0.42]{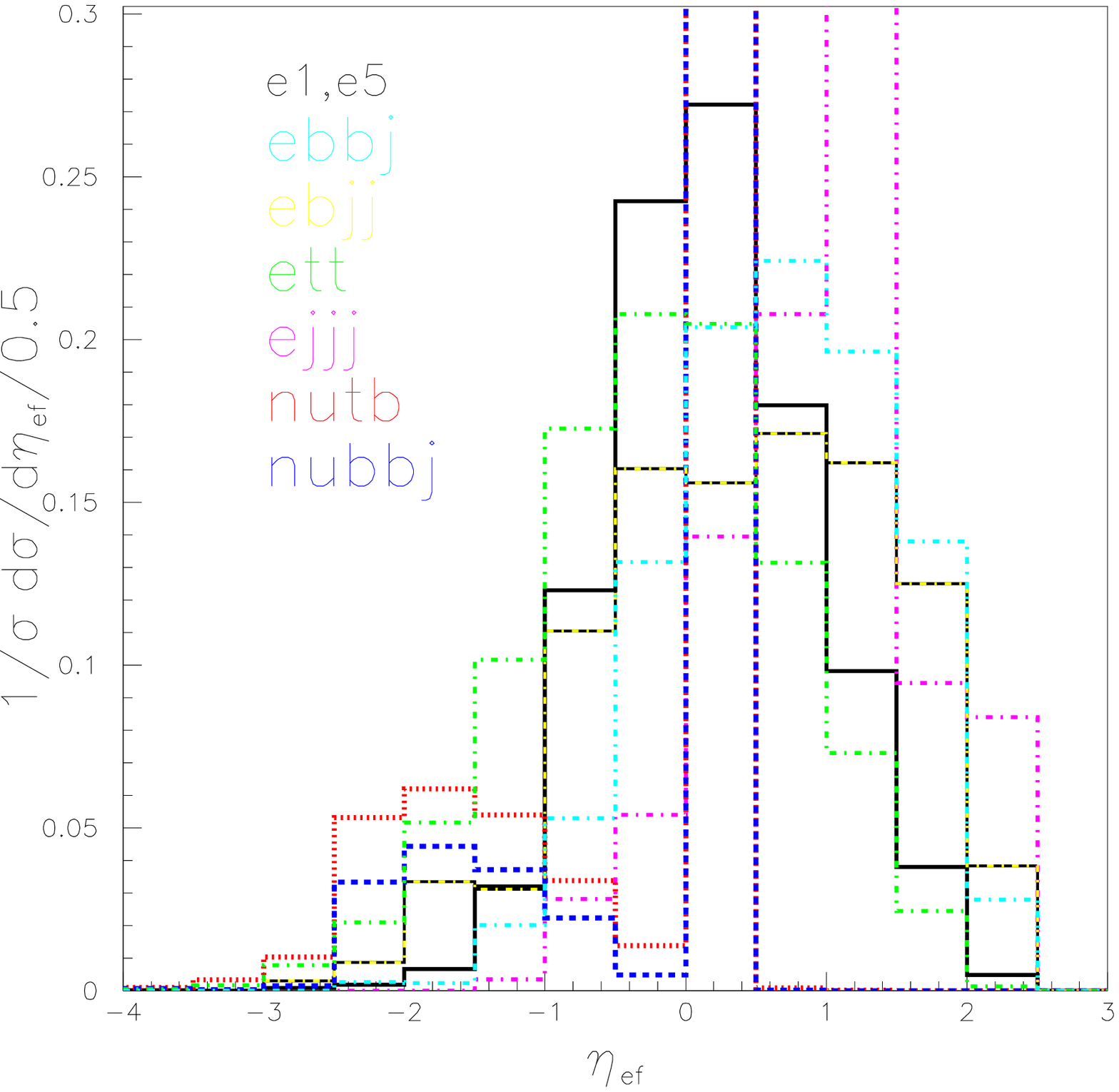}}}
\caption{Left-panel: The rapidity of the forward jet ($\eta_{jf}$) for $e$1 and $\nu$5 signal 
benchmark points. The distributions for other benchmarks lie between  
this distribution profiles. Right-panel: The rapidity of the forward lepton ($\eta_{ef}$) 
for $e$1 and $e$5 benchmark points (somewhat similar for other benchmarks).}
\label{etafjlj}
\end{center}
\end{figure}

\item {\bf e:} We demanded that the remaining leading jet in the events should 
have $Pt_j >$ 15.0 GeV with $-5.5 < \eta < -0.5$ (the values are chosen by 
inspecting the distribution shown in the left panel of Fig. \ref{etafjlj} 
with thick black line for $e$1 benchmark and termed as the forward jet 
($j_f$ or in the histogram as $jf$)). The forward jet (the transverse momentum of the forward jet 
is shown in the left-panel of Fig.\ref{ptfl}) lies very close to the direction 
of the incoming proton --  however it also depends upon the Higgs mass for the given benchmarks. The more massive the Higgs boson is, the smaller the energy which remains for the 
forward jet as it lies close to the proton beam (larger rapidity). Vice versa, 
the less massive the Higgs, the larger the rapidity of the forward jet. With the exception of the $e$2  
benchmark, the survival probability under this selection is larger with heavy Higgs boson masses. 
The backgrounds like $\nu tb$ and $ett$ have generally a large number of jets (shown 
in the left panel of Fig. \ref{njetbtag}). Thus one out of many jets would likely 
pass this selection criterion and hence the efficiencies are as large as 
the signal efficiency. The backgrounds $ebjj$ and $\nu bjj$ have somewhat 
similar efficiencies and are reduced maximally by this selection.

After applying this selection, the dominant remaining backgrounds are from the irreducible 
$ebbj$,  $ebjj$, $\nu tb$ and a part from $ejjj$ as the latter process has a big 
cross-section to begin with.

\item {\bf f:} We demand the magnitude of product of lepton and jet rapidity  
to be negative, i.e., $\eta_j$.$\eta_l$ $<$ 0. This is to say that they lie 
in the opposite hemisphere. For this selection, the efficiencies for the $e$1 to $e$5 
signal benchmarks, vary, as 53.9\%, 55.2\%, 55.9\%, 57.8\% and 55.8\% respectively. 
This selection reduces the high multiplicities backgrounds severely. For example, 
$\nu t b$ ($ett$) survived by approximately 1.5\% (39.5\%).

\item {\bf g:} Like the forward jet cut mentioned above in the item (e), 
we assume the rapidity of the lepton to be in the forward region
since the lepton in the signal source 
is directly from  the $e$-beam\footnote{Please note that the 
lepton(i.e, $e$) and the forward jet ($j_f$) are likely to be 
in the opposite hemisphere, i.e., if the jet is in the forward region then 
the lepton will be in the backward region or vice-versa. This is also reflected 
in  the rapidity distributions in the left and right panel of Fig.\ref{etafjlj}.}. 
The transverse momentum of the forward lepton is shown in the right-panel of Fig.\ref{ptfl}.
By inspecting the rapidity distribution of this lepton shown in the right 
panel of Fig.\ref{etafjlj}, we selected events with lepton 
rapidity within the range $-2.5 <\eta <2.0$. The lepton satisfying this criterion 
is called forward lepton ($e_f$). The signal events which survived this criterion are 
approximately 94.0\% while for $\nu t b$ it reduces to approximately 70\% -- 85\% as 
the source of the lepton in this case is the top quark decay and there is 
no guarantee that the lepton should be in the forward direction with the imposed criterion.
The background $ett$ would be maximally three leptons and it is likely
that one out of three leptons will pass the forward rapidity criterion 
$-2.5 <\eta <2.0$. Thus one would expect large survival probability which indeed is approximately 80\%.
In Table 4 we tabulated the events after imposing the rapidity gap between the forward 
lepton and forward jet, i.e., $\Delta \eta_{jf-ef}$. We have shown the distribution of rapidity of lepton 
in the right-panel of Fig.\ref{etafjlj} and the rapidity differences with jet in 
the left-panel of Fig.\ref{mbbj}. We demanded $-5.5$ $<$ $\Delta \eta_{jf-ef}$ $<$ $0.5$ 
and due to that the signal benchmarks reduced by approximately 2.0 -- 3.0\%. All the 
SM backgrounds remain same except the $ebjj$ changes from 93.3\%, 94.9\%, 95.7\%, 99.2\%  
and 96.1\% for $e$1 to $e$5 signal benchmarks, respectively.

\item {\bf h:} For the di-$b$-jet for which $M_{h_1}$ = $m_{bb}$ as in the selection (d) above 
and with the forward tagged jet ($j_f$), we reconstructed three-jet invariant mass, $m_{h_1 j_f} = m_{bb j_f}$.
This essentially reflects the overall energy scale of the hard scattering. 
The distributions are shown in the right-panel of Fig. \ref{mbbj}. We impose the condition $m_{h_1 j_f}$ $>$ 190 GeV. 
With this selection the signal events for all the benchmarks remain approximately 50.2 -- 54.5\%. 
The most dominant irreducible background $ebbj$ remains approximately 14.0\% -- 24.0\% whereas 
$ebjj$ remains approximately 7.0\% or less. The backgrounds $ett$ remains approximately 
8.0\% or less. The $\nu tb$ becomes zero.

At this stage the most dominant backgrounds which remain are: $ebbj$, $ebjj$ (at the level of 
signal events) and $ejjj$.

\begin{figure}[ht!]
\begin{center}
\raisebox{0.0cm}{\hbox{\includegraphics[angle=0,scale=0.42]{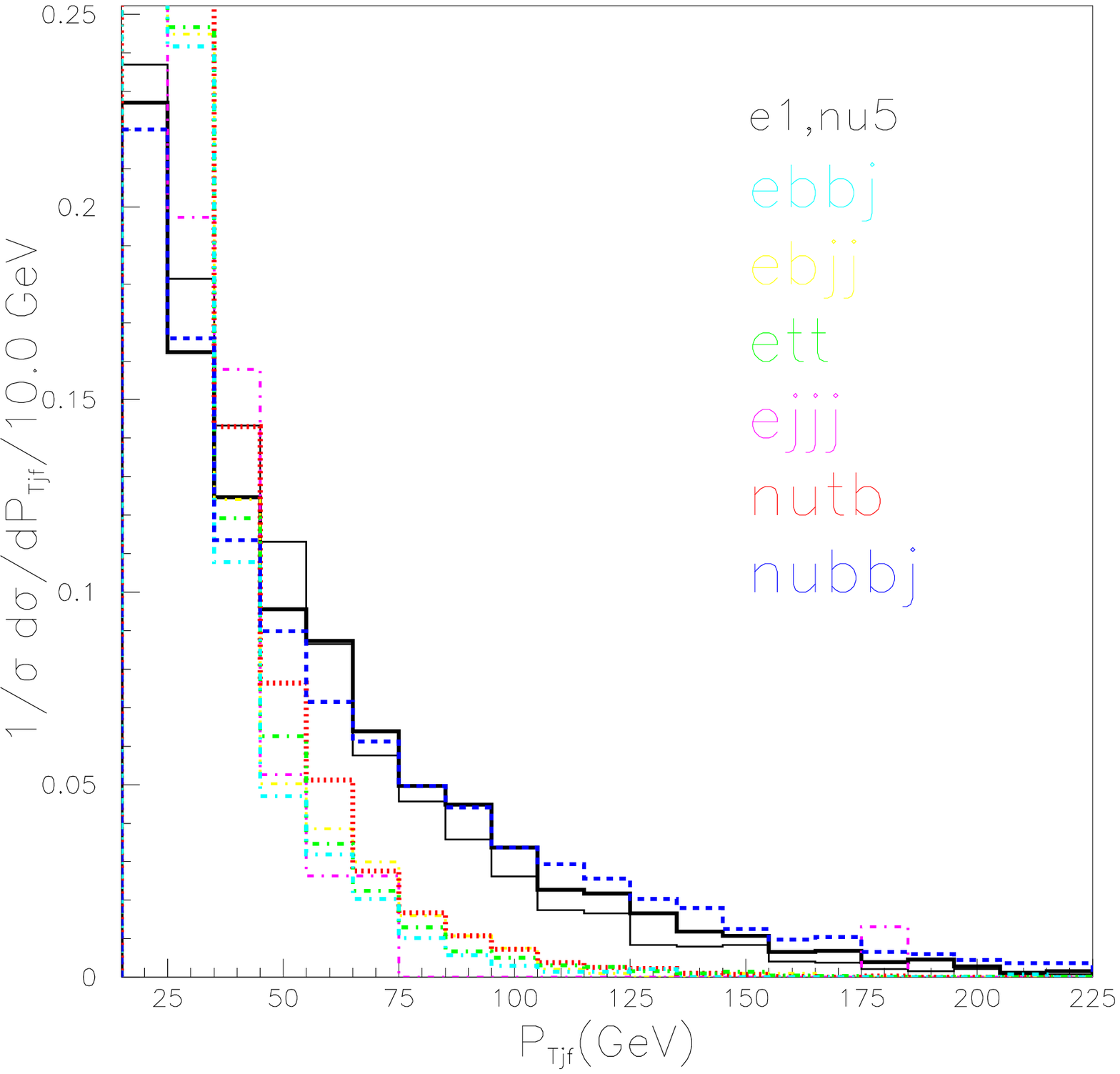}}}
\raisebox{0.0cm}{\hbox{\includegraphics[angle=0,scale=0.42]{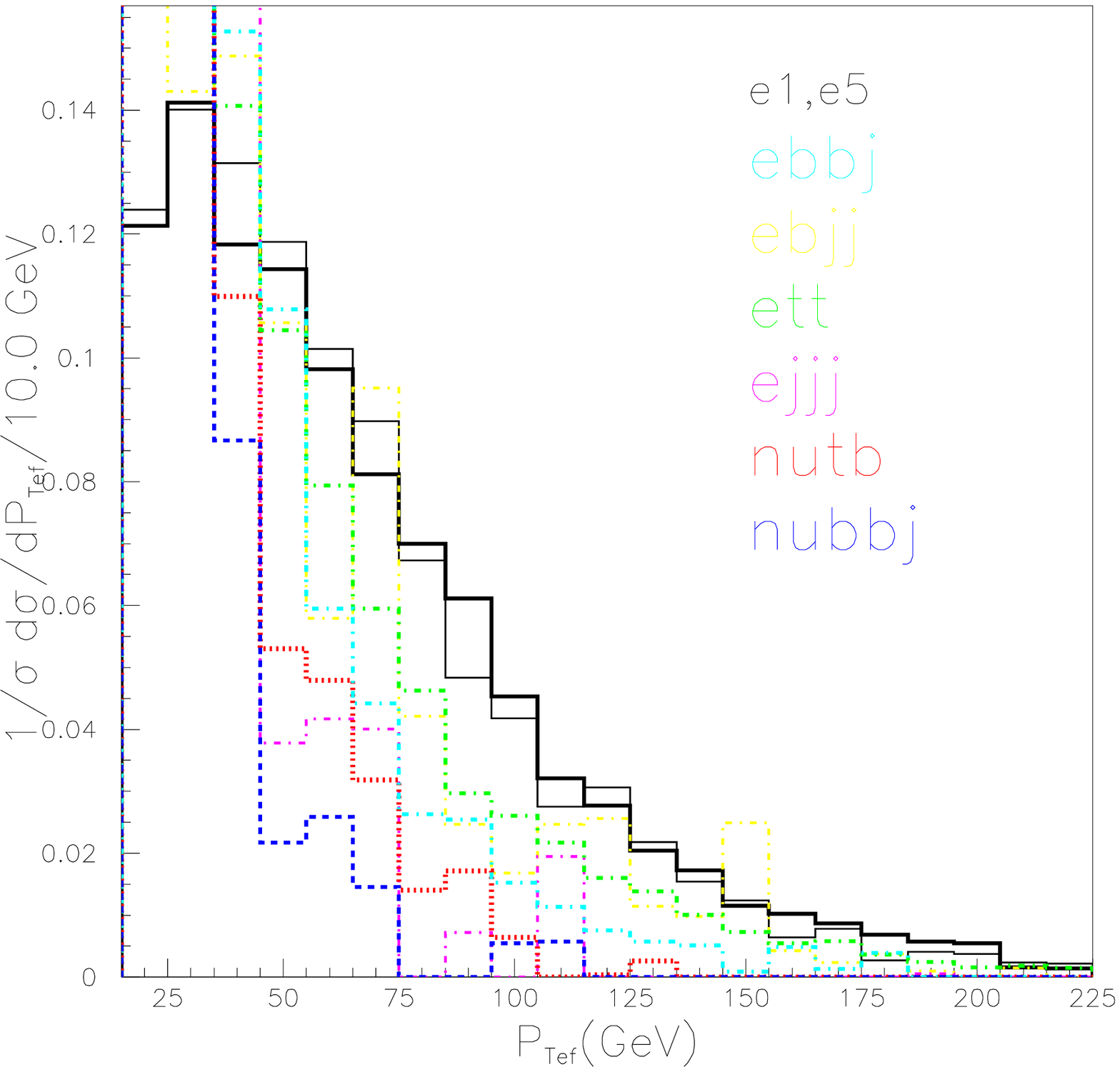}}}
\caption{Left-panel: The transverse momentum ($P_T$) of forward jet ($P_{Tjf}$) for $e$1 and $\nu$5 
signal benchmarks. Right-panel: The same for the forward lepton ($P_{Tef}$) only in the electron channel 
for $e$1 and $e$5 signal benchmarks..}
\label{ptfl}
\end{center}
\end{figure}

\begin{figure}[ht!]
\begin{center}
\raisebox{0.0cm}{\hbox{\includegraphics[angle=0,scale=0.42]{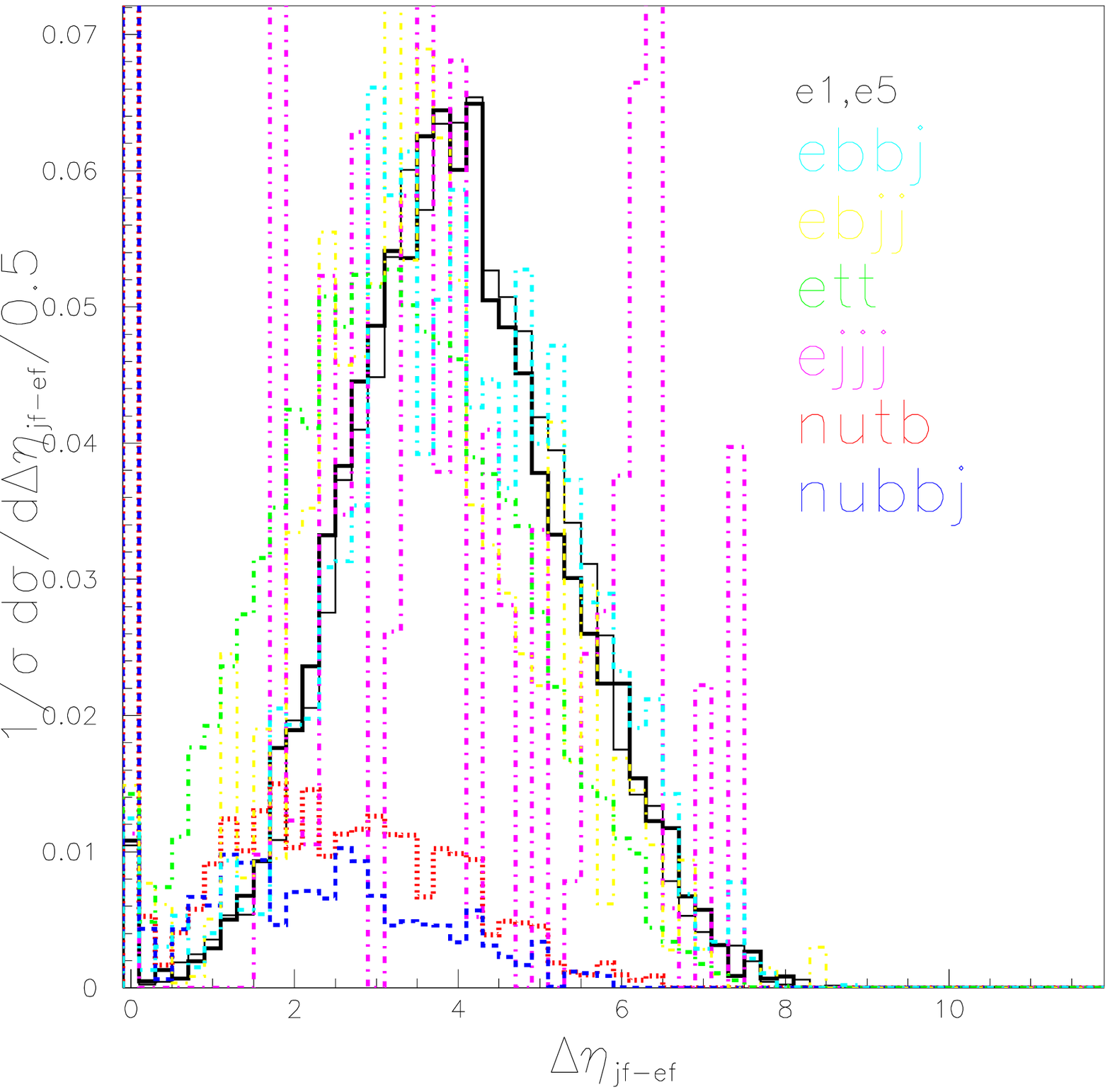}}}
\raisebox{0.0cm}{\hbox{\includegraphics[angle=0,scale=0.42]{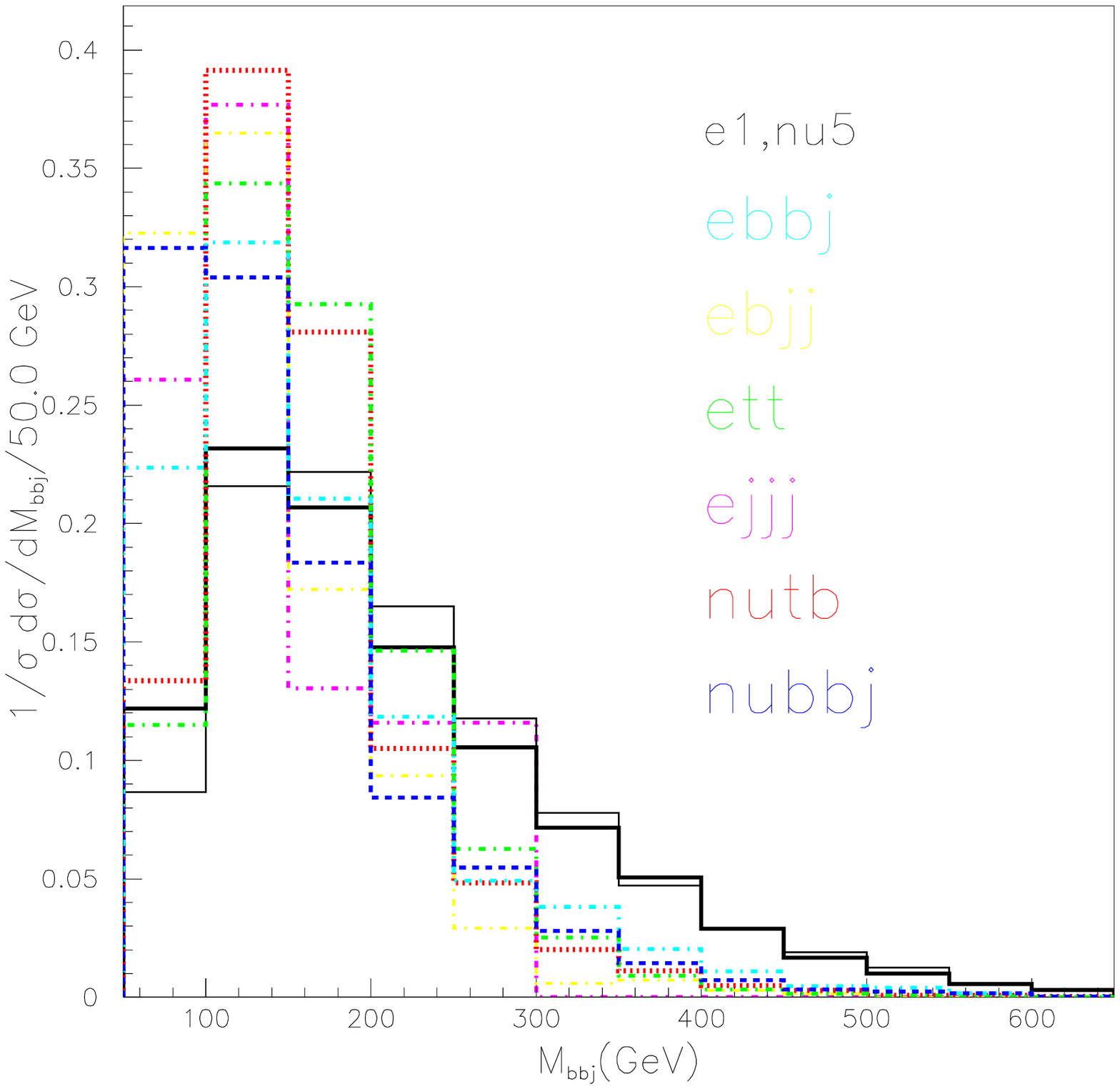}}}
\caption{Left-panel: The rapidity differences between forward jet ($j_f$=$jf$) and  
forward lepton ($e_f$=$ef$), i.e., $\Delta \eta_{jf -- ef}$ only for $e$1 and $e$5 signal benchmarks 
in electron channel. The signal shows large rapidity gap. Right-panel: The invariant masses of the Higgs 
candidate jets, $M_{bb}$ shown in Fig.\ref{mbs}, together with the forward jet, i.e., 
$M_{bb j_f}$ = $m_{h_1 j_f}$ for both the signal channels for $e$1 benchmark ($\nu$5 benchmark) 
using thick(thin) black lines. The signal distributions in both channels 
do not differ as the electron and $\met$ does not have a direct big role to reconstruct the 
three--jet invariant mass. The distributions for other signal benchmarks in both  
the channels are somewhat identical.}
\label{mbbj}
\end{center}
\end{figure}

\item {\bf i:} We devised another set of selection based on the sum of the transverse 
momentum of all jets present in the events, $H_{T}$ = $\sum |Pt_j|$. The distribution 
is shown in the left panel of Fig. \ref{hthtv}. The signal shows the peak around 100 GeV. 
The $\nu t b$ shows a peak around 125 GeV whereas $e t t$ displays it  around 250 GeV --  
the higher value reflects the presence of more number of jets. We demanded a selection of 
$H_T$ $>$ 100 GeV. The number of signal events for all the benchmarks remains approximately 
95.0\% -- 98.0\% (for heavy Higgs boson masses the survival events probablities are more). 
We see that this selection is not reducing much of $e t t$, at most 2.0\%. However 
this background is not big at this stage. The $ebjj$ contribution is also not much, 
it survived by at most 75.0\% for $e$5 benchmark. With this 
selection $ebbj$ is reduced to approximately 52.2\%, 55.6\%, 64.0\%, 65.5\% and 73.9\% 
for $e$1 to $e$5 benchmark points respectively. In spite of large large reduction, 
$ebbj$ is the only dominant contribution at this stage.

\item {\bf j:} Finally, to suppress the $ebbj$ further, we devise a new 
kinematical variable by adding the vector component of three momentum of 
all the jets present in the event. This is defined as: $\vec H_{T}$ = $\sum \vec Pt_j$. 
The distribution is shown in the right panel of Fig. \ref{hthtv} and it is evident that 
here the peaks are at a lower value of $|\vec{H}_{T}|$ as compared to the left panel of the 
same figure ($H_{T}$ is defined in the selection (i) above). In the events having 
more jets,  the jets are naturally distributed symmetrically in the 
$\eta$-$\phi$ plane. This regular arrangements of jets three momentum tend to cancel each other  
which leads to a lower magnitude of $\vec H_{T}$. For example, for 
$e tt$, $\nu tb$ the peaks are around 40 GeV, while for the  $ebbj$, $ebjj$, and $ejjj$ 
around 25 GeV or less. By demanding the magnitude of $|\vec H_{T}|$ $>$ 50 GeV, the signals 
are reduced by approximately 30.0\% -- 35.0\%. The background $ebbj$ is reduced  
severely by approximately 85\% -- 95\%.

\begin{figure}[ht!]
\begin{center}
\raisebox{0.0cm}{\hbox{\includegraphics[angle=0,scale=0.42]{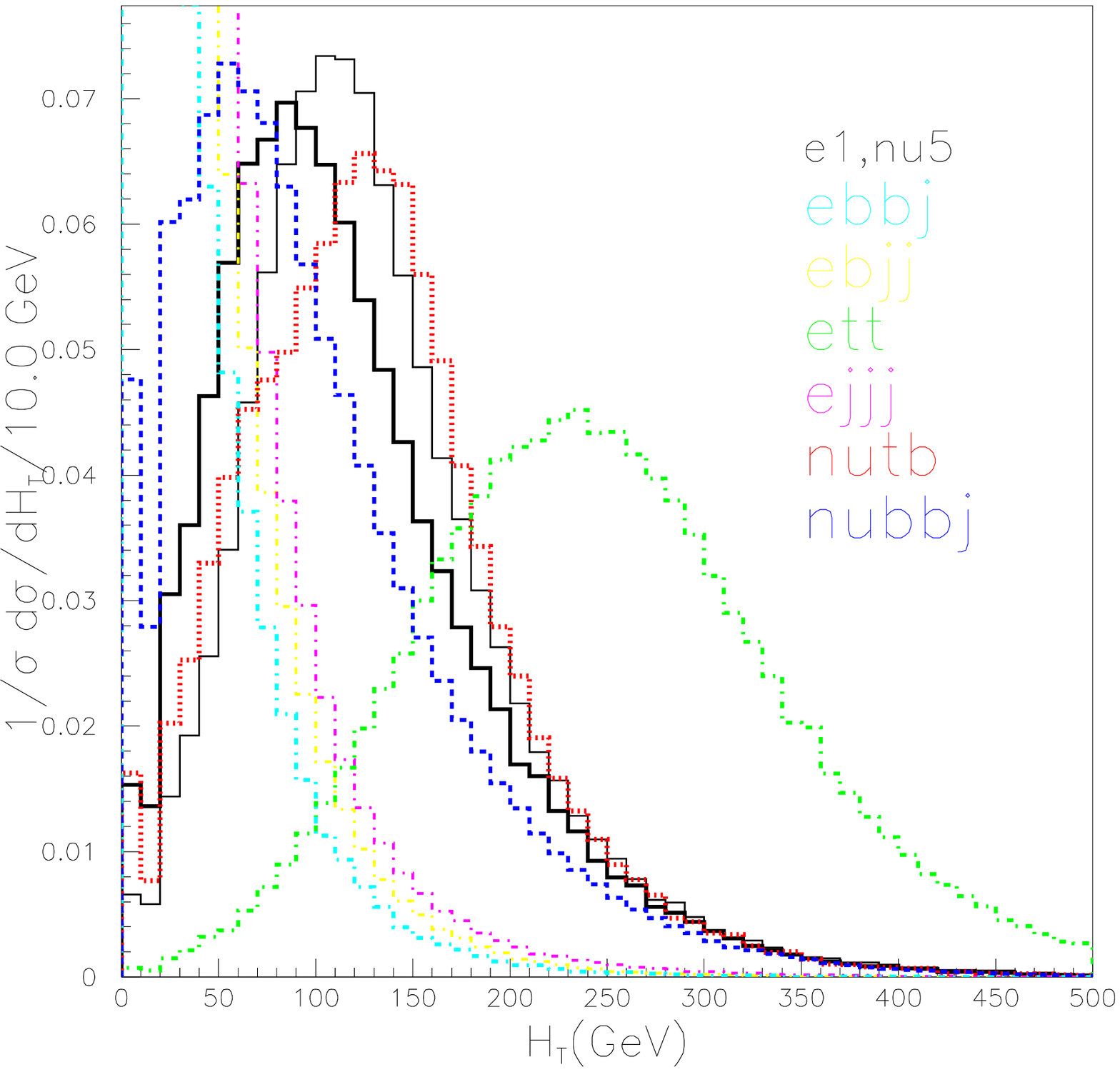}}}
\raisebox{0.0cm}{\hbox{\includegraphics[angle=0,scale=0.42]{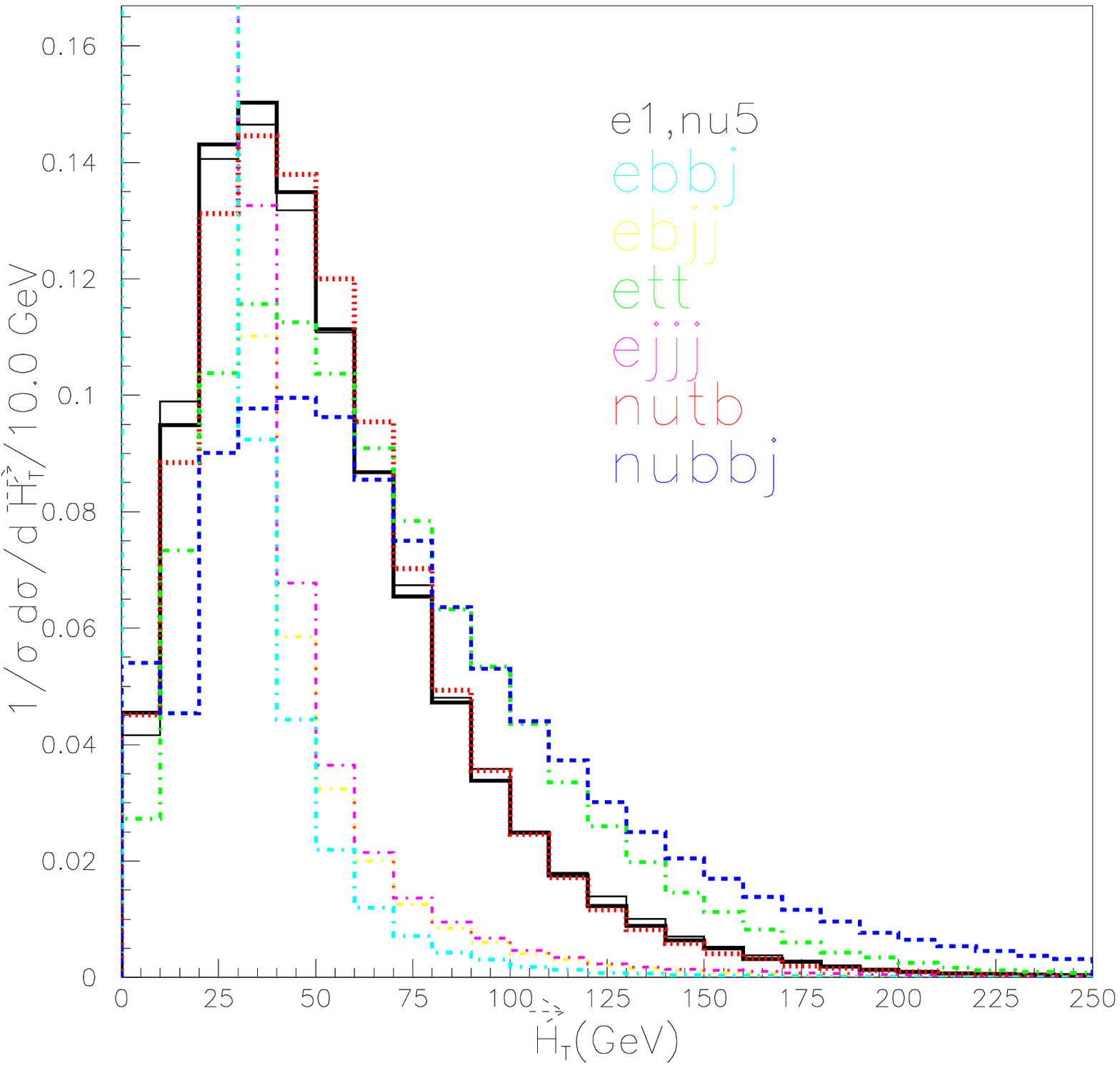}}}
\caption{Left-panel: $H_T$ = $\sum |Pt_j|$ distribution for $e$1 benchmark and 
$\nu$5 benchmark. The distributions for other signal benchmarks in both  
the channels are somewhat identical. Right-panel: $\vec H_T$ = $|\sum \vec Pt_j|$ 
defined (in the figure the magnitude of the vector is naturally implied) in kinematical selection (i), 
for the leptonic and missing energy channels (the presence of lepton or $\met$ do not 
matter directly for the $\vec H_T$ distribution, however the number of jets does.) 
for $e$1 and $\nu$5 benchmark points. The distributions for other benchmarks  
in both the signal channels are somewhat similar.}
\label{hthtv}
\end{center}
\end{figure}

\end{itemize} 

After applying the cumulative selection from (a)--(j), we estimated the total 
SM background events to be found in the final column of Table \ref{tab:lepori}.
The significance for 100  $fb^{-1}$ integrated luminosity is approximately 
0.40 -- 0.12 $\sigma$. This is not good enough to ensure the finding 
of a Higgs boson. However, at the end of LHeC data accumulation with 
approximately 1000 $fb^{-1}$, the significances (quoted in the parenthesis) 
improve to become 1.3 -- 0.4$\sigma$. Notice that the significances 
do not scale accordingly with the Higgs masses. For example, the 
mass of the Higgs boson in $e$2 benchmark is less than the $e$3 benchmark, 
thus one would expect the larger cross-section in $e$2 benchmark. This 
is mainly because of the parameter spaces dependent couplings of the $Z Z h_1$ 
in the gauge boson fusion type of production. For the $e$2 benchmark, 
the values of $Z Z h_1$ coupling is less with respect to $e$3 benchmark. 

It is somewhat clear from Table \ref{tab:lepori} that the significances in the electron-channel   
are not very promising for the cut--based cumulative selections. To find better significances, 
we exploited some kind of optimization technique. First of all, we select the following  
kinematical variable: $\eta_{l}$ (rapidity of the forward lepton); 
$\Delta{\eta_{jl}}$ = $\eta_{j_f} -\eta_{l_f}$  
(the rapidity differences between the forward jet and forward lepton) ; $m_{\phi j}$ (three jet invariant
masses, i.e., two Higgs--boson candidate jets and one forward jet); $H_T$ (the scalar sum of the 
transverse momentum of of jets); $|\vec{H}_T|$ (the vector sum of the transverse momentum of of jets). 
The optimization starts with the events which passed the forward jet criterion (after selection criterion e).
The numerical values of all these kinematical variables are varied within a maximum and minimum ranges 
(by seeing the corresponding distributions). In particular, we varied the following kinematical 
ranges: upper values of $\eta_{l}$ in the range $(1.0,2.5)$ with step-size 0.1 and the lower values 
in $(-2.5,-1.0)$ with step-size 0.1; the upper values of $\Delta {\eta_{jl}}$ in $(0.0,1.5)$ 
with step-size 0.1, while the lower values in $(-6.0,-3.0)$ with step-size 0.1; $m_{\phi j}$ 
in $(80.0,180.0)$ with step-size 10.0; $H_T$ in $(70.0,140.0)$ with step-size 10.0; 
and $|\vec H_T|$  in $(30.0,60.0)$ with step-size 10.0. We have checked approximately $44695552$ 
numbers of combinations and estimated the numbers of signal events and the total SM backgrounds and finally 
the significances.

The kinematical selection configurations and their numerical values for which 
the significances (in the fifth and ninth column) for integrated luminosity of 100 fb $^{-1}$ 
are maximum is shown in Table \ref{tab:optlep}.  It turns out that for Higgs boson masses up to 82.0 GeV, 
the significances reach at approximately $2.1$$\sigma$. In the right panel, we demanded 
number of signal events at least 10 for the low luminosity option (for Higgs masses greater than 80.0 GeV, 
we found approximately 5 signal events) and for that the number of backgrounds is also large. 
It is clear that for the Higgs boson masses of approximately 65.0 GeV, the significances could 
be 0.60 (2.0)$\sigma$ for 100(1000) fb $^{-1}$.

As the above neutral current signal, discussed above, does not have large significances in simple cut-based 
selection, we turn our attention to the charged current mode. It is clear from the right panel 
of Fig.\ref{evtrate} that the events rates are high enough to consider this case.

\subsubsection{$\met +$ 3-jet: missing--energy channel}

We are now looking for the feasibility of finding the Higgs boson in $\met$ $+$ 3-jet. Like in the 
$e$ $+$ 3-jet discussed in the previous sections, we apply kinematical cuts 
to isolate the beyond SM-type Higgs boson signal from the backgrounds. These cuts are, however, slightly different.  
The number of signal and SM background events are tabulated in Table \ref{tab:met}.

\begin{itemize} 

\item {\bf A:} We first selected events containing at least three jets, i.e., $N_{jet} \gsim 3$. 
The distribution of the number of jets ($N_{jet}$) is shown (only for the 
$\nu$5 benchmark) in the left panel of Fig.\ref{njetbtag}. 
The jet reconstruction criterion are same as of the $e +$ 3-jet channel -- thus the 
jet efficiencies of the SM backgrounds are exactly same. The more 
massive the Higgs, the larger are the jet efficiencies as it is evident from the Table \ref{tab:met}.

\item {\bf B:} We required no presence of an electron in our event. If we find an electron  
with $p_T$ $>$ 15.0 GeV and $\eta$ $<$ 3.0, we reject such events, i.e., we are vetoing. 
This selection largely reduced the SM backgrounds as compared to the case with an explicit electron. 
For example, $ebbj$, $ebjj$ and $ejjj$ are reduced by approximately 77\%, 75\% and 80\%, while 
for $ett$ the reduction is close to 86.7\%.  

The $\nu jjj$, $\nu bjj$ and $\nu bbj$ are reduced by approximately 0.2\%, 1.0\% and 3.0\%, respectively. 
For the $\nu bbj$ case having an secondary electron from $B$-meson semileptonic decay is more 
probable than in $\nu bjj$ and much more than in $\nu jjj$. Such an expectation is, indeed, confirmed. 
The efficiencies of this selections for all the five signal benchmarks are nearly identical (approximately 
95.5\%) as this is not related directly with the Higgs boson masses.

\item {\bf C:} Like in the $e +$ 3-jet channel in the previous section, here also we 
demanded at least two $b$-tagged jets with the inclusion of a proper mis-tagging. 
The distributions of the number of $b$-tagged jets ($N_{b-tag}$) are shown in the 
right panel of Fig. \ref{njetbtag}. The efficiencies for the missing-energy signals 
and SM-backgrounds are similar to the $e$ $+$ 3-jet channel -- as the main difference 
between these two channels is the presence of an electron and missing energy. 

\item {\bf D:} After identifying the $b$-tagged jet with proper mis-tagging,  
like in the leptonic channel, we follow exactly similar procedure to reconstruct  
the non-SM Higgs boson ($m_{h_1}$) masses. The reconstructed di-$b$jet invariant mass, 
$m_{bb}$ tends to peak in the lower values than the benchmark values. 
Due to this the mass window selection reduces the number of events more for higher 
Higgs boson mass. As it is clear from Table \ref{tab:met}, for the 
$\nu$1 -- $\nu$5 benchmarks cases the efficiencies are approximately 56.1\%, 49.5\%, 38.0\%, 35.1\% and 26.6\%. They   
are decreasing (as mentioned earlier) with the increase of the Higgs boson masses in the respective benchmark points.
The $\nu bbj$ background survived by approximately, 28.4\%, 14.7\%, 9.9\%, 9.0\% and 6.1\% for the 
$\nu$1 -- $\nu$5 benchmark respectively -- follow similar pattern like the signal benchmarks. 
We also find the same patterns for $ebbj$ background -- 26.2\%, 25.4\%, 19.2\%, 18.3\% and 15.2\% 
(as the mass window selection depends upon the masses of the Higgs boson in the respective benchmarks). 
In addition, the $ebbj$ background has a Z-exchange diagram, i.e., $eZj$ with $Z\to b \bar b$. For the 
benchmark points  where the Higgs boson is close to the Z-boson mass, i.e., for $\nu$4 and $\nu$5, 
the reduction is maximal. For $ebjj$ the events survived are approximately, 27.4\%, 27.7\%, 18.1\%, 19.3\% and 12.0\% a
for $\nu$1 -- $\nu$5 benchmarks respectively, somewhat similar to $ebbj$. For $ebjj$ two more issues are important 
to mention, the $ebZ$ with $Z\to j \bar j$ and mis-tagging of jets. For $\nu tb$, the event survived are 
approximately, 19.0\% for all the benchmark points as many effects play together, e.g., top-quark decay, mistagging, 
differences between Higgs boson and W-boson masses in those benchmark points.

\item {\bf E:} Like in the electron channel selection (e), here we also  
demanded the presence of one forward jet. All the signal benchmarks survived by approximately 99\%. 
This criterion has a mild overall impact -- it rejects the irreducible backgrounds 
$\nu bbj$ ($\nu bjj$) by approximately 34.5\%, 21.0\%, 17.4\%, 16.7\% and 21.7\% 
(22.9\%, 33.0\%, 33.4\%, 33.4\% and 29.0\%) for the $\nu$1 -- $\nu$5 benchmarks respectively. 
The background $\nu jjj$ is reduced by approximately from 16.0\% -- 25.0 \% from $\nu$1 to $\nu$5 benchmark 
points.

\item {\bf F:} The signal contains now a neutrino and this explicitly leads to missing energy
(other than the jet-energy smearing and mis-measurements). The $\met$ is shown in 
the right panel of Fig.\ref{nlepmet}.  We demanded that $\met$ should be larger than 15 GeV. 
All the signal benchmark survived by 91\% or more. This criterion suppressed all the background  
except for $ett$. Most importantly, at this stage $ejjj$ turns out to be zero. 
The $ebjj$ is reduced by 65\% -- 35\%, while $ebbj$ is reduced by 65\% -- 55\% 
for all  the signal benchmarks. The relatively larger reduction in $ebbj$ is again due to 
the semi-leptonic decays of the two bottom quarks. 
The backgrounds $\nu bbj$ survived by 99\% for $\nu$1 benchmarks and for all other 
benchmarks it survived by more than 96\% while $\nu bjj$ ($\nu jjj$) is survived 
by approximately 96\% (99\%).

\item {\bf G:} With the identification of forward jet we reconstructed the 
three jet invariant mass of the $m_{bb j_f}$. The distribution is shown in Fig.\ref{mbbj} 
for the $\nu$5 benchmark. All other benchmarks have similar distributions with the peaks nearly 
at the same values around 120 GeV. Furthermore, we demanded that $m_{bb j_f}$ $>$ 210 GeV. 
The signal reduces by approximately 50\% to 45\% (from $\nu$1 to $\nu$5 benchmark points)  
while the background reduces severely, e.g., $\nu bbj$ reduces by approximately, 80\%.
The other irreducible backgrounds, $\nu tb$, $ebbj$ and $ett$ reduce by approximately  
95\%, 70\%-80\% and 96\%-93\%, respectively for all the signal benchmarks. 

\item {\bf H:} Like in the $e +$ 3-jet channel, we demanded $H_T$ $>$ 100 GeV. The  
event efficiencies for the signal and all backgrounds follow similar 
pattern for the lepton signal channel. This is due to the fact that this 
variable dependents on the number of jets and their transverse momentum 
and does not have an explicit dependence on neither electron nor neutrino. 
Except $ebbj$ (reduced by approximately 30\%) this selection hardly affects 
any other backgrounds.

At this stage, for all the signal benchmarks, the backgrounds of $\nu tb$ and $\nu jjj$ are of similar 
size to the signal events while $ebbj$ is slightly more than factor of two. The signals 
events are quite large and
the SM-backgrounds are manageable. This leads to quite high significances
to observe the Higgs boson signal in the missing energy channel.

\item {\bf I:} However, to reduce the $ebbj$ further we invoked the magnitude of $|\vec H_{T}|$ $>$ 50 GeV. 
The signals and $\nu t b$ are reduced approximately 50\% while $ebbj$ (except the $\nu$1  
benchmark, where it reduced by 85\%) goes to zero.

\end{itemize} 

After applying the cumulative selection from (A)--(I), we estimated the total 
SM background events to be found in the final column of Table \ref{tab:met}.
The significance for 100  $fb^{-1}$ integrated luminosity for $\nu$1, $\nu$2, $\nu$3, $\nu$4 and $\nu$5
benchmark points are is 3.3$\sigma$, 1.8$\sigma$, 1.5$\sigma$, 1.8$\sigma$ and 1.4$\sigma$, respectively. 
Thus except the first benchmark, with Higgs boson mass around 66.0 GeV, all other benchmarks 
are not good enough to be observed with a high confidence level with 
100$fb^{-1}$. We find that with 1000$fb^{-1}$, the significances of 
those benchmarks can be found with 10.6$\sigma$, 5.8$\sigma$, 4.9$\sigma$, 5.5$\sigma$ and 4.5$\sigma$. 
With our choices of simple cut-based selections and at the end of the LHeC, 
a 5.0$\sigma$ discovery would be possible for Higgs boson masses up to 90 GeV. 

\begin{table}[t!]
\centering
{\scriptsize
\begin{tabular}{||c||c|c|c|c||c|c|c|c||}
\hline
\hline
&&&&&&&&\\
BP,$m_h$&$\met$,$m_{\phi j}$,$H_T $,$\vec H_T$&S&B&${\cal S}$&$\met $,$m_{\phi j}$,$H_T$,$\vec H_T$&S&B&${\cal S}$\\
\hline
$\nu$1,65.93&$35,180,70,30$&44.9&90.7&4.7(14.9)&$10,170,70,60$&28.6 &63.9&3.57(11.4)\\  
$\nu$2,71.32&$35,180,70,30$&25.3&83.2&2.8(8.9)&$40,170,70,50$&19.6&75.7&2.3(7.2)\\  
$\nu$3,83.77&$30,180,70,30$&20.1&97.4&2.0(6.5)&$30,180,90,30$& 19.9&96.7&2.0(6.5)\\  
$\nu$4,88.07&$30,180,90,30$&23.5&97.6&2.4(7.6)&$35,180,120,30$& 19.6&92.3&2.0(6.5)\\  
$\nu$5,100.47&$30,180,100,30$&12.3&105.8&1.2(3.8)&$25,180,100,50$& 7.9&45.0&1.2(3.7)\\
\hline
\hline
\end{tabular}
}
\caption{The optimization of the signal channel with different sets of selection cuts, e.g., $\met$, 
$m_{\phi j}$, $H_T$, the magnitude of $\vec H_T$ (see text for details) with the best significance obtained 
for 100 fb $^{-1}$. The significances in the parenthesis are for 1000 fb $^{-1}$.}
\label{tab:metopt} 
\end{table}

Compared to the electron channel, the overall significances in the missing energy channel 
are higher for relatively large Higgs masses. However, likewise in the electron channel, here 
we have exploited the optimization by varying four different kinematical parameters, $\met$, 
$m_{\phi j}$, $H_T$ and the magnitude of $\vec H_T$. The optimization starts after the events have passed the forward 
jet criterion (after selection criterion E). The numerical values of all these kinematical 
variables are varied within a maximum and minimum range (by seeing the corresponding distributions). 
We varied the $\met$ in the ranges $(10.0,40.0)$ with step-size 5.0, together with $m_{\phi j}$, 
$H_T$ and the magnitude of $\vec H_T$ used in the electron channel with the same ranges. We have checked 
approximately $2464$ number of combinations and estimated the number of signal events and 
the total SM backgrounds and finally the significances. The kinematical configurations 
and their numerical values for which the significances (in the fifth and ninth column) 
for 100 fb $^{-1}$ are maximum is shown in Table \ref{tab:metopt}. It turns out that 
one can have $2.4$$\sigma$ up to Higgs boson masses of 88 GeV. The significances in the 
parenthesis are for 1000 fb $^{-1}$. It turns out that the Higgs boson with masses more 
than 90 can have significances approximately $5$$\sigma$.

\section{Conclusions}
\label{sec:conclude}

The discovery of the SM-like Higgs boson at LHC establishes the correct 
pattern of the electroweak symmetry breaking with a doublet. However, to overcome 
theoretical shortcomings, multiple doublets are naturally introduced in models  
beyond the SM -- both without and with supersymmetry. All such multiple 
doublet models predict the presence of more Higgses with different mass ranges. 
Hence looking for any kind of such Higgses in the present and upcoming collision 
experiments is a good opportunity to probe particle physics beyond the SM.

Among the experimental facilities which will soon become operational and potentially 
competitive to look for the intermediate mass Higgs bosons, is the  
LHeC facility (with $ep$ collisions) located at CERN expected to be operational in 2020.

In this work, we have considered the NMSSM model where the non-SM type 
Higgs boson with masses less than the SM-like Higgs boson is naturally 
possible with all the theoretical and most up-to-date 
experimental constraints from the low energy experiments as well as 
the supersymmetry and Higgs searches results from the LHC. 
The model has naturally a low mass lightest neutralino ($\lsp$) which serves as 
the possible candidate of the cold-dark matter. Apart from
the particle physics constraints mentioned above, 
our parameter space respects all the dark matter constraints, 
including WMAP and dark matter searches.

We have considered two different production mechanisms of this 
intermediate Higgs boson, namely, $e h_1 q_f$ and $\nu h_1 q_f$, 
and used the $h_1 \to b \bar b$ decay mode to find the 
Higgs boson signal. In our analysis, we performed a detail parameter 
space scanning using {\tt NMSSMTools v.5.0.1 } assuming that the 
second intermediate mass Higgs boson is of the SM-type. For the allowed parameters 
we estimated the production cross-section and the branching 
ratio to find the event rate at the LHeC facility.

The production processes under consideration are 
$e$ $+$ 3-jets and $\met$ $+$ 3-jets.
Two jets can originate from the Higgs boson decay, $h_1 \to b \bar b$ and 
we demanded both of them to be $b$-tagged (with proper mis-tagging from light-flavor and gluon jets) 
in the central rapidity region. The remaining jet originates
from the remnant of the proton fluxes which is likely to be  with 
large rapidity (in the forward or backward region). We considered the 
reducible and irreducible SM backgrounds (with charge-conjugation 
wherever appropriate) for the charge-current processes: $\nu tb$, $\nu bb j$, $\nu bjj$ and $\nu jjj$ 
and neutral current processes: $ebbj$, $ebjj$, $ett$ and $ejjj$.

We performed a full hadron--level Monte Carlo simulation using {\tt MadGraph/MadEvent} 
followed by {\tt PYTHIA} as the parton shower/hadronization 
event generator and its {\tt PYCELL} toy calorimeter in accordance with the LHeC 
detector parameters. We carefully implemented $b$-tagging, including mis-tagging 
of $c$-jets, light-flavor and gluon jets.

In both of the Higgs signal under consideration, we first applied the basic event 
characteristics, like number of jets, number of lepton, missing energy profile, 
number of $b$-tagged jets.  The kinematics for both the signals are very interesting
due to the fact that the Higgs boson is produced in the central rapidity region 
such that the two $b$-tagged jets are also central. We reconstructed the invariant mass of 
these two (or more) $b$-tagged jets and identified the best combinations where the 
reconstructed Higgs boson mass is close to the benchmark values and thereafter 
selected events only with slightly asymmetric choices with $m_{h_1} - 15 < m_{bb} < m_{h_1} + 5$ GeV. 
This selection reduces the SM backgrounds to a large extent and the invariant mass ensures 
the finding of the Higgs boson.

As a next step of our selection, by seeing the rapidity profiles, we identified the 
most energetic light-flavor forward jet ($j_f$) and demanded that the rapidity product 
of the forward lepton, i.e., electron ($j_e$) and forward jet should be negative, i.e., 
they must lie in the opposite hemisphere. Furthermore, we demanded that the rapidity gap 
between this forward jet and forward electron should satisfy $-5.5$ $<$ $\Delta \eta_{jf-ef}$ $<$ $0.5$. 
After this selection we calculated the three-jet invariant masses, 
$m_{bb j_f}$ which essentially gives the overall energy scale of the hard scattering. 
We demanded that this should be larger than 190 GeV and this selection helps 
to suppress backgrounds coming from $ebbj$, $ejjj$, $\nu tb$ and $ett$. 

After that, for the $e+3j$ signal channel we applied some $H_T$ selection to suppress the 
SM backgrounds further. This is not enough though to ensure a relatively high confidence level.  
Hence finally we exploited the $\vec H_T$ which leads to better 
significances, but not at the discovery level though, for the intermediate non-SM type Higgs boson.

In the $e+3j$ channel, which is naturally event rate limited, we found that to isolate 
the non-SM like Higgs signals, we can attain at most 0.4$\sigma$ with 100 $fb^{-1}$ for 
the $e$1 benchmark. With ten times more luminosity, for the same benchmark with 
$m_h$= 63.59 GeV, we would have the significance approximately 1.3$\sigma$.

In the $\met+3j$ channel with large event rate we can probe a much higher Higgs mass. 
We find that with 100 $fb^{-1}$ for the $\nu$1 benchmark the significance would 
be approximately 3.3$\sigma$. For the $\nu$4 benchmark (same as $e$5 benchmark) we can 
have the  significance approximately 1.8$\sigma$. With 1000$fb^{-1}$ for $\nu$4 benchmark, 
with Higgs masses 88.1 GeV, we would have 5.5$\sigma$. And for $\nu$5 benchmark, 
with Higgs masses of 100.5 GeV we found the significances of 4.5$\sigma$. 
Using a mere interpolation we would expect that one can have 5.0$\sigma$ significance 
up to the Higgs masses of around 90 GeV.

It should be noted that we have adopted a simple cut-based selection. If one would 
instead invoke more complex discriminators and/or use multi-variate analysis, 
we expect that the significance would be larger even with low luminosity option.

We have introduced a simple cut-based optimization method to enhance the Higgs 
boson mass reach in both the channels under consideration.
In the electron channel, by varying the most important kinematical variables,  
$\eta_l$, $\Delta \eta_{jl}$, $m_{\phi j}$, $H_T$ and the magnitude of $\vec {H_T}$ with the 
maximum and minimum ranges (by inspecting the respective distributions) we optimized the significances. We found 
that for 100 (1000) $fb^{-1}$ integrated luminosity, we can have the significances 
of 2.1$\sigma$ (6.6$\sigma$) for the $e$4 benchmark point with the 
Higgs boson masses of 82.2 GeV.

In the missing energy channel, by varying $\met$, $m_{\phi j}$, $H_T$ and magnitude of $\vec {H_T}$, for  
100 $fb^{-1}$ integrated luminosity, we can have the significances of 2.4$\sigma$(1.2$\sigma$) 
for the $\nu$4 ($\nu$5) benchmark point with Higgs boson masses of 88.1(100.5) GeV. 
With high luminosity option one would expect the 5$\sigma$ discovery at--least up to 
Higgs boson masses of 95 GeV.

To conclude, after the first few years of the LHeC run, and using more 
complex discriminators and using multi-variate analyses, we expect that 
in the $e+3j$ channel, non-SM type Higgs boson would be probed up to 85 -- 90 GeV. 
In the $\met+3j$ one can extend the reach at-least up to 95 GeV.

\subsubsection*{Acknowledgments}  

We are very much grateful for discussions with C.~Hugonie and U.~Ellwanger on {\tt NMSSMTools v5.0.1} 
and O. Mattelaer on {\tt MadGraph/MadEvent v2.4.3}. We are acknowledging the High Performance Computing 
(HPC) facility at Uniandes and J.~P.~Mallarino for many useful suggestion regarding HPC facility.


\begin{thebibliography}{500}

\bibitem{ATLASdiscovery}
G. Aad et al. [ATLAS Collaboration], Phys. Lett. B {\bf 716} (2012) 1. 

\bibitem{CMSdiscovery}
S. Chatrchyan et al. [CMS Collaboration], Phys. Lett. B {\bf 716} (2012) 30. 

\bibitem{Wells:2009kq} 
  J.~D.~Wells,
  arXiv:0909.4541 [hep-ph] ;
  I.~P.~Ivanov,
  arXiv:1702.03776 [hep-ph].

\bibitem{Grojean:2006wr} 
  C.~Grojean,
  CERN-PH-TH-2006-172.

\bibitem{Dev:2015vra} 
  P.~S.~Bhupal Dev, C.~H.~Lee and R.~N.~Mohapatra,
  J.\ Phys.\ Conf.\ Ser.\  {\bf 631}, no. 1, 012007 (2015)
  doi:10.1088/1742-6596/631/1/012007
  [arXiv:1503.04970 [hep-ph]].

\bibitem{Seidl:2004xw} 
  G.~Seidl,
  hep-ph/0409162 ; 
  C.~Csaki,
  hep-ph/0412339.


\bibitem{Branco:2011iw}
G.~C.~Branco, P.~M.~Ferreira, L.~Lavoura, M.~N.~Rebelo, M.~Sher and J.~P.~Silva,
  Phys.\ Rept.\  {\bf 516}, 1 (2012)
  [arXiv:1106.0034 [hep-ph]].

\bibitem{Gunion:2002zf}
J.~F.~Gunion and H.~E.~Haber,
  Phys.\ Rev.\ D {\bf 67}, 075019 (2003)
  [hep-ph/0207010].

\bibitem{Cordero-Cid:2013sxa}
  A.~Cordero-Cid, J.~Hernandez-Sanchez, C.~G.~Honorato, S.~Moretti, M.~A.~Perez and A.~Rosado,
  JHEP {\bf 1407}, 057 (2014)
  [arXiv:1312.5614 [hep-ph]];
O.~Felix-Beltran, F.~Gonzalez-Canales, J.~Hernandez-Sanchez, S.~Moretti, R.~Noriega-Papaqui and A.~Rosado,
  Phys.\ Lett.\ B {\bf 742}, 347 (2015)
  [arXiv:1311.5210 [hep-ph]];
 J.~Hernandez-Sanchez, S.~Moretti, R.~Noriega-Papaqui and A.~Rosado,
  JHEP {\bf 1307}, 044 (2013)
  [arXiv:1212.6818];
 J.~Hernandez-Sanchez, S.~Moretti, R.~Noriega-Papaqui and A.~Rosado,
  PoS CHARGED {\bf 2012}, 029 (2012)
  [arXiv:1302.0083];
 J.~Hernandez-Sanchez, C.~G.~Honorato, M.~A.~Perez and J.~J.~Toscano,
  Phys.\ Rev.\ D {\bf 85}, 015020 (2012)
  [arXiv:1108.4074 [hep-ph]];
  J.~L.~Diaz-Cruz and J.~J.~Toscano,
  Phys.\ Rev.\ D {\bf 62}, 116005 (2000)
  [hep-ph/9910233];
A.~G.~Akeroyd, S.~Moretti and J.~Hernandez-Sanchez,
  arXiv:1409.7596 [hep-ph];

\bibitem{Kane:1993td} 
  G.~L.~Kane, C.~F.~Kolda, L.~Roszkowski and J.~D.~Wells,
  Phys.\ Rev.\ D {\bf 49}, 6173 (1994)
  doi:10.1103/PhysRevD.49.6173
  [hep-ph/9312272].

\bibitem{nmssm}
H.-P. Nilles, M. Srednicki and D. Wyler, Phys. Lett. B {\bf 120} (1983) 346;
J.P. Derendinger and C.A. Savoy, Nucl. Phys. B {\bf 237} (1984) 307;
J.R. Ellis, J.F. Gunion, H.E. Haber, L. Roszkowski, and F. Zwirner, Phys. Rev.  D {\bf 39} (1989) 844; 
U. Ellwanger, Phys. Lett. B {\bf 303} (1993) 271. 

\bibitem{Drees:1988fc} 
  M.~Drees,
  Int.\ J.\ Mod.\ Phys.\ A {\bf 4}, 3635 (1989).
  doi:10.1142/S0217751X89001448

\bibitem{Franke:1995tc} 
  F.~Franke and H.~Fraas,
  Int.\ J.\ Mod.\ Phys.\ A {\bf 12}, 479 (1997)
  doi:10.1142/S0217751X97000529
  [hep-ph/9512366].

\bibitem{Maniatis:2009re} 
  M.~Maniatis,
  Int.\ J.\ Mod.\ Phys.\ A {\bf 25}, 3505 (2010)
  doi:10.1142/S0217751X10049827
  [arXiv:0906.0777 [hep-ph]].

\bibitem{review} 
  U.~Ellwanger, C.~Hugonie and A.~M.~Teixeira,
  Phys.\ Rept.\  {\bf 496}, 1 (2010)
  doi:10.1016/j.physrep.2010.07.001
  [arXiv:0910.1785 [hep-ph]].

\bibitem{Miller:2005qua} 
  D.~J.~Miller, S.~Moretti and R.~Nevzorov,
  hep-ph/0501139.

\bibitem{Barbieri:2013hxa} 
  R.~Barbieri, D.~Buttazzo, K.~Kannike, F.~Sala and A.~Tesi,
  Phys.\ Rev.\ D {\bf 87}, no. 11, 115018 (2013)
  doi:10.1103/PhysRevD.87.115018
  [arXiv:1304.3670 [hep-ph]];
  M.~Farina, M.~Perelstein and B.~Shakya,
  JHEP {\bf 1404}, 108 (2014)
  doi:10.1007/JHEP04(2014)108
  [arXiv:1310.0459 [hep-ph]].

\bibitem{Kim:1983dt}
  J.~E.~Kim and H.~P.~Nilles,
  Phys.\ Lett.\  B {\bf 138} (1984) 150.

\bibitem{Ellis:1988er}
  J.~R.~Ellis, J.~F.~Gunion, H.~E.~Haber, L.~Roszkowski and F.~Zwirner,
  Phys.\ Rev.\  D {\bf 39} (1989) 844.

\bibitem{Bechtle:2016kui} For e.g., 
  P.~Bechtle, H.~E.~Haber, S.~Heinemeyer, O.~Stål, T.~Stefaniak, G.~Weiglein and L.~Zeune,
  arXiv:1608.00638 [hep-ph];
  J.~Quevillon,
  arXiv:1405.2241 [hep-ph];
  B.~Dumont, J.~F.~Gunion and S.~Kraml,
Phys.\ Rev.\ D {\bf 89} (2014) 055018.  

\bibitem{Drechsel:2016jdg} 
  P.~Drechsel, L.~Galeta, S.~Heinemeyer and G.~Weiglein,
  arXiv:1601.08100 [hep-ph];
  P.~Drechsel,
  DESY-THESIS-2016-019;
  F.~Staub, P.~Athron, U.~Ellwanger, R.~Grober, M.~Muhlleitner, P.~Slavich and A.~Voigt,
  Comput.\ Phys.\ Commun.\  {\bf 202}, 113 (2016)
  doi:10.1016/j.cpc.2016.01.005
  [arXiv:1507.05093 [hep-ph]].

\bibitem{Goodsell:2016udb} 
  M.~D.~Goodsell and F.~Staub,
  arXiv:1604.05335 [hep-ph].

\bibitem{Ellwanger:2004gz} 
  U.~Ellwanger, J.~F.~Gunion, C.~Hugonie and S.~Moretti,
  hep-ph/0401228;
  V.~Barger, P.~Langacker, H.~S.~Lee and G.~Shaughnessy,
  Phys.\ Rev.\ D {\bf 73}, 115010 (2006)
  doi:10.1103/PhysRevD.73.115010
  [hep-ph/0603247];
  D.~J.~Miller, R.~Nevzorov and P.~M.~Zerwas,
  Nucl.\ Phys.\ B {\bf 681}, 3 (2004)
  doi:10.1016/j.nuclphysb.2003.12.021
  [hep-ph/0304049];
  U.~Ellwanger and M.~Rodriguez-Vazquez,
  JHEP {\bf 1602}, 096 (2016)
  doi:10.1007/JHEP02(2016)096
  [arXiv:1512.04281 [hep-ph]];
  J.~S.~Kim, D.~Schmeier and J.~Tattersall,
  Phys.\ Rev.\ D {\bf 93}, no. 5, 055018 (2016)
  doi:10.1103/PhysRevD.93.055018
  [arXiv:1510.04871 [hep-ph]];
  J.~Rathsman and T.~Rossler,
  Adv.\ High Energy Phys.\  {\bf 2012}, 853706 (2012)
  doi:10.1155/2012/853706
  [arXiv:1206.1470 [hep-ph]]; 
  R.~Enberg, R.~Pasechnik and O.~Stal,
  Phys.\ Rev.\ D {\bf 85}, 075016 (2012)
  doi:10.1103/PhysRevD.85.075016
  [arXiv:1112.4699 [hep-ph]];
  J.~F.~Gunion, Y.~Jiang and S.~Kraml,
Phys.\ Lett.\ B {\bf 710} (2012) 454;
  S.~Liebler, H.~Mantler and M.~Wiesemann,
  arXiv:1608.02949 [hep-ph];
  M.~Carena, H.~E.~Haber, I.~Low, N.~R.~Shah and C.~E.~M.~Wagner,
  Phys.\ Rev.\ D {\bf 93}, no. 3, 035013 (2016)
  doi:10.1103/PhysRevD.93.035013
  [arXiv:1510.09137 [hep-ph]];
  E.~Conte, B.~Fuks, J.~Guo, J.~Li and A.~G.~Williams,
  JHEP {\bf 1605}, 100 (2016)
  doi:10.1007/JHEP05(2016)100
  [arXiv:1604.05394 [hep-ph]];
  S.~F.~King, M.~Muhlleitner, R.~Nevzorov and K.~Walz,
Nucl.\ Phys.\ B {\bf 870} (2013) 323 ;
  S.~F.~King, M.~Muhlleitner and R.~Nevzorov,
Nucl.\ Phys.\ B {\bf 860} (2012) 207 ;
  N.~D.~Christensen, T.~Han, Z.~Liu and S.~Su,
JHEP {\bf 1308} (2013) 019 ; 
  J.~Cao, F.~Ding, C.~Han, J.~M.~Yang and J.~Zhu,
JHEP {\bf 1311} (2013) 018;
  G.~Belanger, U.~Ellwanger, J.~F.~Gunion, Y.~Jiang, S.~Kraml and J.~H.~Schwarz,
JHEP {\bf 1301} (2013) 069;
  U.~Ellwanger, J.~F.~Gunion, C.~Hugonie and S.~Moretti,
  hep-ph/0305109;
  D.~Das, U.~Ellwanger and A.~M.~Teixeira,
  JHEP {\bf 1204}, 067 (2012)
  doi:10.1007/JHEP04(2012)067
  [arXiv:1202.5244 [hep-ph]];
  U.~Ellwanger and C.~Hugonie,
  arXiv:1203.5048 [hep-ph];
  K.~S.~Jeong, Y.~Shoji and M.~Yamaguchi,
  JHEP {\bf 1411}, 148 (2014)
  doi:10.1007/JHEP11(2014)148
  [arXiv:1407.0955 [hep-ph]];
  B.~Dutta, Y.~Gao and B.~Shakya,
  Phys.\ Rev.\ D {\bf 91}, no. 3, 035016 (2015)
  doi:10.1103/PhysRevD.91.035016
  [arXiv:1412.2774 [hep-ph]]; 
  S.~Baum, K.~Freese, N.~R.~Shah and B.~Shakya,
  arXiv:1703.07800 [hep-ph].

\bibitem{Dreiner:2012ec} 
  H.~K.~Dreiner, F.~Staub and A.~Vicente,
  Phys.\ Rev.\ D {\bf 87}, no. 3, 035009 (2013)
  doi:10.1103/PhysRevD.87.035009
  [arXiv:1211.6987 [hep-ph]].

\bibitem{Ellwanger:2013rsa} 
  U.~Ellwanger,
  JHEP {\bf 1311}, 108 (2013)
  doi:10.1007/JHEP11(2013)108
  [arXiv:1309.1665 [hep-ph]].

\bibitem{Domingo:2015eea} 
  F.~Domingo and G.~Weiglein,
  JHEP {\bf 1604}, 095 (2016)
  doi:10.1007/JHEP04(2016)095
  [arXiv:1509.07283 [hep-ph]].

\bibitem{Barducci:2015zna} 
  D.~Barducci, G.~Belanger, C.~Hugonie and A.~Pukhov,
  JHEP {\bf 1601}, 050 (2016)
  doi:10.1007/JHEP01(2016)050
  [arXiv:1510.00246 [hep-ph]].

\bibitem{cpodd} 
  N.~E.~Bomark, S.~Moretti, S.~Munir and L.~Roszkowski,
  PoS EPS {\bf -HEP2015}, 162 (2015)
  [arXiv:1510.02661 [hep-ph]];N.~E.~Bomark, S.~Moretti, S.~Munir and L.~Roszkowski,
  JHEP {\bf 1502}, 044 (2015)
  doi:10.1007/JHEP02(2015)044
  [arXiv:1409.8393 [hep-ph]];
  S.~Moretti and S.~Munir,
  Adv.\ High Energy Phys.\  {\bf 2015}, 509847 (2015)
  doi:10.1155/2015/509847
  [arXiv:1505.00545 [hep-ph]]; 
  M.~M.~Almarashi and S.~Moretti,
  Phys.\ Rev.\ D {\bf 85}, 017701 (2012)
  doi:10.1103/PhysRevD.85.017701
  [arXiv:1109.1735 [hep-ph]];
  M.~Almarashi and S.~Moretti,
  Phys.\ Rev.\ D {\bf 84}, 015014 (2011)
  doi:10.1103/PhysRevD.84.015014
  [arXiv:1105.4191 [hep-ph]];
  N.~E.~Bomark, S.~Moretti, S.~Munir and L.~Roszkowski,
  PoS Charged {\bf 2014}, 029 (2015)
  [arXiv:1412.5815 [hep-ph]];
  N.~E.~Bomark, S.~Moretti and L.~Roszkowski,
  J.\ Phys.\ G {\bf 43}, no. 10, 105003 (2016)
  doi:10.1088/0954-3899/43/10/105003
  [arXiv:1503.04228 [hep-ph]];
  M.~M.~Almarashi and S.~Moretti,
  Phys.\ Rev.\ D {\bf 83}, 035023 (2011)
  doi:10.1103/PhysRevD.83.035023
  [arXiv:1101.1137 [hep-ph]].


\bibitem{Potter:2015wsa} 
  C.~T.~Potter,
  Eur.\ Phys.\ J.\ C {\bf 76}, no. 1, 44 (2016)
  doi:10.1140/epjc/s10052-015-3867-x
  [arXiv:1505.05554 [hep-ph]].
 
\bibitem{Guchait:2015owa} 
  M.~Guchait and J.~Kumar,
  Int.\ J.\ Mod.\ Phys.\ A {\bf 31}, no. 12, 1650069 (2016)
  doi:10.1142/S0217751X1650069X
  [arXiv:1509.02452 [hep-ph]];
  M.~Guchait and J.~Kumar,
  arXiv:1608.05693 [hep-ph].

\bibitem{Khachatryan:2015nba} 
  V.~Khachatryan {\it et al.} [CMS Collaboration],
  JHEP {\bf 1601}, 079 (2016)
  doi:10.1007/JHEP01(2016)079
  [arXiv:1510.06534 [hep-ex]].

\bibitem{cern:lhec} https://lhec.web.cern.ch

\bibitem{heraphys} See for e.g., M.Klein, R.Yoshida: Collider Physics at HERA
Prog.Part.Nucl.Phys. 61(2008)343-393.

\bibitem{Han:2009pe} T.~Han and B.~Mellado,
 Phys.\ Rev.\ D {\bf 82}, 016009 (2010)
 [arXiv:0909.2460 [hep-ph]].

\bibitem{Sarmiento-Alvarado:2014eha}
  I.~A.~Sarmiento-Alvarado, A.~O.~Bouzas and F.~Larios,
  arXiv:1412.6679 [hep-ph] ;
  A.~O.~Bouzas and F.~Larios,
  J.\ Phys.\ Conf.\ Ser.\  {\bf 651}, no. 1, 012004 (2015).
  doi:10.1088/1742-6596/651/1/012004

\bibitem{Das:2015kea} 
  S.~P.~Das, J.~Hernandez-Sanchez, S.~Moretti, A.~Rosado and R.~Xoxocotzi,
  Phys.\ Rev.\ D {\bf 94}, no. 5, 055003 (2016)
  doi:10.1103/PhysRevD.94.055003
  [arXiv:1503.01464 [hep-ph]].

\bibitem{Drees:2012ji} 
  M.~Drees and G.~Gerbier,
  arXiv:1204.2373 [hep-ph].

\bibitem{Sanabria:2014yva} 
  J.~C.~Sanabria,
  Rev.\ Acad.\ Colomb.\ Cienc.\  {\bf 38}, 34 (2014).


\bibitem{Planck:2015xua}
{\bf Planck} Collaboration, P.~Ade et~al., {\it Planck 2015 results. XIII.
  Cosmological parameters},  http://arxiv.org/abs/1502.01589 {{\tt
  arXiv:1502.01589}}.

\bibitem{Belanger:2005kh} 
  G.~Belanger, F.~Boudjema, C.~Hugonie, A.~Pukhov and A.~Semenov,
  JCAP {\bf 0509}, 001 (2005)
  doi:10.1088/1475-7516/2005/09/001
  [hep-ph/0505142];
  J.~Cao, Y.~He, L.~Shang, W.~Su, P.~Wu and Y.~Zhang,
  JHEP {\bf 1610}, 136 (2016)
  doi:10.1007/JHEP10(2016)136
  [arXiv:1609.00204 [hep-ph]];
  J.~Cao, Y.~He, L.~Shang, W.~Su and Y.~Zhang,
  JHEP {\bf 1608}, 037 (2016)
  doi:10.1007/JHEP08(2016)037
  [arXiv:1606.04416 [hep-ph]];
  D.~Das and U.~Ellwanger,
  JHEP {\bf 1009}, 085 (2010)
  doi:10.1007/JHEP09(2010)085
  [arXiv:1007.1151 [hep-ph]];
  Q.~F.~Xiang, X.~J.~Bi, P.~F.~Yin and Z.~H.~Yu,
  arXiv:1606.02149 [hep-ph];
  J.~Kozaczuk and S.~Profumo,
  Phys.\ Rev.\ D {\bf 89}, no. 9, 095012 (2014)
  doi:10.1103/PhysRevD.89.095012
  [arXiv:1308.5705 [hep-ph]].

\bibitem{Badziak:2015exr} 
  M.~Badziak, M.~Olechowski and P.~Szczerbiak,
  JHEP {\bf 1603}, 179 (2016)
  doi:10.1007/JHEP03(2016)179
  [arXiv:1512.02472 [hep-ph]].

\bibitem{Horiuchi:2016tqw} 
  S.~Horiuchi, O.~Macias, D.~Restrepo, A.~Rivera, O.~Zapata and H.~Silverwood,
  JCAP {\bf 1603}, no. 03, 048 (2016)
  doi:10.1088/1475-7516/2016/03/048
  [arXiv:1602.04788 [hep-ph]].

\bibitem{Alvares:2012qv} 
  J.~D.~Ruiz-Alvarez, C.~A.~de S.Pires, F.~S.~Queiroz, D.~Restrepo and P.~S.~Rodrigues da Silva,
  Phys.\ Rev.\ D {\bf 86}, 075011 (2012)
  doi:10.1103/PhysRevD.86.075011
  [arXiv:1206.5779 [hep-ph]].

\bibitem{Das:2014kwa} 
  S.~P.~Das, M.~Guchait and D.~P.~Roy,
  Phys.\ Rev.\ D {\bf 90}, no. 5, 055011 (2014)
  doi:10.1103/PhysRevD.90.055011
  [arXiv:1406.6925 [hep-ph]].

\bibitem{Florez:2016lwi} 
  A.~Florez, L.~Bravo, A.~Gurrola, C.~Avila, M.~Segura, P.~Sheldon and W.~Johns,
  Phys.\ Rev.\ D {\bf 94}, no. 7, 073007 (2016)
  doi:10.1103/PhysRevD.94.073007
  [arXiv:1606.08878 [hep-ph]].

\bibitem{Diessner:2015iln} 
  P.~Diessner, J.~Kalinowski, W.~Kotlarski and D.~Stoeckinger,
  JHEP {\bf 1603}, 007 (2016)
  doi:10.1007/JHEP03(2016)007
  [arXiv:1511.09334 [hep-ph]].

\bibitem{Hagimoto:2015tua} 
  K.~Hagimoto, T.~Kobayashi, H.~Makino, K.~i.~Okumura and T.~Shimomura,
  JHEP {\bf 1602}, 089 (2016)
  doi:10.1007/JHEP02(2016)089
  [arXiv:1509.05327 [hep-ph]].

\bibitem{JeanLouis:2009du} 
  C.-C.~Jean-Louis and G.~Moreau,
  J.\ Phys.\ G {\bf 37}, 105015 (2010)
  doi:10.1088/0954-3899/37

\bibitem{Ellwanger:2004xm} 
  U.~Ellwanger, J.~F.~Gunion and C.~Hugonie,
  JHEP {\bf 0502} (2005) 066.

\bibitem{micr43} D. Barducci, G. Belanger, J. Bernon, F. Boudjema, J. DaSilva, S. Kraml, U. Laa, 
A. Pukhov,  arXiv:1606.03834 [hep-ph]

\bibitem{Belanger:2013oya} 
  G.~Belanger, F.~Boudjema, A.~Pukhov and A.~Semenov,
  Comput.\ Phys.\ Commun.\  {\bf 185}, 960 (2014)
  doi:10.1016/j.cpc.2013.10.016
  [arXiv:1305.0237 [hep-ph]].

\bibitem{Zheng:2014loa} 
  S.~Zheng,
  Eur.\ Phys.\ J.\ C {\bf 75}, no. 5, 195 (2015)
  doi:10.1140/epjc/s10052-015-3416-7
  [arXiv:1405.6907 [hep-ph]].

\bibitem{Sanabria:2015mxh} 
  J.~C.~Sanabria [ATLAS and CMS Collaborations],
  Nucl.\ Part.\ Phys.\ Proc.\  {\bf 267-269}, 25 (2015).
  doi:10.1016/j.nuclphysbps.2015.10.078

\bibitem{Khachatryan:2016vau} 
  G.~Aad {\it et al.} [ATLAS and CMS Collaborations],
  JHEP {\bf 1608}, 045 (2016)
  doi:10.1007/JHEP08(2016)045
  [arXiv:1606.02266 [hep-ex]].

\bibitem{Khachatryan:2016whc} 
  V.~Khachatryan {\it et al.} [CMS Collaboration],
  JHEP {\bf 1702}, 135 (2017)
  doi:10.1007/JHEP02(2017)135
  [arXiv:1610.09218 [hep-ex]].

\bibitem{Aad:2015pla} 
  G.~Aad {\it et al.} [ATLAS Collaboration],
  JHEP {\bf 1511}, 206 (2015)
  doi:10.1007/JHEP11(2015)206
  [arXiv:1509.00672 [hep-ex]].

\bibitem{Butter:2015fqa} 
  A.~Butter, T.~Plehn, M.~Rauch, D.~Zerwas, S.~Henrot-Versillé and R.~Lafaye,
  Phys.\ Rev.\ D {\bf 93}, 015011 (2016)
  doi:10.1103/PhysRevD.93.015011
  [arXiv:1507.02288 [hep-ph]].


\bibitem{ALEPH:2005ab} {\bf SLD Electroweak Group, DELPHI, ALEPH, SLD, SLD Heavy Flavour Group, OPAL,
  LEP Electroweak Working Group, L3} Collaboration, S.~Schael et~al., {\it
  Precision electroweak measurements on the $Z$ resonance},  {\em Phys. Rept.}
  {\bf 427} (2006) 257--454. 

\bibitem{Bennett:2004pv}
{\bf Muon g-2} Collaboration, G.~W. Bennett et~al., {\it {Measurement of the
  negative muon anomalous magnetic moment to 0.7 ppm}},  {\em Phys. Rev. Lett.}
  {\bf 92} (2004) 161802.

\bibitem{Amhis:2014hma}
{\bf Heavy Flavor Averaging Group (HFAG)} Collaboration, Y.~Amhis et~al., {\it
  {Averages of $b$-hadron, $c$-hadron, and $\tau$-lepton properties as of
  summer 2014}}, {{\tt arXiv:1412.7515}}.

\bibitem{Lees:2012ju}
{\bf BaBar} Collaboration, J.~P. Lees et~al., {\it Evidence of $B^+ \to
  \tau^+\nu$ decays with hadronic B tags},  {\em Phys. Rev.} {\bf D88} (2013),
  no.~3 031102.

\bibitem{Aaboud:2016tnv} 
  M.~Aaboud {\it et al.} [ATLAS Collaboration],
  Phys.\ Rev.\ D {\bf 94}, no. 3, 032005 (2016)
  doi:10.1103/PhysRevD.94.032005
  [arXiv:1604.07773 [hep-ex]].


\bibitem{Alwall:2014hca} 
  J.~Alwall {\it et al.},
  JHEP {\bf 1407}, 079 (2014)
  doi:10.1007/JHEP07(2014)079
  [arXiv:1405.0301 [hep-ph]].


\bibitem{Bruening:2013bga}
  O.~Bruening and M.~Klein,
  Mod.\ Phys.\ Lett.\ A {\bf 28}, no. 16, 1330011 (2013)
  [arXiv:1305.2090 [physics.acc-ph]].

\bibitem{AbelleiraFernandez:2012ty}
  J.~L.~Abelleira Fernandez {\it et al.}  [LHeC Study Group Collaboration],
  ``On the Relation of the LHeC and the LHC,''
  arXiv:1211.5102 [hep-ex].

\bibitem{AbelleiraFernandez:2012cc}
  J.~L.~Abelleira Fernandez {\it et al.}  [LHeC Study Group Collaboration],
  ``A Large Hadron Electron Collider at CERN: Report on the Physics and Design Concepts for Machine and Detector,''
  J.\ Phys.\ G {\bf 39}, 075001 (2012)
  [arXiv:1206.2913 [physics.acc-ph]].

\bibitem{lheclumi} See for e.g. R B Appleby et al 2013 J.Phys.G: Nucl.Part.Phys.40 125004.

\bibitem{Ball:2013hta} 
  R.~D.~Ball {\it et al.} [NNPDF Collaboration],
  Nucl.\ Phys.\ B {\bf 877}, 290 (2013)
  doi:10.1016/j.nuclphysb.2013.10.010
  [arXiv:1308.0598 [hep-ph]];
  R.~D.~Ball {\it et al.} [NNPDF Collaboration],
  JHEP {\bf 1504}, 040 (2015)
  doi:10.1007/JHEP04(2015)040
  [arXiv:1410.8849 [hep-ph]].

\bibitem{Pumplin:2002vw} 
  J.~Pumplin, D.~R.~Stump, J.~Huston, H.~L.~Lai, P.~M.~Nadolsky and W.~K.~Tung,
  JHEP {\bf 0207}, 012 (2002)
  [hep-ph/0201195].

\bibitem{pythia}
T.~Sjostrand, S.~Mrenna and P.~Z.~Skands,
  ``PYTHIA 6.4 Physics and Manual,''
  JHEP {\bf 0605}, 026 (2006)
  [hep-ph/0603175].

\bibitem{effw} 
R.~N. Cahn and S.Dawson,  Phys. Lett. B {\bf 136} (1984) 196;
M.~S. Chanowitz and  M.~K.Gaillard, Phys. Lett. B {\bf 142} (1984) 85; 
G.~L.Kane and W.~W.Repko and W.~B.Rolnick, Phys. Lett. B {\bf 148} (1984) 367.

\bibitem{Agrawal:2013qka} 
  P.~Agrawal, S.~Bandyopadhyay and S.~P.~Das,
  arXiv:1308.6511 [hep-ph].

\bibitem{Agrawal:2013owa} 
  P.~Agrawal, S.~Bandyopadhyay and S.~P.~Das,
  Phys.\ Rev.\ D {\bf 88}, no. 9, 093008 (2013)
  doi:10.1103/PhysRevD.88.093008
  [arXiv:1308.3043 [hep-ph]].

\bibitem{Das:2010ds}
  S.~P.~Das and M.~Drees,
  Phys.\ Rev.\ D {\bf 83}, 035003 (2011)
  [arXiv:1010.3701 [hep-ph]];
  S.~P.~Das and M.~Drees,
  J.\ Phys.\ Conf.\ Ser.\  {\bf 259}, 012071 (2010)
  [arXiv:1010.2129 [hep-ph]];
  S.~P.~Das, A.~Datta and M.~Drees,
  AIP Conf.\ Proc.\  {\bf 1078}, 223 (2009)
  [arXiv:0809.2209 [hep-ph]].  

\end{thebibliography}
\end{document}